\documentclass[review,3p,doublespacing,sort&compress,12pt]{elsarticle}

\usepackage{graphicx}
\usepackage{subfigure}

\usepackage{amssymb}
\usepackage{stmaryrd}
\usepackage{amsmath}
\usepackage{bm}
\usepackage{tabularx}
\usepackage{setspace}
\usepackage{longtable}
\usepackage[export]{adjustbox}
\usepackage{color}
\usepackage[flushleft]{threeparttable}

\usepackage[colorlinks=true]{hyperref}

\newcommand{\abs}[1]{\lvert#1\rvert}
\newcommand{\grad}{\nabla}
\usepackage{xcolor}
\newcommand{\firstreviewer}[1]{{\color{red}#1}}
\renewcommand{\firstreviewer}[1]{#1}
\newcommand{\secondreviewer}[1]{{\color{blue}#1}}
\renewcommand{\secondreviewer}[1]{#1}
\newcommand{\changes}[1]{{\color{green}#1}}
\renewcommand{\changes}[1]{#1}

\newcommand{\removeDetails}[1]{}

\makeindex

\begin{document}

\begin{frontmatter}



\title{Recurrent Neural Networks (RNNs) with \secondreviewer{dimensionality reduction} and break down in computational mechanics; application to multi-scale localization step}  
\author[ulg]{Ling Wu }\ead{L.Wu@ulg.ac.be}
\author[ulg]{Ludovic Noels \corref{cor}}\ead{L.Noels@ulg.ac.be}

\cortext[cor]{Corresponding author}
\address[ulg]{University of Liege, Department of Mechanical and Aerospace Engineering, Computational \& Multiscale Mechanics of Materials, All\'ee de la d\'ecouverte 9, B-4000 Li\`ege, Belgium\\
  \vspace{12pt}
 Preprint submitted to Computer Methods in Applied Mechanics and Engineering. (C) 2021; Licensed under the Creative Commons (CC-BY-NC-ND); formal publication on: \href{doi.dx.org/10.1016/j.cma.2021.114476}{10.1016/j.cma.2021.114476}}
\begin{abstract}

Artificial Neural Networks (NNWs) are appealing functions to substitute high dimensional and non-linear history-dependent problems in computational mechanics since they offer the possibility to drastically reduce the computational time.
This feature has recently been exploited in the context of multi-scale simulations, in which the NNWs serve as surrogate model of micro-scale finite element resolutions.
Nevertheless, in the literature, mainly the macro-stress-macro-strain response of the meso-scale boundary value problem was considered and the micro-structure information could not be recovered in a so-called localization step.
In this work, we develop Recurrent Neural Networks (RNNs) as surrogates of the RVE response while being able to recover the evolution of the local micro-structure state variables for complex loading scenarios. 
The main difficulty is the high dimensionality of the RNNs output which consists in the internal state variable distribution in the micro-structure.
We thus propose and compare several surrogate models based on a \secondreviewer{dimensionality reduction}: i) direct RNN modeling with implicit NNW \secondreviewer{dimensionality reduction}, ii) RNN with PCA \secondreviewer{dimensionality reduction}, and iii) RNN with PCA \secondreviewer{dimensionality reduction} and \secondreviewer{dimensionality break down}, \emph{i.e.} the use of several RNNs instead of a single one.
Besides, we optimize the sequential training strategy of the latter surrogate for GPU usage in order to speed up the process. 
Finally, through RNN modeling of the principal components coefficients, the connection between the physical state variables and the hidden variables of the RNN is revealed, and exploited in order to select the hyper-parameters of the RNN-based surrogate models in their design stage.\\
\end{abstract}

\begin{keyword}
Recurrent neural networks \sep Multi-scale \sep \secondreviewer{Dimensionality Reduction} \sep Localization step \sep History-dependence \sep High dimensionality
\end{keyword}

\end{frontmatter}

\section{Introduction}

\begin{figure}[!htb]
	\centering
	\subfigure[]{\includegraphics[scale=0.8]{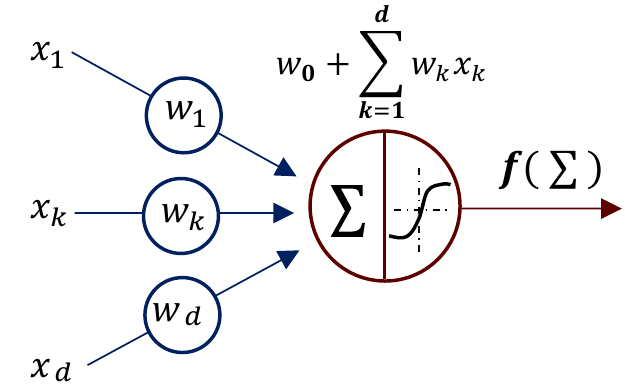}\label{fig:Neural}}\quad
	\subfigure[]{\includegraphics[scale=0.7]{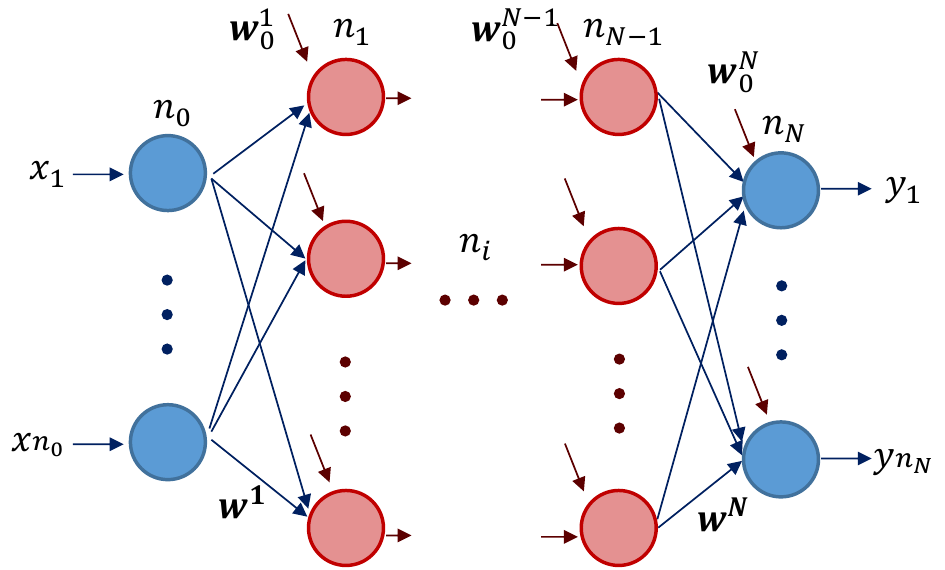}\label{fig:NNW}}
	\caption{(a) An artificial neuron; (b) An artificial Neural Network with the weights ${w}^i_{kj}$ and biases ${w}^i_{0_j}$ defining the weights matrix $\mathbf{W}$ of the feed-forward neural network.}\label{fig:ANNW}
\end{figure}

With the rapid development of data science in the recent years, Artificial Neural Networks (NNWs) have increasingly attracted attention in the field of computational mechanics, especially in the cases of complex material behavior and/or when intensive computation is required like in multi-scale \firstreviewer{analyses}.
A NNW is actually a network of artificial neurons, which perform a weighted sum operation on the $d$ input values ($w_0+\sum_{k=1}^{d} w_k x_k$) in order to produce an output value through an activation function of the weighted sum $f(w_0+\sum_{k=1}^{d} w_k x_k)$, see Fig. \ref{fig:Neural}.
The most commonly used architecture is the feed-forward neural network in which the information moves only along the forward direction, from the input nodes and to the output nodes \emph{via} $N-1$ hidden layers, see Fig. \ref{fig:NNW}.
The weights ${w}^i_{kj}$, with $i=1,\,\ldots,\,N;\; k=1,\,\ldots,\,n_{i-1} \text{ and }j= 1,\,\ldots,\,n_i$, where $n_{i-1}$ and $n_i$ are respectively the numbers of entries and outputs of layer $i$, and biases ${w}^i_{0_j}$, with $i=1,\,\ldots,\,N \text{ and }j= 1,\,\ldots,\,n_i$, see the notations in Fig. \ref{fig:NNW}, are obtained through the so-called training of the NNW. The weights ${w}^i_{kj}$ and biases ${w}^i_{0_j}$ define the weights matrix $\mathbf{W}$ of the feed-forward neural network.

NNWs can be regarded as a general non-linear mapping with high computational efficiency.
In computational mechanics, computationally expensive physical based modeling can be substituted by efficient NNWs to accelerate the numerical \firstreviewer{analyses}, such as \changes{non-linear history-dependent evolution laws like visco-plasticity \cite{furukawa1998implicit}, cyclic plasticity \cite{FURUKAWA2004195}, constitutive laws of interface \cite{WANG2019216, fernandez2020application} or constitutive laws of non-linear material behaviors \cite{LEFIK20033265,Hashash2004,ZHANG2020102732,JUNG2006608}.}
In \cite{WU2020112693}, NNWs were adopted to substitute complex material homogenization constitutive laws to accelerate the massive micro-mechanics modelings involved in Bayesian inference\changes{. In} \cite{LEFIK20021699} NNWs served as a surrogate for an elasto-plastic material model for parameters identification.
NNW was also adopted as a surrogate of the damaged-elastic response of meso-scale volumes of bones in \cite{hambli2011multiscale}.

\begin{figure}[htb]
	\centering
	\includegraphics[scale=0.5]{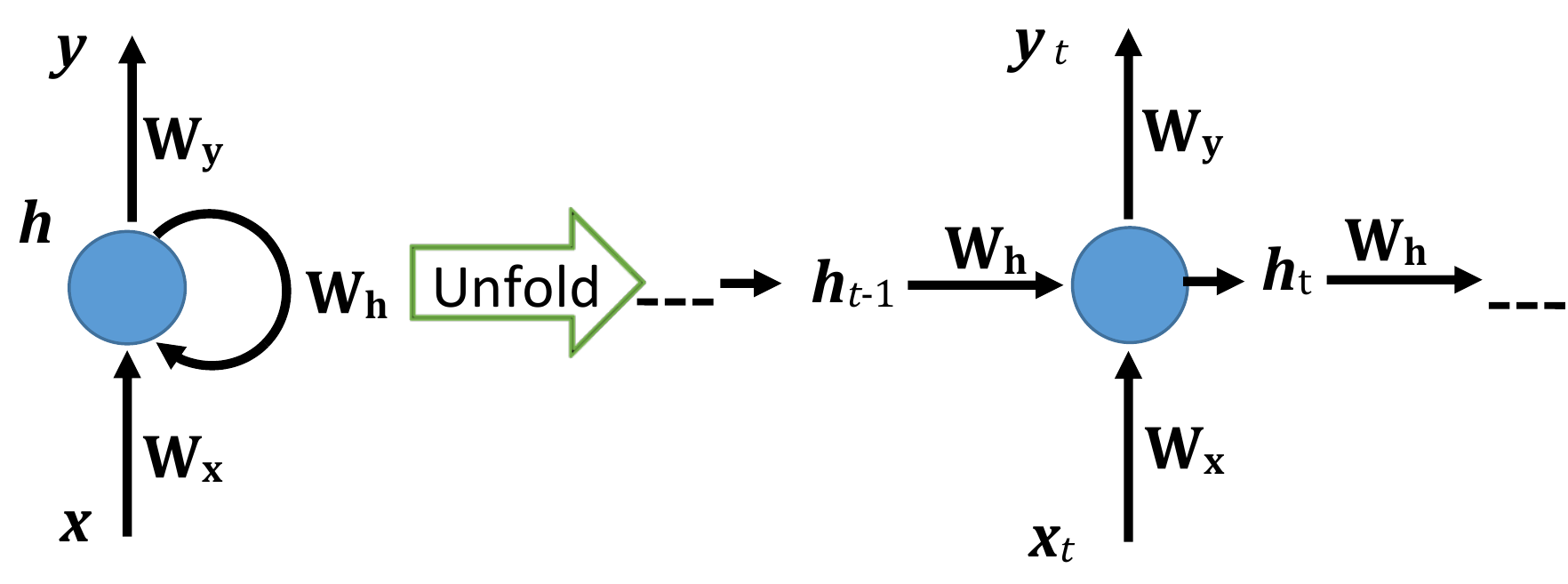}
	\caption{Recurrent Neural Network, with hidden variables passing trough the input series; $\mathbf{W_x}$, $\mathbf{W_y}$ and $\mathbf{W_h}$ are the weights matrix related to the linear operations on the input vector $\bm{x}$, output vector $\bm{y}$ and hidden variables $\bm{h}$, respectively. In the recurrent neural network, the operations at step $t$ use the hidden variables evaluated at step $t-1$.}\label{fig:RNN} 
\end{figure}

In a large variety of engineering and industrial fields, multi-scale methods \cite{NOELS2016,Yvonnet2019} were extensively developed as an alternative to the direct one-scale \firstreviewer{analyses} of structures made of heterogeneous materials.
Among the existing different multi-scale methods, the multi-scale analysis based on computational homogenization, which is usually called $\text{FE}^2$ analysis \cite{FEYEL1999344, Miehe1999, TERADA20002285, Kouznetsova2001}, is the most versatile method.
However, the tremendous demand in numerical resources in terms of time and memory limits the applicability of the $\text{FE}^2$ multi-scale simulations to reduced size problems. 
\changes{In order to improve the computational efficiency, feed-forward neural networks have been used as surrogate models in $\text{FE}^2$ \firstreviewer{analyses}, either by approximating the strain energy density surfaces as suggested in \cite{Le2015, BESSA2017633}, the stress-strain responses as achieved in \cite{Fritzen2019, UNGER20081994, SETTGAST2020102624}, or functions of both the current stress and the plastic dissipation density \cite{ZHANG2020102732}}.
Although NNWs \changes{have been shown} to be reliable surrogate models in elasticity and in non-linear elasticity \cite{Le2015}, when it comes to irreversible behaviors the loading history plays an important role in RVE response involving more difficulties both in the NNW architecture definition and in its training.
Some state variables are needed to account for the loading history at the meso-scale level, \emph{i.e.} on the RVE, and serve as a part of the input, beside the strain variables. In \cite{UNGER20081994}, state variables were defined for the decision of loading/unloading and a feed-forward network was used to extract meso-scale resolution for multi-scale failure analyses, but only 1D loading conditions were considered at the macro-scale.
In \cite{SETTGAST2020102624}, meso-scale plastic strains were used as state variables and were updated at each loading step through an empirical model or in combination with another feed-forward neural network.
Free energy and dissipation rate were represented by feed-forward NNWs in \cite{MASI2021104277} in order to obtain a thermo-dynamically consistent surrogate of an elasto-plastic law, but the method was not applied in the context of multi-scale analysis.
We refer  to the recent review by \cite{ROCHA2020103995} for a critical comparison in the context of composite materials.
Besides, the computational efficiency of the \changes{NNWs} paves the way to stochastic multi-scale resolutions as they can be used to build stochastic surrogate models which account for uncertainties in the micro-structure. Three-dimensional deep convolution neural networks (3D-CNN) have been trained in \cite{RAO2020109850} using a database consisting of spherical inclusion micro-structure images and of the predicted homogenized elasticity tensors obtained by computational homogenization. In the context of non-linear electrical conduction, a hybrid neural-network-interpolation was developed in \cite{ma14112875} by interpolating database obtained at given volume fractions, allowing uncertainties related to non-homogeneous distributions of volume fractions to be accounted for at the higher scale.

Instead of using feed-forward networks to reproduce state variables at each numerical time step, a special kind of NNWs, \emph{i.e.} Recurrent Neural Networks (RNNs), can be considered.
The idea behind RNNs is to make use of sequential information: RNNs are called recurrent because they perform the same task on every step input of a sequence, with the output being dependent on the previous evaluations.
A typical illustration of RNNs is presented in Fig. \ref{fig:RNN}, in which $\bm{x}$ and $\bm{y}$ are respectively the input and output variables and $\bm{h}$ are called hidden variables. The hidden variables are comparable to the state variables in history-dependent constitutive laws, and capture information about what has been calculated so far.
When using RNNs to substitute computational homogenization, the difficulty inherent to the definition of history-dependent state variables \changes{at the meso-scale} can be avoided, and the historical variables can be extracted automatically during the RNNs training with sequential data.
In \cite{GHAVAMIAN2019112594}, Long Short Term Memory networks (LSTMs) were used to study cyclic loading of elasto-visco-plastic micro-structures. In \cite{KOEPPE2020113088}, RNNs were used to predict the inside and boundary history-dependent responses of sub-structures.
 RNNs with Gated Recurrent Unit (GRU) were used as \changes{surrogate models} for the history-dependent response of RVEs in \cite{Mozaffar26414,GORJI2020103972}. In the context of $\text{FE}^2$ multi-scale simulations, LSTM was used to substitute finite element resolution of the meso-scale \secondreviewer{Boundary Value Problem (BVP)} in \cite{LOGARZO2021113482} for infinitesimal strain problem. In \cite{WU2020113234}, a GRU-based RNN has been used as an accurate surrogate of the meso-scale BVP during $\text{FE}^2$ multi-scale simulations in a finite strain setting.
The training of these \changes{RNNs} requires a synthetic database which covers enough possible loading history paths.
The generation of this database thus requires to introduce some stochasticity in the loading paths, either under the form of Gaussian process as suggested in \cite{Mozaffar26414}, cubic-spline interpolations as conducted in \cite{GORJI2020103972}, or more generally under the form of a random walk process as developed in \cite{WU2020113234}.

In \cite{WU2020113234,LOGARZO2021113482}, only the homogenized strain and stress were collected from direct finite element \changes{resolutions on RVEs} in order to conduct the RNN training, while the micro-scale physical information arising \changes{at the level of the RVEs was} discarded.
In this context the RNN could only serve as a surrogate of the stress-strain response of the RVE and the micro-structure information could not be recovered in a so-called localization step.
\firstreviewer{
However, this information, which was kept in the state variables during the direct finite element analyses of the RVEs, is of interest when conducting multi-scale analyses since it brings insight on the micro-structure loading condition.
For example it can be used for posterior estimation of the structure integrity in the following cases: i) Predicting the damage initiation at the design stage with the aim of preventing/delaying it, in which case fast methods can be favored to a complex detailed analysis; This is of particular interest for engineered micro-structures such as lattices; ii) Obtaining a first estimation of the life of composite materials; indeed, in the case of polymeric materials, the initiation stage takes most the fatigue life as discussed in \cite{Janssen2008,KRAIRI2016179}, and approximations of the life can be obtained using fatigue failure criteria based on stress tensor invariants \citep{BERREHILI20101389}; although such \changes{models} can handle neither complex loading conditions nor progressive failure, see the discussion in \cite{KRAIRI2016179}, they can be used for a first estimation, while a progressive failure analysis would require the enrichment of the micro-scale model.}

In this work, it is intended to develop a RNN as surrogate of the RVE response while being able to recover the evolution of the local micro-structure state variables along with the macro stress-strain history. 

\begin{figure}[htb] 
\centering
		\includegraphics[scale=0.6]{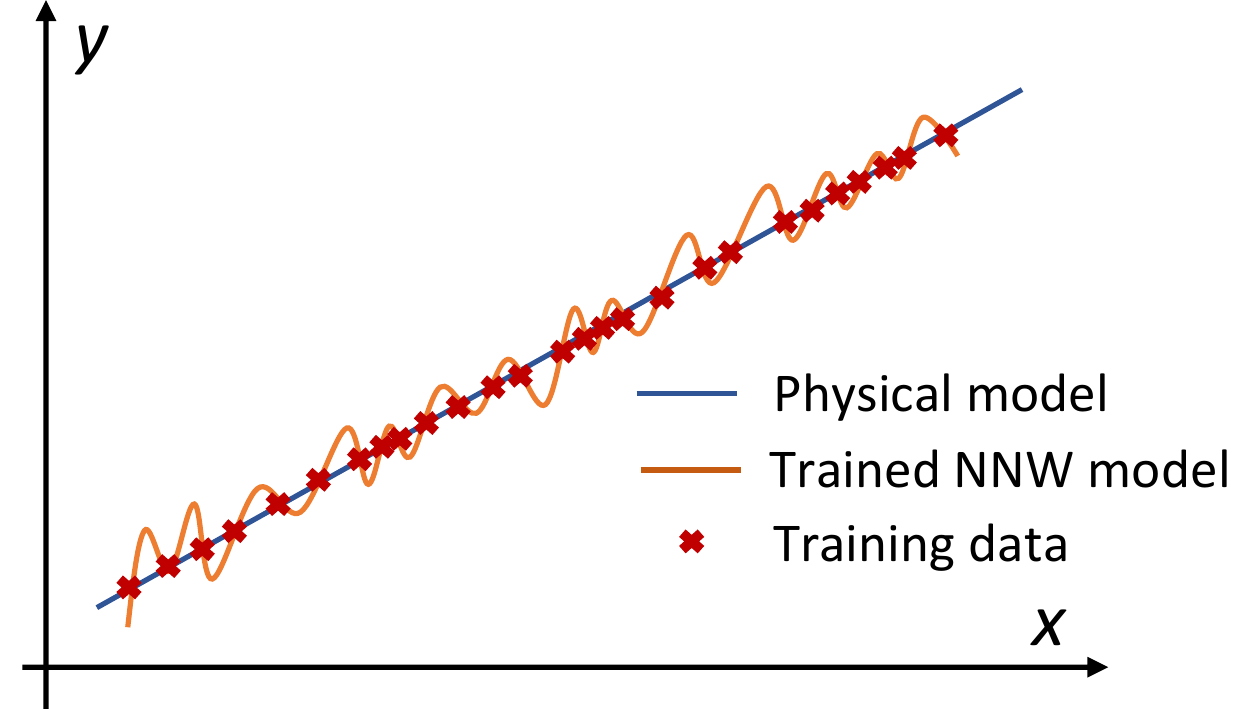}
		\caption{Over-fitting of NNWs.}\label{fig:overfit}
\end{figure}
The application of NNWs in computational mechanics seems simple at a first look since NNWs serve as versatile non-linear mapping black boxes without any physical interpretations.
Theoretically speaking, a NNW should be able to approximate any multi-dimensional non-linear function accurately under the conditions that its structure is properly designed and that it has been adequately trained, which implies to have enough data.
However, because of the shortage of physical \changes{interpretation}, the design of a NNW structure is not always straightforward for given mechanical problems.
The structure of a NNW includes the depth (layers of neurons), width (number of neurons on each layer), and number of hidden variables for the RNNs in particular.
In general, for a simple mechanical problem, in terms of non-linearity and number of state variables, a simple NNW structure should be adopted to limit the number of learnable parameters and avoid \changes{the issue of over-fitting}.
Fig. \ref{fig:overfit} illustrates the problem of using a NNW with high flexibility to fit a simple linear function. \changes{This} kind of over-fitting can be alleviated by using a high weight decay parameter.
However, in order to substitute a complex high dimensional non-linear mapping, a NNW needs to be enough flexible, which means that the NNW needs to be deep and/or wide enough with an increasing number of learnable parameters.
It is convenient to start a structure design from an existing NNW structure which was proved to be accurate for a physical model of similar complexity.
If the reference NNW structure is not available, for simple and \secondreviewer{low dimensional} problems, a \secondreviewer{trial and error approach}, which involves progressively increasing the depth and width of the NNW, is feasible, since the training of a simple NNW, in terms of depth and width, can be fast.
However, this \secondreviewer{trial and error approach} is not always applicable for NNWs with \secondreviewer{an enormous number of learnable parameters}, which is often the case for high dimensional and non-linear problems.
It is even more difficult for history-dependent problems which will involve RNNs of high dimensional hidden variables and sequential training data.
In order to train a NNW with a huge number of learnable parameters, not only a huge amount of training data is required, but the requirement of computer memory is also high.
Because of limited computational resources in practice, this will lead to a heavy and slow training process.
This will be the case when considering the evolution of the state variables in computational homogenization, which is a typical high dimensional and non-linear history-dependent problem.
\secondreviewer{Dimensionality reduction} is always an option when dealing with high dimensional problems.
In fact, it is \firstreviewer{possible, as it will be shown,} to use NNWs for non-linear \secondreviewer{dimensionality reduction} since the dimensionality of the hidden variables of a RNN could be much lower than \changes{the one} of the final output.
Nevertheless this will remain \secondreviewer{computationally} expensive to apply a \secondreviewer{trial and error approach} to determine the required number of hidden variables in a RNN because of the high dimensionality of the final output. 

Principal components analysis (PCA) \cite{Jolliffe1986}, as a classical technique for \secondreviewer{dimensionality reduction}, is usually called Proper Orthogonal Decomposition (POD) in the applications of computational mechanics \cite{YVONNET2007341}.
PCA provides the best linear \secondreviewer{dimensionality reduction} to high-dimensional observations.
However, PCA is less efficient for data \secondreviewer{dimensionality reduction} than some non-linear \secondreviewer{dimensionality reduction} algorithms, such as kernel PCA, locally linear embedding, Laplacian eigen-maps and iso-maps.
Nevertheless, the reconstruction of data from reduced dimensionality into its original dimensionality can be carried out easily by PCA, which it is not always the case for non-linear \secondreviewer{dimensionality reduction}.
PCA has been used in some NNWs based surrogate models \cite{Cao2016,Bamer2017,Vijayaraghavan2021}. 

In the present work, because the aim is to develop a surrogate of a RVE response, not only of the macro stress-strain history but also of the evolution of the local micro-structure state variables, the number of output variables is high.
Besides, since the synthetic database consists of the state variables evolution of the micro-structure subjected to random walk processes \cite{WU2020113234}, the \changes{resulting} patterns of state variables are various and the \changes{equired} PCA reduced dimensionality of the state variables is still quite high. 
In order to overcome the limitation of computational resource and accelerate the \secondreviewer{trial and error approach} of the RNNs structure design, the outputs with PCA reduced dimensionality are further broken down into a few groups, and the evolution of each output group is substituted by an independent RNN.
\changes{In this work, the mapping of RVE macro-strain to the state variables, \emph{i.e.} the localization step of the multi-scale analysis, is successively obtained by three RNNs based surrogate models.}
\begin{itemize}
	\item Surrogate I: Direct RNN modeling without \secondreviewer{dimensionality reduction} on the output;
	\item Surrogate II: RNN modeling with PCA \secondreviewer{dimensionality reduction} on the output;
	\item Surrogate III: RNN modeling with PCA \secondreviewer{dimensionality reduction} and break down on the output.	
\end{itemize}
\secondreviewer{The dimensionality break down consists in splitting the observation output into groups of smaller dimensionality before submitting them to different independent neural networks for training. As a result, the neural networks have a smaller output dimensionality, which eases their training and improves their efficiency. However, it is not possible to divide the original output data without reducing the accuracy of the independent different neural networks since the output data are correlated to some extent. By applying a PCA process beforehand, the correlations among the output data is removed and then dimensionality break down can be carried out in a more accurate way.}

In this study, the considered micro-structure is a continuous fiber reinforced elasto-plastic matrix material, and the studied 2D RVE contains several randomly arranged fibers.
The equivalent plastic strain distributions in the RVE are adopted as studied state variables for the comparison of the three surrogate models.
\firstreviewer{In general, linear dimensionality reduction like PCA is shown to be less efficient than the non-linear one constructed using NNWs.
However, PCA remains useful to extract the amount of reduced information that ought to be represented for the surrogate model to be accurate. The RNNs structure can then be designed by considering this reduced information, which allows speeding up the trial and error approach inherent to a NNW design, since this could be carried out on a problem of lower dimensionality.
In particular, the number of RNN hidden variables can be estimated by training and evaluating a RNN with a unique output, which is obtained from PCA and \secondreviewer{dimensionality break down}.}
During the training process of the RNNs of high dimensional hidden variables, the problem arising due to the limitation of computational resources is also pointed out and a solution based on mini-batches is suggested.

The paper is organized as follows.
Section \ref{sec:FE2} summarizes the homogenization-based multi-scale framework with a particular attention \changes{paid to} the role of the state variables at both scales. The design of the three different surrogate models of the state variables distribution with \secondreviewer{dimensionality reduction} is presented in Section \ref{sec:NNW}.
The strategy used to generate the synthetic database using random walk processes is then described in Section \ref{sec:trainingData}.
Finally, the different developed surrogate models are studied by reconstructing the distributions of the equivalent plastic strain and of the equivalent von Mises stress\footnotemark in the RVE for different loading scenarios in Section \ref{sec:compositeRVE}.
An efficient design methodology of the RNN structure is also proposed from the drawn observations.

\footnotetext[\value{footnote}]{Rigorously speaking, the von Mises stress is not a state variable of the constitutive behavior at the micro-structural level but it can be handled as such for the distribution reconstruction purpose.}
\section[State variables in FE2]{State variables in $\text{FE}^2$}\label{sec:FE2}

The notation $\text{FE}^2$ refers to multi-scale simulations with finite-element analyses being carried out concurrently at both the meso-scale and the macro-scale.
In this context, the BVPs needed to be solved using the finite-element method at \changes{the} two scales are connected by a specially defined scale-transition according to the Hill-Mandel condition.
In this section, we restrict \secondreviewer{ourselves} to the case of finite-strain mechanics and introduce the studied problem through a brief summary of $\text{FE}^2$. \secondreviewer{The notations used in this manuscript are summarized in \ref{app:notations}.}

\subsection{The BVPs at the two scales and the scale-transition}

\subsubsection{Definition of the macro-scale BVP}

Assuming no dynamical effects, the macro-scale linear momentum equation of the body $\Omega$ stated in its reference configuration reads
\begin{equation}\label{eq:equForceMacro}
\textbf{P}_\text{M}(\bm{X})\cdot\grad_0 +\bm{b}_{\text{M}}=0 \quad\forall \bm{X} \in \Omega\,,\\
\end{equation}
where the subscript ``$\text{M}$'' refers to the values at the macro-scale, $\textbf{P}_\text{M}$ is the first Piola-Kirchhoff stress tensor, $\grad_{0}$ is the gradient operator with respect to the reference configuration, and
$\bm{b}_\text{M}$ is the load per unit reference volume.
The boundary conditions read
\begin{eqnarray}
\bm{U}_{\text{M}}(\bm{X}) =  \bar{\bm{u}}_{\text{M}} &&\forall \bm{X} \in \partial_D\Omega\,,\text{ and}\label{eq:equBCDMacro}\\
\textbf{P}_\text{M}(\bm{X})\cdot \bm{n}_\text{M} =\bar{\bm{t}}_\text{M} && \forall \bm{X} \in \partial_N \Omega\,,\label{eq:equBCNMacro}
\end{eqnarray}
where $\bar{\bm{t}}_\text{M}$ is the surface traction, per unit reference surface, on the Neumann boundary $\partial_N\Omega$, $\bm{n}_\text{M}$ is the outward unit normal in the reference configuration, and $ \bar{\bm{u}}_{\text{M}}$ is the constrained displacement on the Dirichlet boundary $\partial_D\Omega$.

As a difference with the \secondreviewer{use of traditional} constitutive laws in a single-scale problem, the BVP is completed by casting the relation between the macro-scale stress tensor $\textbf{P}_\text{M}$ and the deformation gradient $\textbf{F}_{\text{M}}=\bm{U}_{\text{M}}\otimes\grad_{0}$, through state variables $\bm{Z}_\text{M}$,
\begin{equation}\label{eq:equConstitutiveMacro}
\textbf{P}_\text{M}\left(\bm{X},\,t\right)= \textbf{P}_\text{M}\left(\textbf{F}_\text{M}\left(\bm{X},\,t\right); \bm{Z}_\text{M}\left(\bm{X},\,\tau\right),\, \tau\in[0,\,t]\right)\,,
\end{equation}
into the resolution of a meso-scale BVP \emph{via} a scale-transition.
\secondreviewer{As it will be discussed in Section \ref{sec:FtoStateVar}, in the context of $\text{FE}^2$ multi-scale simulation, the state variables $\bm{Z}_\text{M}\left(\bm{X},\,\tau\right)$ arise from the definition and resolution of the meso-scale BVP defined at the macro-scale material points $\bm{X}\in \Omega$.}

\subsubsection{Definition of the meso-scale BVP}\label{sec:defMesoScaleBVP}

The meso-scale BVP is usually defined on Representative Volume Elements (RVEs), which are parallelepiped (rectangular in 2D) domain $\omega(\bm{X})$ in the reference configuration, with planar boundary faces $\partial\omega$, and defined at the macro-scale material points $\bm{X}\in \Omega$.
It is assumed that the classical continuum mechanics equations hold and that the time for a stress wave to propagate in the meso-scale volume element remains negligible.
Considering the material points $\bm{x} \in \omega$, in the absence of dynamical effects, the equilibrium equations read
\begin{eqnarray}
\textbf{P}_\text{m}\cdot\grad_0 =0 &&\forall \bm{x} \in \omega\,,\label{eq:equForceMicro}\\
\textbf{P}_\text{m}\cdot \bm{n}_\text{m} = {\bm{t}}_\text{m} &&\forall \bm{x} \in \partial \omega\label{eq:equTractionMicro}\,,
\end{eqnarray}
where the subscript ``$\text{m}$'' refers to the local value at the micro-scale,  and ${\bm{t}}_\text{m}$ is the surface traction, per unit reference surface, on the boundary $\partial\omega$ of outward unit normal $\bm{n}_\text{m}$ in the reference configuration.

The micro-scale problem is completed by the local constitutive laws of the different material phases at a given time $t$ and material point $\bm{x}$, which are written as  
\begin{equation}\label{eq:equConstitutiveMicro}
\textbf{P}_\text{m}\left(\bm{x},\,t\right)= \textbf{P}_\text{m}\left(\textbf{F}_\text{m}\left(\bm{x},\,t\right); \bm{Z}_\text{m}\left(\bm{x},\,\tau\right),\, \tau\in[0,\,t]\right)\,,
\end{equation}
where $\textbf{F}_\text{m}(\bm{x})=\bm{U}_{\text{m}}\otimes\grad_{0}$ is the micro-scale deformation gradient tensor evaluated in terms of the micro-scale displacement $\bm{U}_\text{m}$, and $\bm{Z}_\text{m}$ is a set of state variables tracking  history-dependent processes.

\subsubsection{The scale-transition}\label{sec:scaletransition}

The scale-transition connects the macro-scale deformation gradient $\textbf{F}_{\text{M}}$ and stress tensor $\textbf{P}_\text{M}$ to the averages of the micro-scale deformation gradient tensor $\textbf{F}_\text{m}(\bm{x})$ and of stress tensor $\textbf{P}_{\text{m}}(\bm{x})$ over the meso-scale volume element $\omega$.
In the context of homogenization theories one thus has\footnote{\secondreviewer{As pointed out by Michel et al. \cite{MICHEL1999109}, these relations require particular attention when voids are present and intersect the volume boundary. If the domain $\omega=\omega_S\cup \omega_V$ is the union of the solid domain $\omega_S$ and of the voids domain $\omega_V$, one has $\textbf{F}_{\text{M}} =\frac{1}{V(\omega)}\int_{\omega_S \cup \omega_V} \textbf{F}_{\text{m}} d\omega$. When applying the Gauss theorem on both $\omega_V$ and $\omega_S$, since their respective surface normals on solid/void interfaces are in opposite direction, and since the surface normal on the intersection of voids with the boundary is in the same direction as the normal to the surface $\partial \omega$ the expression simplifies in $\textbf{F}_{\text{M}}=\frac{1}{V(\omega)}\oint_{\partial \omega} \bm{U}_{\text{m}}\otimes\bm{n}_{\text{m}} d\partial\omega$, and similarly for the stress averaging equation.}}
\begin{eqnarray}
\textbf{F}_{\text{M}}\left(\bm{X},\,t\right)&=&\frac{1}{V(\omega)}\int_{\omega} \textbf{F}_{\text{m}}(\bm{x},\,t) d\omega \,, \text{ and }\label{eq:FM}\\
\textbf{P}_{\text{M}}\left(\bm{X},\,t\right)&=&\frac{1}{V(\omega)}\int_{\omega} \textbf{P}_{\text{m}}(\bm{x},\,t) d\omega \,.\label{eq:PM}
\end{eqnarray}
The requirement of energy consistency between the different scales, which corresponds to the Hill-Mandel condition, reads 
\begin{equation}\label{eq:hill_mandell_energy}
\textbf{P}_{\text{M}}:\delta\textbf{F}_{\text{M}}= \frac{1}{V(\omega)}\int_{\omega} \textbf{P}_{\text{m}}:\delta\textbf{F}_{\text{m}} d\omega\,.
\end{equation}

\secondreviewer{The solution of the meso-scale BVP (\ref{eq:equForceMicro}-\ref{eq:equTractionMicro}) needs to satisfy Eqs. (\ref{eq:FM}) and (\ref{eq:hill_mandell_energy}). These constraints are enforced by applying specially defined boundary conditions on the RVEs. In this work, the Periodic Boundary Conditions (PBCs) are adopted and the detailed implementation of the homogenization process with PBCs can be found in \cite{NGUYEN2017}.
The way of applying periodic boundary conditions on non periodic micro-structures can be found in \cite{WIPPLER20116029,Nguyen2012390}.}

\subsection[The role of the state variables in a FE2 resolution]{The role of the state variables in a $\text{FE}^2$ resolution}\label{sec:FtoStateVar}

In general, the macro-scale constitutive equation (\ref{eq:equConstitutiveMacro}) in a $\text{FE}^2$ framework is a finite element analysis which computes the meso-scale responses, \emph{i.e.} the first Piola-Kirchhoff stress tensor $\textbf{P}_{\text{M}}$ and \secondreviewer{the fourth order macro-scale material tensor $\mathbb{C}_\text{M}=\frac{\partial \textbf{P}_\text{M}}{\partial \textbf{F}_\text{M}}$} of a given RVE under prescribed $\mathbf{F}_\text{M}$. The resolution of the constrained micro-scale finite element problem is called computational homogenization \secondreviewer{and follows the different steps}:
\begin{itemize}
	\item A finite element discretization of a chosen RVE represents the micro-structure of the heterogeneous material. 
	\item From the macro-scale deformation gradient tensor $\mathbf{F}_\text{M}$, the micro-scale finite element problem is formulated in its weak form under certain boundary conditions satisfying  Eqs. (\ref{eq:FM}) and (\ref{eq:hill_mandell_energy}) such as PBCs.
	\item This micro-scale finite element problem is solved by considering the micro-scale phases constitutive models (\ref{eq:equConstitutiveMicro}); iterations are needed for the non-linear cases. 
	\item The extraction of the meso-scale response includes the homogenized first Piola-Kirchhoff stress tensor $\textbf{P}_{\text{M}}=\frac{1}{V(\omega)}\sum_{e}\int_{\omega^e} \textbf{P}_{\text{m}} d\omega$, and the fourth order macro-scale material tensor $\mathbb{C}_\text{M}$ required to perform the Newton-Raphson iterations of the macro-scale analysis. \secondreviewer{The evaluation of $\mathbb{C}_\text{M}$ directly results from the RVE finite element resolution following a constraint elimination method \cite{NGUYEN2017}.}
\end{itemize}

From the above process, it can be seen that the traditional state variables are not used at the macro-scale. However, the deformation history of a macro-scale material point corresponds to \changes{the one} of a RVE, and is recorded by the set of state variables of the meso-scale volume element.
Comparing the constitutive laws at the macro-scale and at the micro-scale, Eqs. (\ref{eq:equConstitutiveMacro}) and (\ref{eq:equConstitutiveMicro}), yields,
\begin{equation}\label{eq:Inv}
	\bm{Z}_\text{M}\left(\bm{X},\,t\right)=\{\bm{Z}_\text{m}\left(\bm{x},\,t\right)\,:\; \bm{x}\in\omega(\bm{X})\}\,,
\end{equation}
where \firstreviewer{the notation $\{\bullet\,:\; \bm{x}\in\omega(\bm{X})\}$ refers to the set of local data defined at all material points of the meso-scale volume element $\omega(\bm{X})$ associated to the macro-scale material point $\bm{X}$}. \secondreviewer{In this context $\bm{Z}_\text{M}\left(\bm{X},\,t\right)$ is not seen as a classical state variable of a constitutive model: its definition results from the $\text{FE}^2$ formalism.} 

It has been shown that RNNs can provide an accurate history-dependent non-linear mapping from $\textbf{F}_\text{M}$ to $\textbf{P}_\text{M}$ in \cite{WU2020113234}, in which case the hidden variables $\bm{h}_t$ of the RNN play the role of $\bm{Z}_\text{M}\left(\bm{X},\,t\right)$ --the hidden variables $\bm{h}_t$ are actually a reduced order version of the state variables but without physical interpretation.
In that case, the information kept in the state variables during the training data generation is discarded by the surrogate model.
However, beside the macro-scale stress-strain history $\textbf{F}_\text{M}(t)$-$\textbf{P}_\text{M}(t)$, the information provided by the state variables distribution resulting from a direct finite element \changes{analysis} of the RVE is of interest when conducting multi-scale \firstreviewer{analyses} since it brings insight on the micro-structure loading condition and can be used to assess failure or fatigue.
The history-dependent non-linear mapping from $\textbf{F}_\text{M}(t)$ to $\bm{Z}_\text{M}(t)$ should thus also be recovered by the surrogate model.

\section{\secondreviewer{Data-driven surrogate modeling}}\label{sec:NNW}

In this section, Recurrent Neural Networks (RNNs) and some basic data operations related to the Principal Component Analysis (PCA) are briefly recalled.
The effectiveness of the PCA being limited by its global linearity, the use of NNWs for non-linear \secondreviewer{dimensionality reduction} is also discussed.
Then a data-driven based surrogate model is proposed for the history-dependent non-linear mapping from $\textbf{F}_\text{M}$ to $\bm{Z}_\text{M}$. We note that the  data-driven based surrogate model for the history-dependent non-linear mapping from $\textbf{F}_\text{M}$ to $\textbf{P}_\text{M}$ was previously developed in \cite{WU2020113234}.
Finally, the training strategy using GPU resources is briefly explained.

\subsection{The Recurrent Neural Networks (RNNs)}\label{sec:RNNGen}

\begin{figure}[htb] 
\centering
		\includegraphics[scale=0.45]{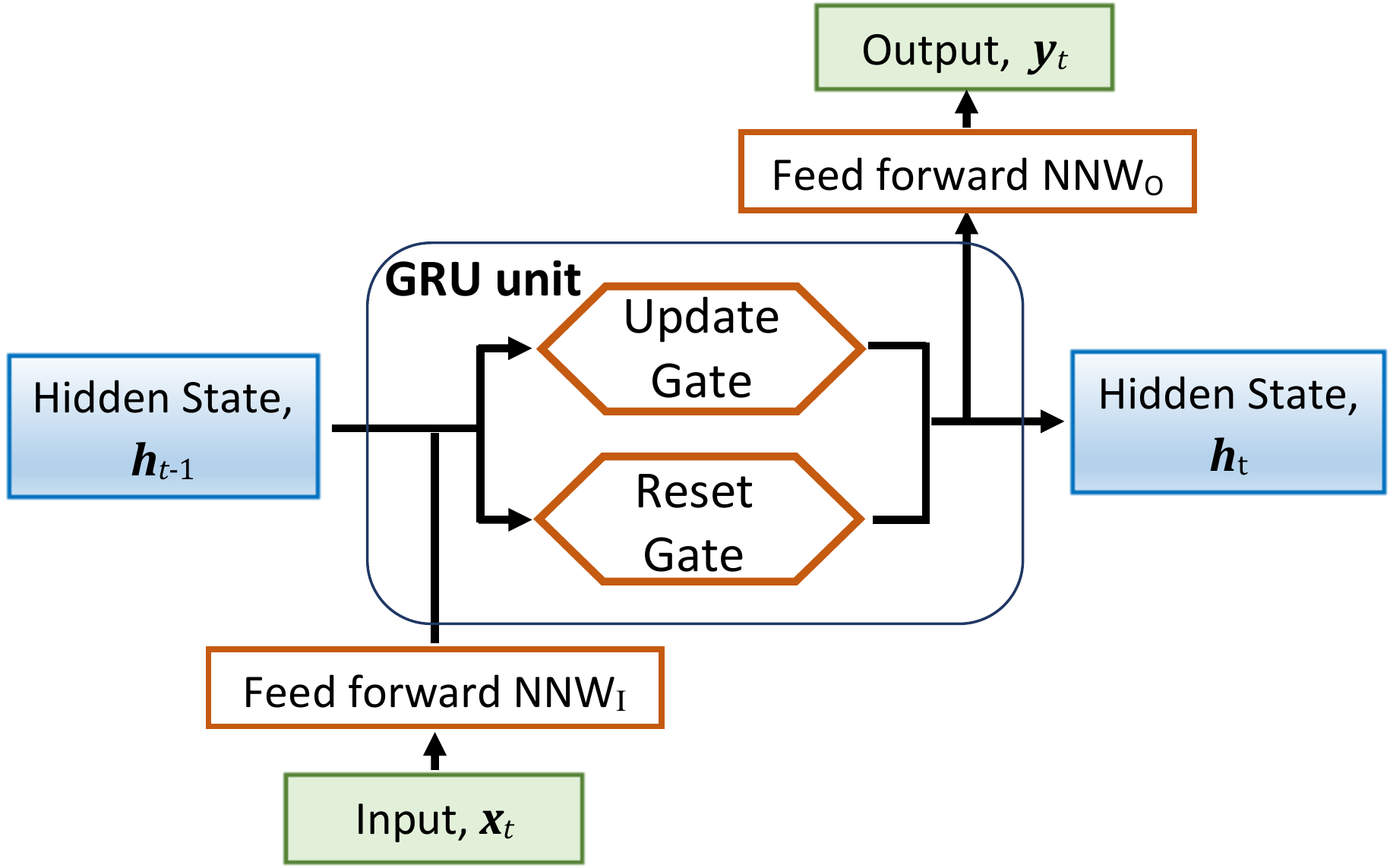}
		\caption{Recurrent Neural Network with Gated Recurrent Unit.}\label{fig:GRU}
\end{figure}

RNNs have a 'memory' defined through the hidden variables, $\bm{h}_t$, which allows them to trace the input history.
Long Short Term Memory networks (LSTMs) are ones of those kinds of specially designed RNNs and are now widely used in a large variety of problems.
As a variation of the LSTM, the Gated Recurrent Unit (GRU) has been chosen as surrogate for the direct finite element analysis on RVEs in \cite{WU2020113234}.
The typical functional character of GRU is illustrated by the block of GRU unit in Fig. \ref{fig:GRU}.
In the GRU unit, the previous hidden variables information and the current input need to pass through the so-called reset and update gates, which are respectively used to decide how much of the past information \changes{can be forgotten} and how much of the past information needs to be passed along for the future use, \secondreviewer{see the detail provided in \ref{app:gru}.}
Two feed-forward NNWs are then respectively added in the paths of input and output of the GRU, see $\text{NNW}_\text{I}$ and $\text{NNW}_\text{O}$ in Fig. \ref{fig:GRU}\changes{. These NNWs} make the model more flexible than a single GRU unit so that it can be adapted to complex problems.
\firstreviewer{The output vector $\bm{x}'_{t}$ of the first feed-forward $\text{NNW}_\text{I}$ acts as input of the GRU and the GRU output vector $\bm{y}'_{t}$, which also corresponds to the updated hidden variables vector $\bm{h}_{t}$, constitutes the input of the second feed-forward $\text{NNW}_{\text{O}}$.}
This architecture was used to construct the surrogate model for the history-dependent non-linear mapping from $\textbf{F}_\text{M}$ to $\textbf{P}_\text{M}$ in \cite{WU2020113234}.

In order \changes{for the NNWs to become surrogate models, their learnable parameters} have to be identified by training them using a database.
\firstreviewer{To this end, a loss function is defined to measure the difference between the predicted output $\hat{\bm{y}}(\bm{x})$ and observation output $\bm{y}(\bm{x})$ of the training database for the same input $\bm{x}$. In this work we consider as a loss function the Mean Square Error (MSE), which reads for an output of dimensionality $n$,
\begin{equation}\label{eq:MSE}
\text{MSE}=\frac{1}{n}\sum_{i=1}^{n}(y_i-\hat{y}_i)^2\,.
\end{equation}
A back-propagation algorithm is used to conduct the deep learning of the NNWs, and it iteratively updates the learnable parameters, or weights, in order to minimize the loss function $\text{MSE}$.}
However, before performing the training of NNWs, all the input and \secondreviewer{observation} output features must be standardized. If not, the NNWs will not be well trained because the input/output features may not have the same scale, and the input of activation functions might be out of their active range. 
Normalization by feature is used in this work: for each input and \secondreviewer{observation} output feature, noted with $\chi$, a simple linear operation is performed, and reads
\begin{equation}\label{eq:normalization}
\underline{\chi}=\frac{(\chi-\chi_\mu)}{\chi_\text{s}}\,,
\end{equation}
with
\begin{equation}\label{eq:mean}
\chi_\mu=\frac{(\chi_\text{min}+\chi_\text{max})}{2}\quad\text{and}\quad\chi_\text{s}=\frac{(\chi_\text{max}-\chi_\text{min})}{2}\,,
\end{equation}
where $\chi_\text{max}$ and $\chi_\text{min}$ are the maximum and minimum values of this feature among all the available data, so that the \secondreviewer{normalized features $\underline{\chi}$ are mapped to the range $[-1,1]$; the notation $\underline{\bullet}$ holds for normalized values.}  
Since all the used basic training operations of the NNWs, such as computation and optimization of the loss function, update of weights of NNWs, \emph{etc.} are all provided by the PyTorch library \cite{Torch}, they are not detailed in this paper.

For the RNN designed in Fig. \ref{fig:GRU}, the learnable parameters are divided in three parts which include a GRU together with two feed-forward NNWs, $\text{NNW}_\text{I}$ and $\text{NNW}_\text{O}$.
The number of learnable parameters depends on the adopted number of layers and the dimensionality of each part.
In general, for a feed-forward network, the learnable parameters include the weights ${w}^i_{kj}$ and biases ${w}^i_{0_j}$ defining the weights matrix $\mathbf{W}$, see notations in Fig. \ref{fig:ANNW}. 
\firstreviewer{In that case, the number of learnable parameters between two neural layers of the feed-forward neural networks $\text{NNW}_\text{I}$ and $\text{NNW}_\text{O}$ is
\begin{equation}\label{eq:NP_NNW}
	{N}_\text{NNW}^{i,\,i+1}=(n_i+1)\times n_{i+1}\,,
\end{equation}
where $n_i$ and $n_{i+1}$ are the dimensionalities of the two sequential neural layers $i$ and $i+1$.}
The number of learnable parameters in a GRU of one layer reads,
\begin{equation}\label{eq:NP_GRU}
{N}_\text{GRU}=\secondreviewer{3 n_{\text{h}}\times(n_{\text{h}}+n_\text{I}+2)}\,,
\end{equation}
where \secondreviewer{$n_{\text{h}}$} and $n_\text{I}$ are respectively the dimensionalities of hidden variables and input of the GRU.
\firstreviewer{Following Eqs. (\ref{eq:NP_NNW}-\ref{eq:NP_GRU}), for the RNN presented in Fig. \ref{fig:GRU}, with low input and high output dimensionalities, the total number of learnable parameters is dominated by the number of hidden variables, \secondreviewer{$n_{\text{h}}$}, of the GRU and by the number of outputs of the $\text{NNW}_\text{O}$.}
The architecture presented in Fig. \ref{fig:GRU} was used in \cite{WU2020113234} to construct the surrogate model for the history-dependent non-linear mapping from $\textbf{F}_\text{M}$ to $\textbf{P}_\text{M}$.
Such an architecture can be used herein in order to construct the surrogate model for the history-dependent non-linear mapping from $\textbf{F}_\text{M}$ to $\bm{Z}_\text{M}$. However, because of the increase in the output dimensionality in that case, and since the complexity of the state variables evolution implies the use of a higher number of hidden variables \secondreviewer{$n_{\text{h}}$}, the number of learnable parameters increases drastically.

\firstreviewer{\subsection{Principal Component Analysis (PCA)}\label{sec:PCA}}

Principal components analysis (PCA) \cite{Jolliffe1986} is a classical and one of the most popular techniques for \secondreviewer{dimensionality reduction}.
It provides the best linear approximations to given high-dimensional observations.
Considering data $\bm{a}$ with a high-dimensionality $d$, the goal of the PCA is to find a linear subspace of lower dimensionality $p$ $(p\leq d)$, such that the maximum variance is retained in this lower-dimensional space.
In other words, PCA reduces the dimensionality of the data, while minimizing the loss of information which comes from the variation.

We define the data matrix ${\mathbf{A}}=\left[{\bm{a}}_1-\bm{a}_{\mu}\;{\bm{a}}_2-\bm{a}_{\mu}\;\ldots\;{\bm{a}}_n-\bm{a}_{\mu}\right]\in\mathbb{R}^{d\times n}$, in which the mean vector $\bm{a}_{\mu}$ is introduced for ${\mathbf{A}}$ to be row-wise zero-mean, and the square matrix $\mathbf{M}={\mathbf{A}}{\mathbf{A}}^\text{T}\in\mathbb{R}^{d\times d}$.
The $d$ eigenvalues of $\mathbf{M}$ are denoted in a descending order as $\Lambda_1\geq\Lambda_2\geq\ldots\geq\Lambda_{d}$, and their corresponding eigenvectors, which are called the ``Principal Components'', as $\underline{\bm{v}}_1$, $\underline{\bm{v}}_2$, \ldots, $\underline{\bm{v}}_{d}$.
A criterion is then defined to retain only some components of interest, such as  
\begin{equation}\label{eq:reduce_accuracy}
1.0-\frac{\sum_{i=1}^p \Lambda_i}{\sum_{k=1}^{d} \Lambda_k}\leq \delta\quad \text{with}\quad p\leq d\,,
\end{equation}
where $\delta$, a small value close to zero, controls the accuracy of the \secondreviewer{dimensionality reduction} --\emph{e.g.} if we set $\delta = 0$ either we keep the original dimensionality $p=d$ or $\mathbf{M}$ has at least one zero eigen  value and $p<d$.
Finally we define
\begin{equation}
\mathbf{V}=\left[\underline{\bm{v}}_1\;\underline{\bm{v}}_2\; \ldots\; \underline{\bm{v}}_p\right]_{d\times p}\,,
\end{equation}
and the dimensionaly reduced data matrix $\mathbf{B}\in\mathbb{R}^{p\times n}$
\begin{equation}
\mathbf{B} = \mathbf{V}^\text{T}{\mathbf{A}}\,.
\end{equation}
The data can be reconstructed from the following relation  
\begin{equation}\label{eq:PCA_reconstruction}
\hat{\mathbf{A}}_{d\times n} = \mathbf{V}\mathbf{B}\,.
\end{equation}
Finally the approximate data $\left[\hat{\bm{a}}_1\;\hat{\bm{a}}_2\;\ldots\;\hat{\bm{a}}_n\right]$ are obtained by adding the mean $\bm{a}_{\mu}$ to the columns of $\hat{\mathbf{A}}_{d\times n}$.

The effectiveness of the PCA is limited by its global linearity.
A series of non-linear \secondreviewer{dimensionality reduction} algorithms, which can be cast as general kernel PCA, have thus been developed, such as kernel PCA, locally linear embedding, Laplacian eigen-maps and iso-maps.
Although these methods are efficient for data \secondreviewer{dimensionality reduction}, because of the non-linear \secondreviewer{dimensionality reduction} methods, the reconstruction of data in its original dimensionality is not a trivial task and can even sometimes be impossible.
The use of NNWs for non-linear \secondreviewer{dimensionality reduction} is now discussed.
 
\subsection{\secondreviewer{Dimensionality reduction} by Feed-forward Neural Network (NNW)}\label{sec:DNN}

\begin{figure}[!htb]
	\centering
	\subfigure[\secondreviewer{NNW for dimensionality reduction}]{\includegraphics[scale=0.7]{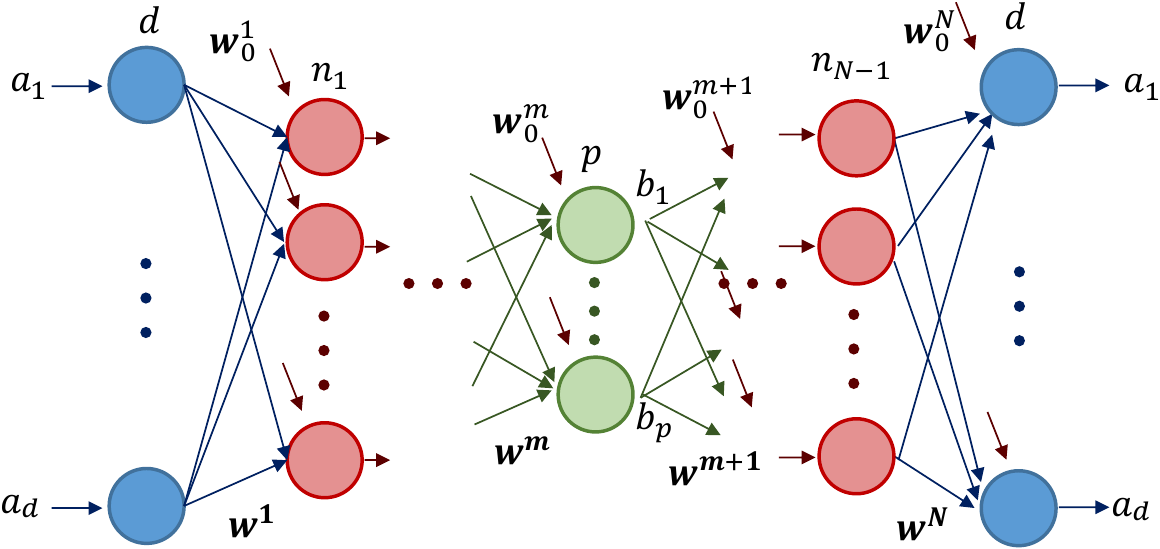}\label{fig:NNW_ord}}\quad\quad
	\subfigure[Details of $\text{NNW}_\text{O}$]{\includegraphics[scale=0.7]{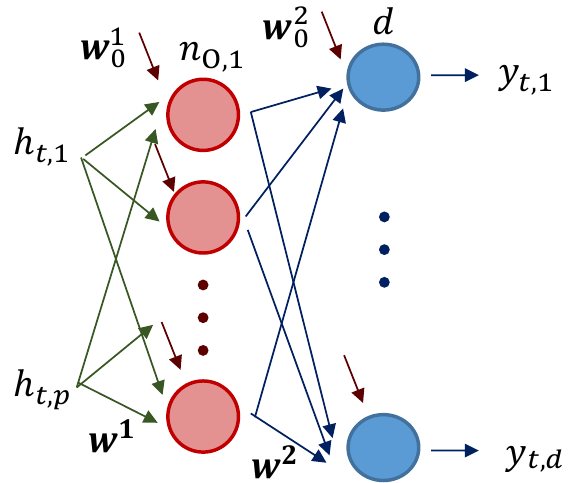}\label{fig:NNW_out}}
	\caption{Different feed-forward NNW architectures: (a) An artificial Neural Network for \secondreviewer{dimensionality reduction} and data reconstruction; (b) Feed-forward $\text{NNW}_\text{O}$ as output of the Gated Recurrent Unit as depicted in Fig. \ref{fig:GRU} used for data reconstruction (we note that the updated hidden state vector of a GRU is the same as its output vector).}\label{fig:ANNWRED}
\end{figure}

Low dimensional representation of high dimensional state variables in a RVE can also be constructed through NNWs.
\firstreviewer{NNWs applied to dimensionality reduction define a function $F: \bm{a}\rightarrow\bm{b}\rightarrow\hat{\bm{a}}$, see Fig. \ref{fig:NNW_ord}, in which $\bm{a}=[a_1,\,a_2,\,\ldots,\,a_{d}]$, where $\hat{\bullet}$ is used to indicate the approximation.}
Because of the bottleneck shaped NNW, lower-dimensional data $\bm{b}=[b_1,\,b_2,\,\ldots,\,b_{p}]$, ($p< d$), can be extracted at the central $m^{\text{th}}$ layer of the NNW. 

Based on a given set of data ${\bm{A}}=\left\lbrace{\bm{a}}_1,\,{\bm{a}}_2,\,\ldots,\,{\bm{a}}_n\right\rbrace$, supervised learning refers to training the weights ${w}^i_{kj}$, $(i=1,\,\ldots,\,N;\; k=1,\,\ldots,\,n_{i-1},\,j= 1,\,\ldots,\,n_i,\, n_0=n_N=d \text{ and }n_m=p)$ and biases ${w}^i_{0_j}$, $(i=1,\,\ldots,\,N \text{ and }j= 1,\,\ldots,\,n_i,\,n_0=n_N=d \text{ and }n_m=p)$, see the notations in Fig. \ref{fig:NNW_ord}, and making the NNW being able to map the input $\bm{a}$ to $\bm{b}$ and then back to $\hat{\bm{a}}$, an approximation of $\bm{a}$. After training, the first $m$ layers of the NNW, $\bm{a}\rightarrow\bm{b}$, are used for \secondreviewer{dimensionality reduction} and the remaining $N-m$ layers, $\bm{b}\rightarrow\hat{\bm{a}}$, for data reconstruction. \secondreviewer{This nonlinear dimensionality reduction is similar to the concept of autoencoder \cite{HINTON06}.} \firstreviewer{In Section \ref{sec:RNN_direct}, the hidden variables of RNNs will be used as the lower dimensional data and, therefore, only the data reconstruction part depicted in Fig. \ref{fig:NNW_out} needs to be trained.}

\subsection{Design of several surrogate models for state variables}\label{sec:Surrogate}

\secondreviewer{\subsubsection{Input and output variables definition}\label{sec:inputoutput}

Since the resolution of the meso-scale BVP respects the frame indifference, rigid rotation modes can be eliminated by using the macro-scale \secondreviewer{Green-Lagrange} strain tensor $\mathbf{E}_\text{M}$ as input, instead of using the deformation gradient $\mathbf{F}_\text{M}$. Hence, the mapping from $\mathbf{F}_\text{M}$ to $\bm{Z}_\text{M}$ becomes a mapping from $\mathbf{F}_\text{M}$ to $\mathbf{E}_\text{M}$ first through
\begin{equation}\label{eq:EGL}
\mathbf{E}_\text{M} = \frac{1}{2}\left( {\mathbf{F}_\text{M}}^\text{T}\cdot{\mathbf{F}}_\text{M}-\mathbf{I} \right)\,,
\end{equation}
and then from $\mathbf{E}_\text{M}$ to $\bm{Z}_\text{M}$.
Therefore, we use the strain measure $\mathbf{E}_\text{M}$ as the input of RNN and the state variables of the meso-scale volume element $\bm{Z}_\text{M}$ for output. Particularly for 2D RVE under plane strain condition, we have $\bm{x}_t=\{E_{\text{M}{XX}}(t),\,E_{\text{M}{YY}}(t), E_{\text{M}{XY}}(t)\}$, and $\bm{y}_t= \bm{Z}_\text{M}(t)=\{\bm{Z}_\text{m}(\bm{x},\,t)\,:\; x\in\omega({\bm{X}})\}$, see notations in Fig. \ref{fig:GRU}.
We note that the output $\bm{y}_t$ corresponds to the spatial distribution of the RVE state variables. }

\subsubsection{Surrogate I: RNN for state variables without \secondreviewer{dimensionality reduction}}\label{sec:RNN_direct}

It is possible to use the hidden variables of the RNN architecture depicted in Fig. \ref{fig:GRU} as a low order dimensional representation of the high dimensional state variables.
Since the hidden variables are generated automatically during the supervised training, there is no need to carry out a separate order reduction process with NNW, see Section \ref{sec:DNN}.
A feed-forward $\text{NNW}_\text{O}$ for output, see Fig. \ref{fig:GRU}, with \changes{an} input dimensionality ``$p$'' lower than the output dimensionality ``$d$'' is adopted to perform the data reconstruction and is detailed in Fig. \ref{fig:NNW_out}. 

\begin{figure}[htb] 
\centering
		\includegraphics[scale=0.8]{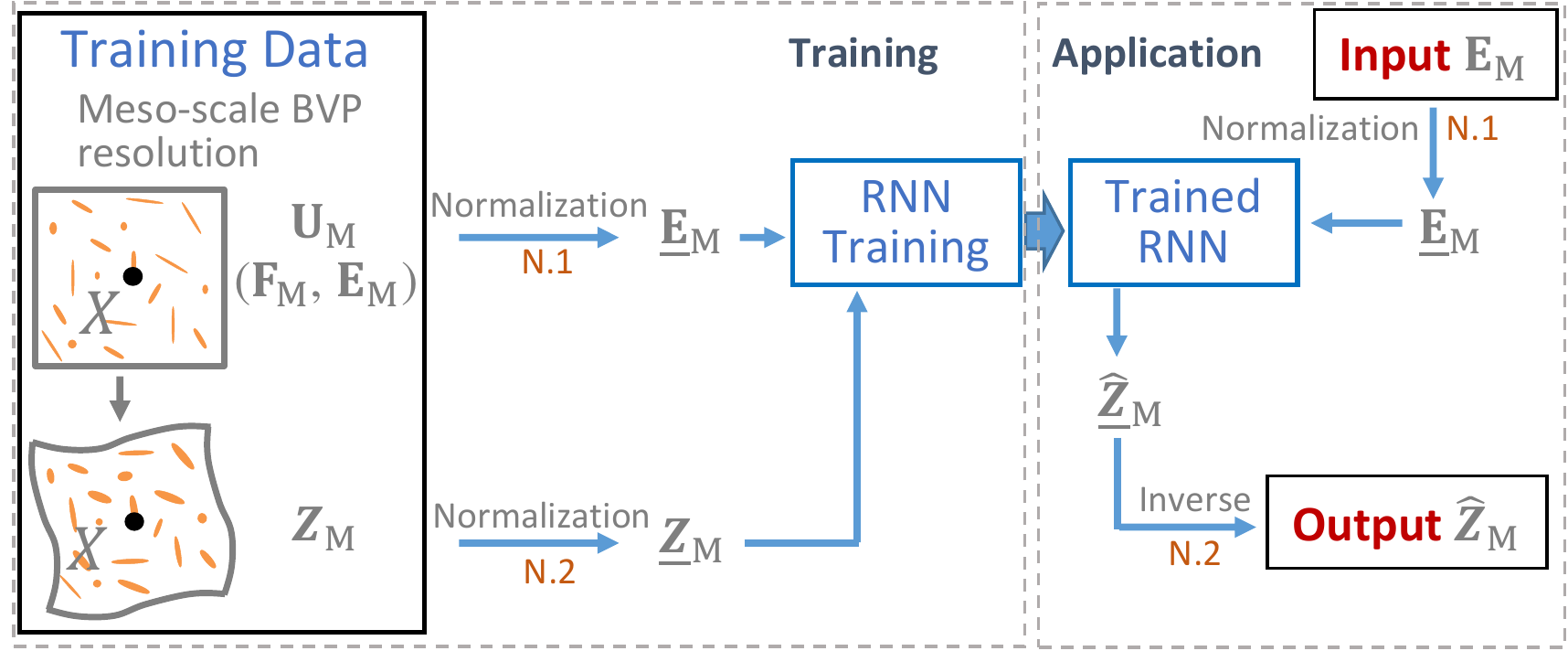}
		\caption{Data-driven surrogate model I without \secondreviewer{dimensionality reduction}}\label{fig:SurrogateDR}
\end{figure}

The corresponding data-driven modeling process is presented in Fig. \ref{fig:SurrogateDR} with its training and on-line phases.
\begin{itemize}
\item The training phase includes the training data generation and preparation and the RNN training itself.
 The training data are collected from direct finite element simulations conducted on the RVE, as detailed in the next Section \ref{sec:trainingData}. 
 The input data, \emph{i.e.} the Green-Lagrange strain $\mathbf{E}_\text{M}$ tensor, are normalized by feature, ``N.1'', with Eq. (\ref{eq:normalization}).
 For the observation output data, $\bm{Z}_\text{M}$, a normalization by feature, ``N.2'', is also applied before using them during the training.
 Finally, the RNN is trained with the normalized \secondreviewer{Green-Lagrange} strain $\underline{\mathbf{E}}_\text{M}$ as input and $\underline{\bm{Z}}_\text{M}$ as observation output.
\item In the on-line phase, the NNW is used as a surrogate. The input \secondreviewer{Green-Lagrange} strain $\mathbf{E}_\text{M}$, is first normalized by feature, ``N.1'' (using bounds $\chi_{\text{min}}$ and $\chi_{\text{max}}$ \changes{determined during} the training stage), before being sent to the trained RNN.
Then, the output of the RNN, $\underline{\hat{\bm{Z}}}_\text{M}$, goes through the inverse operations of normalization ``N.2'' to yield the sought state variables, $\hat{\bm{Z}}_\text{M}$, where $\hat{\bullet}$ is used to indicate the approximations provided by the surrogate model.
\end{itemize}

Because the feed-forward neural network $\text{NNW}_\text{O}$ has a high dimensionality in all the layers, the number of learnable parameters that have to be trained increases dramatically, see Eq. (\ref{eq:NP_NNW}), and not only the requirement of computer memory, but also the amount of training data needed to avoid over-fitting, become critical.
These will lead to a heavy training process even on a \secondreviewer{computer cluster}.
The high demand of computational resources can be relieved by reducing the dimensionality of the output.

\subsubsection{Surrogate II: RNN for state variables with \secondreviewer{dimensionality reduction}}\label{sec:RNN_PCA}

\begin{figure}[htb] 
	\begin{center}
		\includegraphics[scale=0.8]{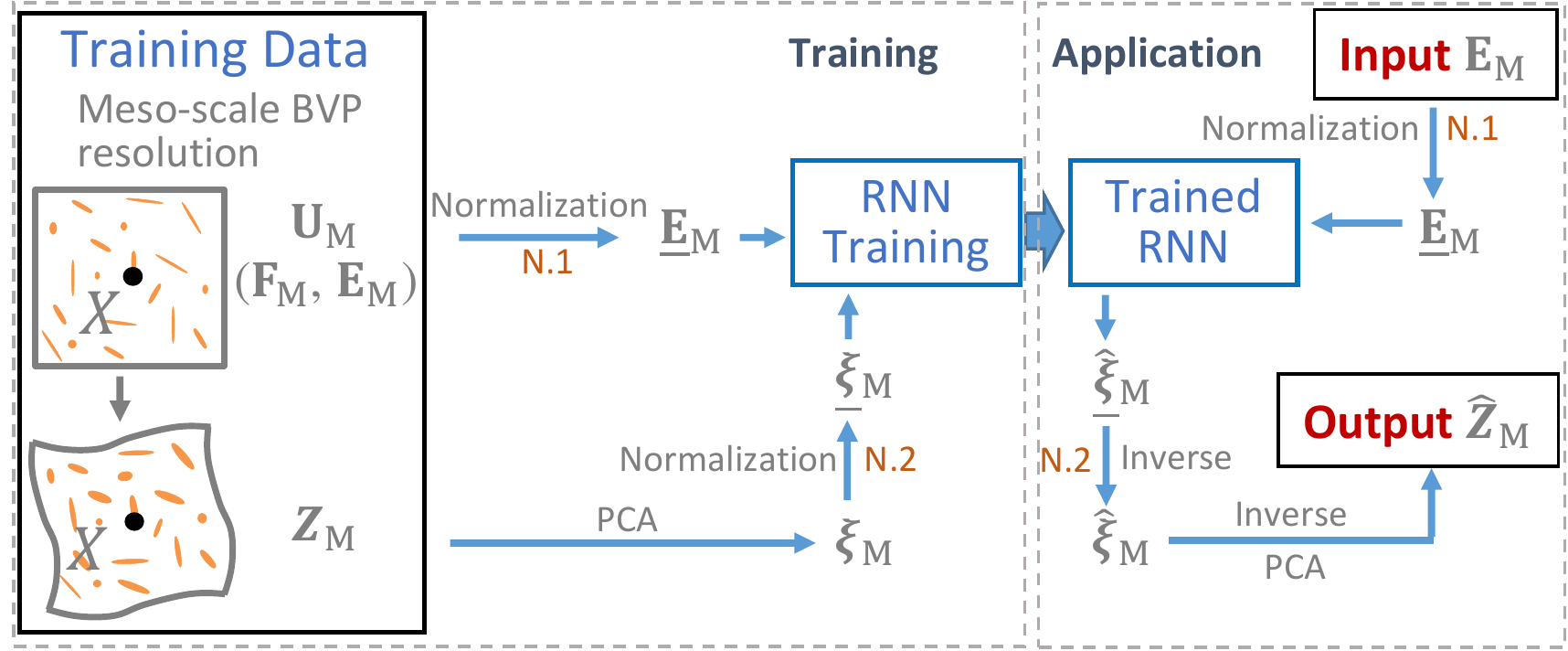}
		\caption{Data-driven surrogate model II with \secondreviewer{dimensionality reduction}}\label{fig:Surrogate}
	\end{center}
\end{figure}

In this process, an extra PCA operation is applied on the state variables, see Fig. \ref{fig:Surrogate}.
\begin{itemize}
\item For the training phase, a PCA is performed on the \secondreviewer{observation} output data, $\bm{Z}_\text{M}$, in order to obtain its low dimensional representation, $\bm{\xi}_\text{M}$. The low dimensional data, $\bm{\xi}_\text{M}$, is then normalized by feature, ``N.2'', yielding $\underline{{\bm{\xi}}}_\text{M}$, before being used for training.
  Then, the RNN is trained with the normalized \secondreviewer{Green-Lagrange} strain $\underline{\mathbf{E}}_\text{M}$ as input and $\underline{\bm{\xi}}_\text{M}$ as \secondreviewer{observation} output.
	\item In the on-line phase, the trained RNN is used as a surrogate. The output of \changes{the RNN}, $\underline{\hat{\bm{\xi}}}_\text{M}$, goes through, in sequence, the inverse operations of normalization (``N.2'') and PCA to yield the desired state variables, $\hat{\bm{Z}}_\text{M}$.
\end{itemize}

\subsubsection{Surrogate III: RNN for state variables with \secondreviewer{dimensionality reduction} and break down}\label{sec:RNN_PCAsplit}

\begin{figure}[htb] 
	\begin{center}
		\subfigure[Global view]{\includegraphics[scale=0.8]{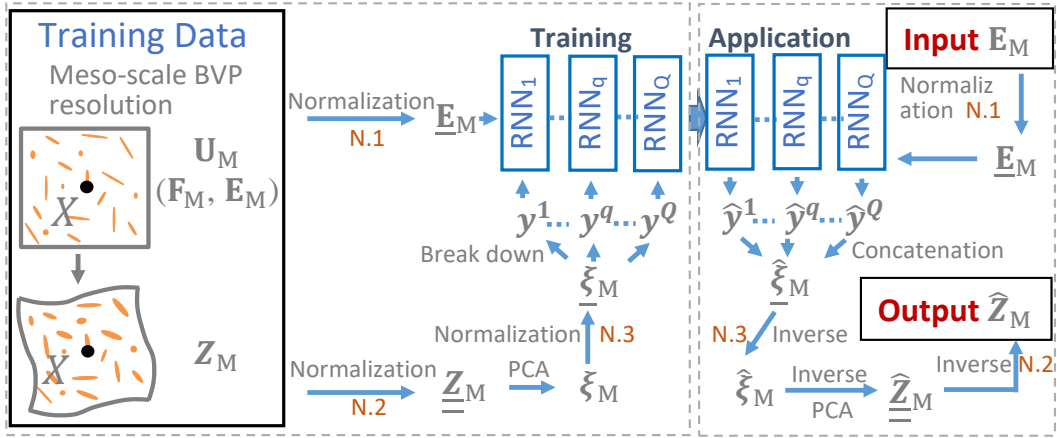}
		\label{fig:SurrogateSplit}}\\
		\subfigure[Detail of the RNNs]{\includegraphics[scale=0.7]{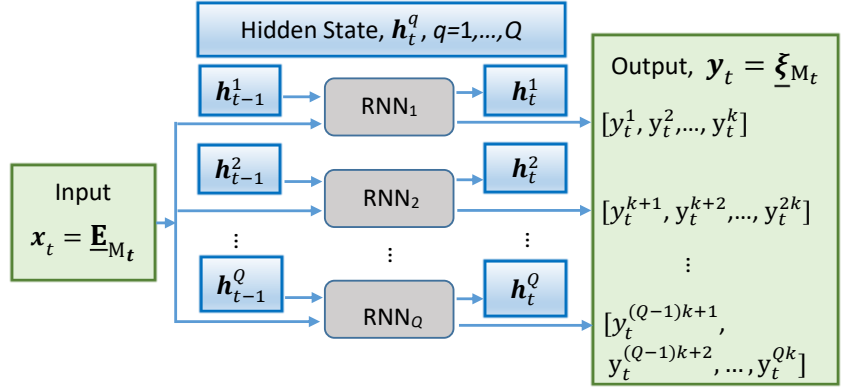}\label{fig:SplitRNN}}
	\end{center}\caption{Data-driven surrogate model III with \secondreviewer{dimensionality reduction and break down}: (a) Training and on-line data flows; (b) Data flow in the $Q$ RNN models $\text{RNN}_q$.}\label{fig:DataSurrogateIII}
\end{figure}

Based on the PCA \secondreviewer{dimensionality reduction} of the output, the RNN training task can be further relieved by considering a \secondreviewer{dimensionality break down}.
The \secondreviewer{high dimensional} output is divided into $Q$ groups, and one RNN is defined per group to reproduce only a part of the total output.
Accordingly, the data flow structure of RNNs presented in Fig. \ref{fig:SurrogateSplit} is adopted.
\changes{The outputs $\bm{y}=\underline{{\bm{\xi}}}_\text{M}$ to be split correspond to the PCA low dimensional data, $\bm{\xi}_\text{M}$, having been normalized by feature. Considering that the first $p$ eigen-values have been retained by the PCA, the dimensionality of $\underline{{\bm{\xi}}}_\text{M}$ is also $p$. This outputs vector is then split into $Q$ sub-vectors either evenly, as in Fig. \ref{fig:SplitRNN}, or unevenly. In this work we opt for an event split following the order of the $p$ retained eigen-values $\Lambda_1\geq\Lambda_2\geq\ldots\geq\Lambda_{p}\geq\ldots\geq\Lambda_{d}$. Considering that the normalized coefficients vector $\underline{{\bm{\xi}}}_\text{M}$ follows the same order, the $Q$ reduced dimensionality outputs $\bm{y}^q$, with $q=1,\,\ldots,\,Q$ gather the components $(q-1)k+1$ to $qk$ of the original outputs vector, \emph{i.e.} $\bm{y}^q=[\underline{{\xi}}^{(q-1)k+1}_\text{M},\,\ldots,\, \underline{{\xi}}^{qk}_\text{M}]$, where $k=p/Q$ is the dimensionality of the $Q$ outputs vectors after dimensionality break down.}
In Fig. \ref{fig:SplitRNN}, each block of ``$\text{RNN}_q$'', $q=1,\,2,\,\ldots,\, Q$, has the structure presented in Fig. \ref{fig:GRU}, and is independent of the other RNNs.
Therefore, they can be trained separately and the low dimensional output of each RNN leads to a reduced number of learnable parameters in its respective feed-forward network $\text{NNW}_\text{O}$.


\subsection{RNNs training with mini-batches using graphics processing unit (GPU)}\label{sec:minibatch}

\secondreviewer{The training data of RNNs are sequential data and as a result it is impossible to use the whole synthetic database at once for training.
Therefore, the training data are divided into a number of mini-batches, and the RNNs are trained with data shifting among these mini-batches.
The number of sequences in each mini-batch depends on the sequence length and dimensionality, and on the accessible computer memory.}
\firstreviewer{For large scale RNNs, the training can be accelerated using GPUs as they can process multiple computations simultaneously. In that case, the mini-batch size is constrained by the memory of the GPU instead of the CPU memory.}

\begin{figure}[htb] 
	\begin{center}
		\includegraphics[scale=1.0]{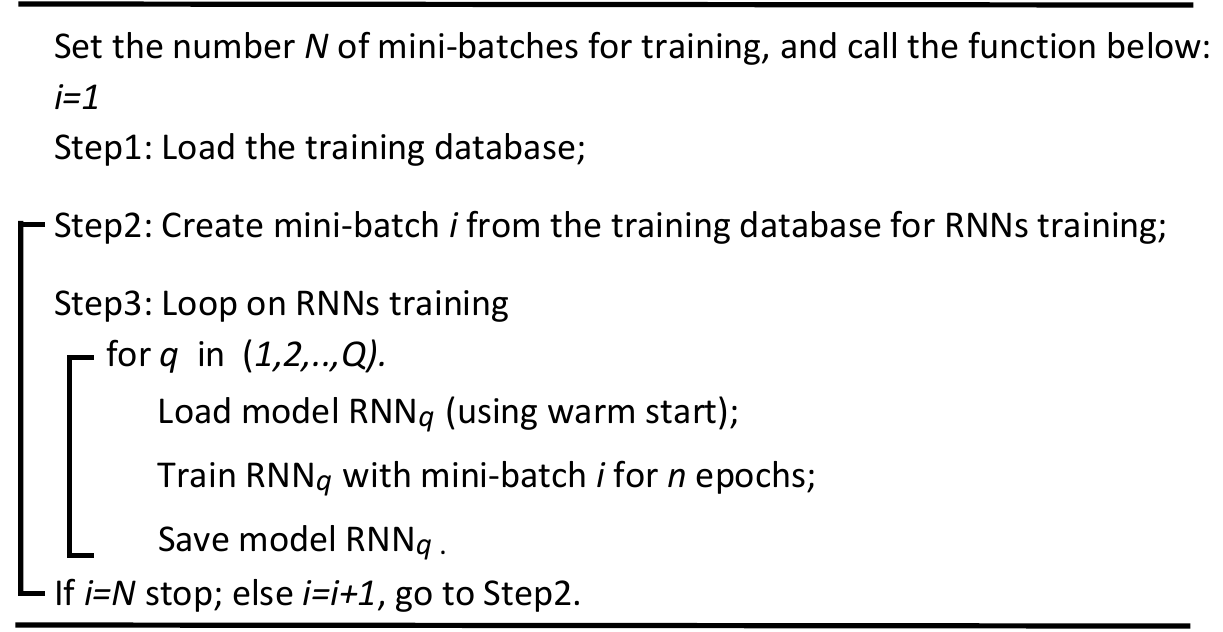}
		\caption{Flow chart of RNNs training with mini-batches.}\label{fig:flowchart}
	\end{center}
\end{figure}

The training flow chart is given in Fig. \ref{fig:flowchart}.
\begin{itemize}
	\item
	  Practically, the data in each mini-batch are randomly picked from the training database, and the number of mini-batches $N$ is normally set with a high value such as a few thousands.
	\item 
	The training epoch for each mini-batch, $n-$epoch, is set with a value within $10$ to avoid over-fitting because of the small dataset. 
	\item
	The training process presented in Section \ref{sec:RNNGen} is applicable for all the RNNs presented in Section \ref{sec:Surrogate}. If the \secondreviewer{dimensionality break down} is used, the model $\text{RNN}_q$ $(q=1,\ldots, Q)$ is only loaded at index $q$ of the loop, during its training stage, and saved afterward, because the parameters of each $\text{RNN}_q$ take a lot of memory. \secondreviewer{Since the creation of the random mini-batches from a large database is more time consuming than loading and saving each \changes{$\text{RNN}_q$ model}, a given mini-batch is used to train all the $\text{RNN}_q$ models, with the outputs being split accordingly to the RNNs.}
\end{itemize} 

We note
\begin{itemize}
  \item It appeared that the creation of the random mini-batches is quite time consuming, especially for \secondreviewer{high dimensional training data} for which the synthetic database may take tens of Gigabyte.
    In this case, it would be better to reload the database before the shift of mini-batch instead of keeping it in memory continuously and the ``go to Step 2'' in Fig. \ref{fig:flowchart} would be replaced by ``go to Step 1''.
  \item The applicable size of mini-batches can be increased in the case of \secondreviewer{dimensionality break down}, which means that the number of mini-batches, $N$, can be reduced. \firstreviewer{Because of the reduction in the number of mini-baches and since the training using GPU with small $n-$epoch is fast, the training time is not necessarily longer than the training of the RNN using \secondreviewer{full dimensional output} of PCA despite the fact that the \secondreviewer{dimensionality break down} results in more RNNs to be trained.}
\end{itemize}

\section{Synthetic database generation for training and testing of the RNNs}\label{sec:trainingData}

In \cite{WU2020113234} and in Section \ref{sec:NNW}, we have designed RNNs which provide history-dependent non-linear mappings, respectively, from $\textbf{E}_\text{M}(t)$ to $\textbf{S}_\text{M}(t)=\textbf{F}_\text{M}^{-1}(t)\cdot\textbf{P}_\text{M}(t)$, the macro-scale second Piola-Kirchhoff stress tensor from which $\textbf{P}_\text{M}(t)$ can be directly deduced, and from $\textbf{E}_\text{M}(t)$ to $\bm{Z}_\text{M}(t)$, the state variables.
Once trained these RNNs constitute surrogate models of a meso-scale BVP in $\text{FE}^2$ analyses, to replace costly micro-scale finite element simulations.
In this section, the generation of the synthetic dataset required for the training and testing of the RNNs, which has already been detailed in \cite{WU2020113234} in the context of the non-linear mapping $\textbf{E}_\text{M}(t)$ to $\textbf{S}_\text{M}(t)$, is reframed here in the context of the non-linear mapping $\textbf{E}_\text{M}(t)$ to $\bm{Z}_\text{M}(t)$ for completeness.
\firstreviewer{The training and use in multi-scale simulations of the surrogate models acting as non-linear mappings from $\textbf{E}_\text{M}(t)$ to $\textbf{S}_\text{M}(t)$ and from $\textbf{E}_\text{M}(t)$ to $\bm{Z}_\text{M}(t)$ are then briefly explained.}

\subsection{Data collection from micro-scale finite element simulations}

The training and test data are generated by solving the meso-scale BVP on a RVE controlled by the history of $\mathbf{E}_\text{M}$.
Since $\mathbf{E}_\text{M}$ cannot be used directly to define the boundary condition on the RVE, the right stretch tensor ${\mathbf{U}}_\text{M}$ is used instead of $\mathbf{E}_\text{M}$: 
\begin{itemize}
	\item
	Firstly, ${\mathbf{F}}_\text{M}$ has a unique decomposition which reads,
	\begin{eqnarray}\label{eq:UtoF}
	{\mathbf{F}}_\text{M} = {\mathbf{R}}_\text{M}\cdot{\mathbf{U}}_\text{M}\,,
	\end{eqnarray}
	where ${\mathbf{R}}_\text{M}$ is the rotation tensor, which can be defined arbitrarily because of the frame indifference of the meso-scale BVP. Therefore, ${\mathbf{R}}_\text{M} = \mathbf{I}$ for no rotation is used for simplicity.
	\item 
	Secondly, $\mathbf{E}_\text{M}$ is obtained from
	\begin{equation}\label{eq:UtoE}
	{\mathbf{E}}_\text{M} = \frac{1}{2}\left({\mathbf{U}}_\text{M}^2-\mathbf{I}\right)\,.
	\end{equation}
\end{itemize}

For a given $\mathbf{U}_\text{M}$, we can constrain the meso-scale BVP deformation according to $\mathbf{F}_\text{M}$ computed by Eq. (\ref{eq:UtoF}) and we can obtain the input of the RNN, $\mathbf{E}_\text{M}$, through Eq. (\ref{eq:UtoE}).
The state variables of \changes{the} meso-scale volume element, $\bm{Z}_\text{M}$, are then collected from each resolved meso-scale BVP by considering the state variables distribution $\{\secondreviewer{\bm{Z}_\text{m}(\bm{x},\,t)}\,:\; x\in\omega({\bm{X}})\}$ of the RVE $\omega$ and serve as \secondreviewer{observation} output of the RNN.

\subsection{Random loading paths} \label{sec:RandomPath}

The loading history-dependency of the RNNs arises from their architecture and their training requires using sequential data.
Considering the involved dimensionality of $\mathbf{U}_\text{M}$ and the variety of possible loading/unloading paths in a multi-scale simulation, using proportional loading paths is not \secondreviewer{an optimal choice to cover all the possible changes} of loading directions.
Therefore, random loading paths, which have been shown to be efficient in \cite{WU2020113234}, are adopted to generate a sequential synthetic database of $\mathbf{U}_\text{M}(t)$.

A loading path is defined by a sequence of right stretch tensors, such as $\{ \mathbf{U}_{\text{M}\,0},\, \mathbf{U}_{\text{M}\,1},\, \ldots,\,\mathbf{U}_{\text{M}\,N}\}$, where $\mathbf{U}_{\text{M}\,0}=\mathbf{I}$ is the starting stage of the loading process.
Random increments of loading, $\Delta\mathbf{U}_{\text{M}\,n}=\mathbf{U}_{\text{M}\,n}-\mathbf{U}_{\text{M}\,n-1}$, are then generated and permit the loading path to be changed in any possible direction at each increment.
As a symmetric second-order tensor, each increment $\Delta\mathbf{U}_\text{M}$ has its spectral decomposition form expressed by its eigenvalues and eigenvectors as,
\begin{eqnarray}
\Delta\mathbf{U}_{\text{M}\,n} = \Delta\lambda_1 \mathbf{n}_1\otimes \mathbf{n}_1 + \Delta\lambda_2 \mathbf{n}_2\otimes \mathbf{n}_2 +\Delta\lambda_3 \mathbf{n}_3\otimes \mathbf{n}_3 \,,
\end{eqnarray}
where the eigenvectors $\mathbf{n}_1$, $\mathbf{n}_2$, and $\mathbf{n}_3$ control the direction of the loading path increment, and the eigenvalues $\Delta\lambda_1$, $\Delta\lambda_2$, and $\Delta\lambda_3$ control the increment size. 
The increment of loading, $\Delta\mathbf{U}_{\text{M}\,n}$, is obtained through randomly generated orthogonal vectors $\mathbf{n}_1$, $\mathbf{n}_2$, and $\mathbf{n}_3$, and three eigenvalues which satisfy $\sqrt{\Delta\lambda_1^2+\Delta\lambda_2^2+\Delta\lambda_3^2}\leq \Delta R$, where $\Delta R$ is a defined upper bound of the increment size.
Practically, $\Delta\lambda_1^2$, $\Delta\lambda_2^2$ and $\Delta\lambda_3^2$ can be easily generated by a random split of a random variable $\mathcal{R} \in(0,\Delta R^2]$, or $\in(\Delta R_\text{min}^2,\,\Delta R^2]$ in order to avoid tiny steps, see \cite{WU2020113234} for more details.
 \secondreviewer{Because of this way of generating the deformation increment, a wide range of loading step sizes is considered when generating the random loading paths.}
We note that during the RVE resolution, for a given $\Delta\mathbf{U}_{\text{M}\,n}$, sub-steps may be needed to guarantee the numerical convergence of the meso-scale BVP resolution.

The random walk or path generation process is terminated when the eigenvalues of $\mathbf{U}_{\text{M}\,N}$ reach $\max_i\{\abs{\lambda_i}\}>R_\text{max}$, where $R_\text{max}$ is a given critical value characterizing the training range of the surrogates.
In this work, we consider 2D loading cases with $\Delta R=5\times 10^{-3}$ and $R_\text{max}=0.1$ when generating the loading paths.
Beside thousands of simulations following a random loading path, hundreds of proportional loading paths are also used to increase the coverage of loading paths in the strain space when training the RNNs. \firstreviewer{The proportional cyclic loading paths are created by considering a random loading direction and several random reversal points.}

\secondreviewer{For each random loading path, there is a succession of volume dilatations and compressions --depending on the sign of the trace of $\mathbf{E}_{\text{M}}$-- during the sequential loading, while for each cyclic loading path, there is a reversal in the volume dilatation or compression during the unloading/reverse loading stage. The amplitude of the dilatation or compression varies for each generated path.}

\subsection{Pre-trimming and data padding/trimming}\label{sec:NNW_train}

\secondreviewer{
\begin{figure}[!htb]
	\centering
	\subfigure[]{\includegraphics[scale=0.5]{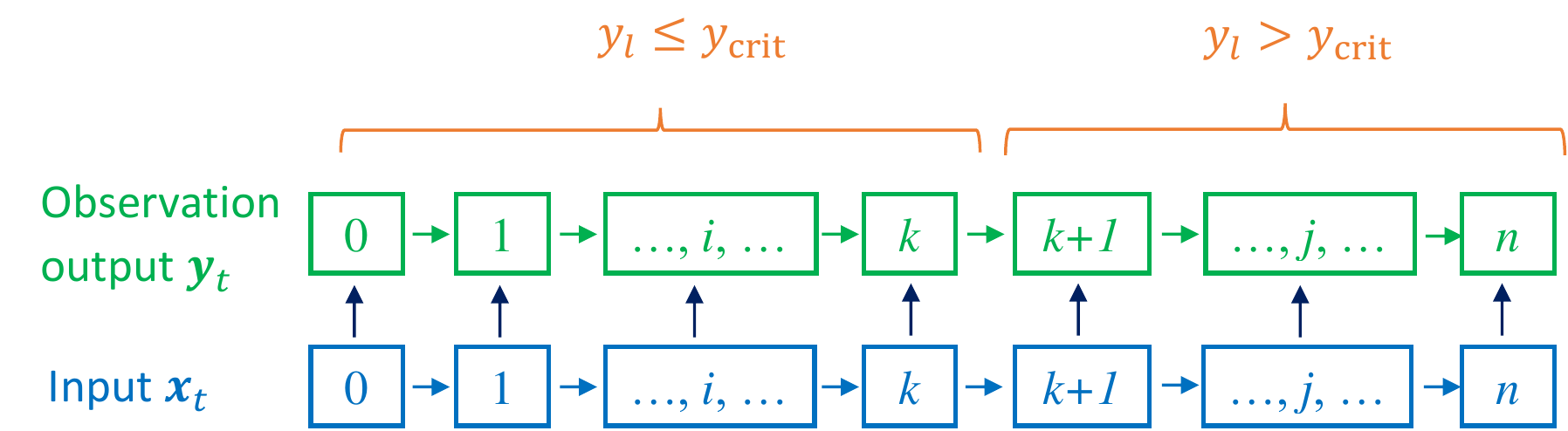}\label{fig:OrigSequenceCrit}}\\
	\subfigure[]{\includegraphics[scale=0.5]{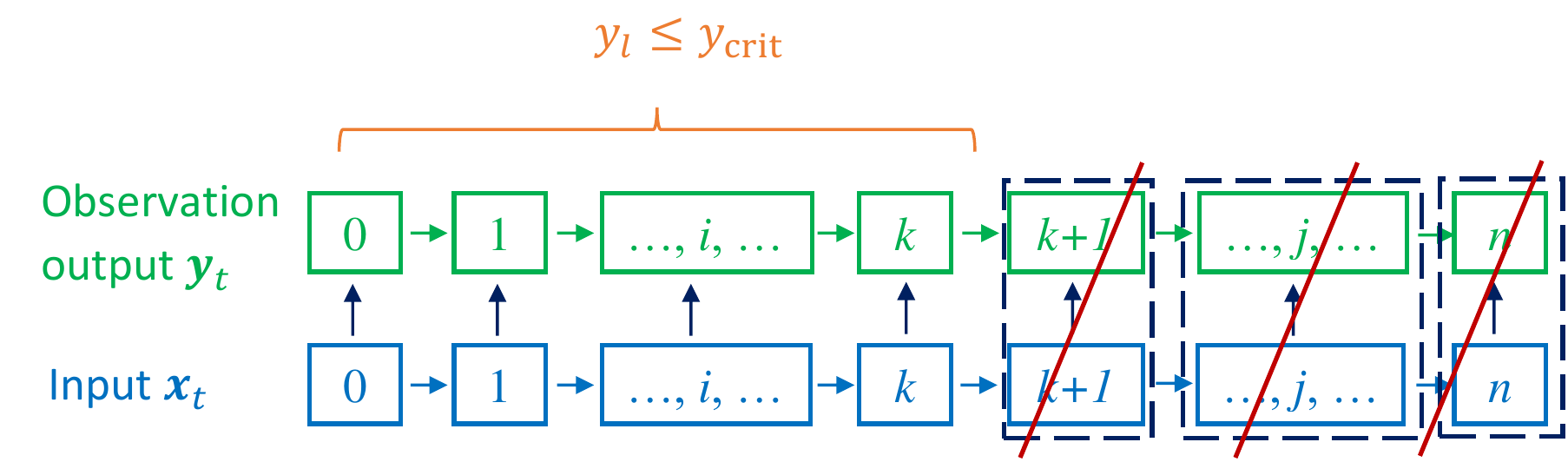}\label{fig:TrimCrit}}
	\caption{Trimming of a sequence when observation data are out of range: (a) Original sequence with observation data component $l$ within the critical value $y_{\text{crit}}$ up to index $k$; and (b) Trimmed sequence so that observation data component $l$ remains always within the critical value $y_{\text{crit}}$.}\label{fig:criticalTrimming}
\end{figure}
}

Since random loading paths are applied on the RVEs, this could yield some extreme values for the considered state variables, such as local equivalent plastic strain higher than $20$ in the RVE.
On the one hand, such extreme values are not physical, and on the other hand, this occurs only for a few training sequences, whose state variables evolve in such an extreme range that the data covering this extreme range are not enough to train a NNW within this range.
Therefore, a pre-trimming is applied on the sequential data, to remove the subsequent steps in each sequence when an element of a considered state variables vector reaches a critical value, $y_\text{crit}$, \secondreviewer{as illustrated in Fig. \ref{fig:criticalTrimming}.}

\secondreviewer{
\begin{figure}[!htb]
	\centering
	\subfigure[]{\includegraphics[scale=0.5]{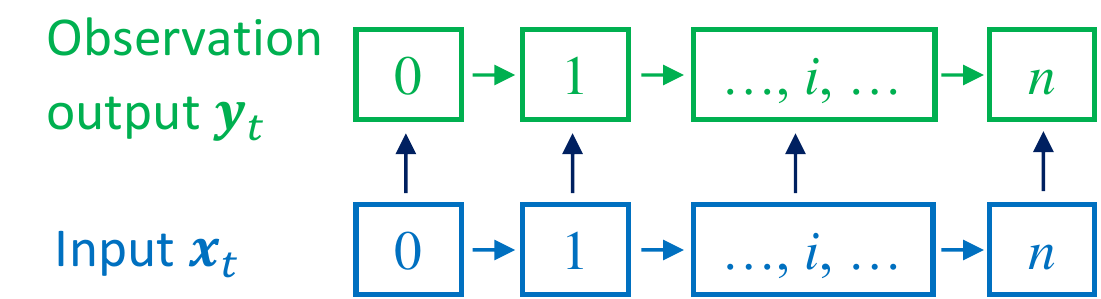}\label{fig:OrigSequence}}\\
	\subfigure[]{\includegraphics[scale=0.5]{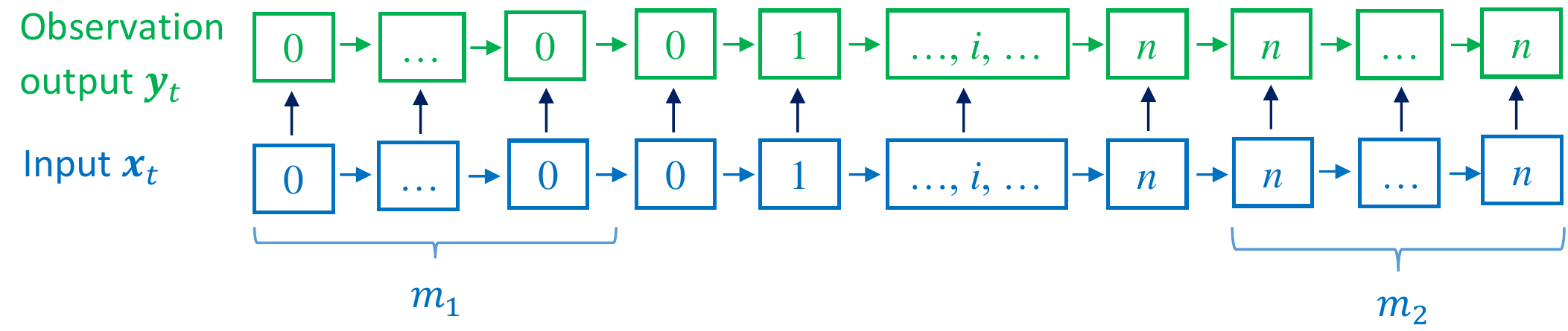}\label{fig:Pad}}\\
	\subfigure[]{\includegraphics[scale=0.5]{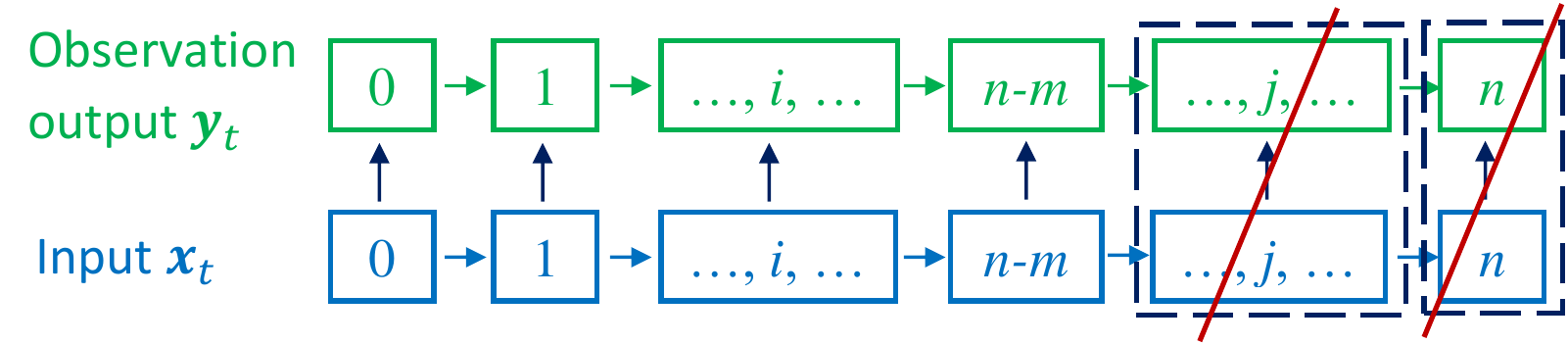}\label{fig:Trim}}
	\caption{Padding and trimming of a sequence in order to reach the targeted length: (a) Original sequence of size $n+1$; (b) Padding at the beginning of the sequence by repeating the entry number 0 $m_1$ times and the last entry number $n$ $m_2$ times in order to reach a sequence \changes{of} length $n+1+m_1+m_2$; and (c) Trimming of the sequence by removing the last $m$ entries to reach a length $n+1-m$.}\label{fig:trimmingPadding}
\end{figure}
}

RNNs are typically able to take in variable size inputs, but training data of the same sequential length will usually be used to feed the RNNs in batches to speed up the training process.
However, the lengths of the training data sequences obtained by a \secondreviewer{random walk process} are different from each other, either because a \secondreviewer{random walk process} is terminated when a critical strain measure is reached as discussed in Section \ref{sec:RandomPath} or because a sequence is pre-trimmed according to a critical value of state variable as discussed here above.
In order to use batches to train the RNNs, it is needed to ensure that each sequence within the input database is of equal size.
Therefore, both zero padding at the beginning and repeatedly adding the last element at their end are used for short sequences, while long sequences are trimmed from their end\secondreviewer{, see Fig. \ref{fig:trimmingPadding}}. 

\firstreviewer{
\subsection{Use of the surrogate models in multi-scale analyses}\label{sec:practicalApplication}

In \cite{WU2020113234} the authors have developed a surrogate model of the history-dependent non-linear mapping from $\textbf{E}_\text{M}$ to $\textbf{S}_\text{M}$ by training a recurrent neural network whose architecture is depicted in Fig. \ref{fig:GRU}. This surrogate model is used on-line during multi-scale simulations allowing accelerating the resolution process.

In Section \ref{sec:NNW}, we have developed a similar recurrent neural network, possibly enhanced by dimensionality reduction, to act as a surrogate model for the history-dependent non-linear mapping from $\textbf{E}_\text{M}$ to $\bm{Z}_\text{M}$, the micro-scale state variables distribution.

Both surrogate models are trained off-line, and once trained can be used to conduct multi-scale simulations. Their use is however different. On the one hand, the surrogate model of the history-dependent non-linear mapping from $\textbf{E}_\text{M}(\bm{X},\,t)$ to $\textbf{S}_\text{M}(\bm{X},\, t)$ has to be used on-line at each macro-scale material point $\bm{X}\in\Omega$ during the multi-scale process since there exists a strong non-linear coupling between the two scales, see the application example in \cite{WU2020113234}. On the other hand, the history-dependent non-linear mapping from $\textbf{E}_\text{M}(\bm{X},\,t)$ to $\bm{Z}_\text{M}(\bm{X},\,t)$ can be used \emph{a posteriori} from the recorded strain history $\textbf{E}_\text{M}(\bm{X},\,t)$ at one macro-scale material point $\bm{X}\in\Omega$ since this does not affect the homogenized material response. Also it is usually not meaningful to recover micro-scale state variables distributions at all macro-scale material points $\bm{X}\in\Omega$, but only in regions of interest such as stress concentration ones.}

\section{Application on a composite RVE}\label{sec:compositeRVE}

In this section, the finite element simulations performed on a composite RVE to provide the training and testing database of the history-dependent non-linear mapping from $\textbf{E}_\text{M}(\bm{X},\,t)$ to $\bm{Z}_\text{M}(\bm{X},\,t)$ are first described.
The considered state variables and their surrogate modeling are then specified.
The \secondreviewer{dimensionality reduction} method presented in Section \ref{sec:PCA} is then applied and the error measure that can be used to compare the different surrogate models of the micro-scale state variables distribution is defined.
The accuracy and performance of the different surrogate models developed in Section \ref{sec:Surrogate} are then studied by reconstructing the distributions of the equivalent plastic strain $\bm{\gamma}=\{\gamma(\bm{x}):\,\bm{x}\in\Omega\}$ and of the equivalent von Mises stress $\bm{\tau}_\text{eq}=\{{\tau}_\text{eq}(\bm{x}):\,\bm{x}\in\Omega\}$ in the RVE $\Omega$ for different loading scenarios\footnotemark[1].
Finally, an efficient design methodology of the RNN structure is proposed from the drawn observations.

\subsection{Micro-scale simulations}

\begin{figure}[htb] 
	\begin{center}
		\includegraphics[scale=0.23]{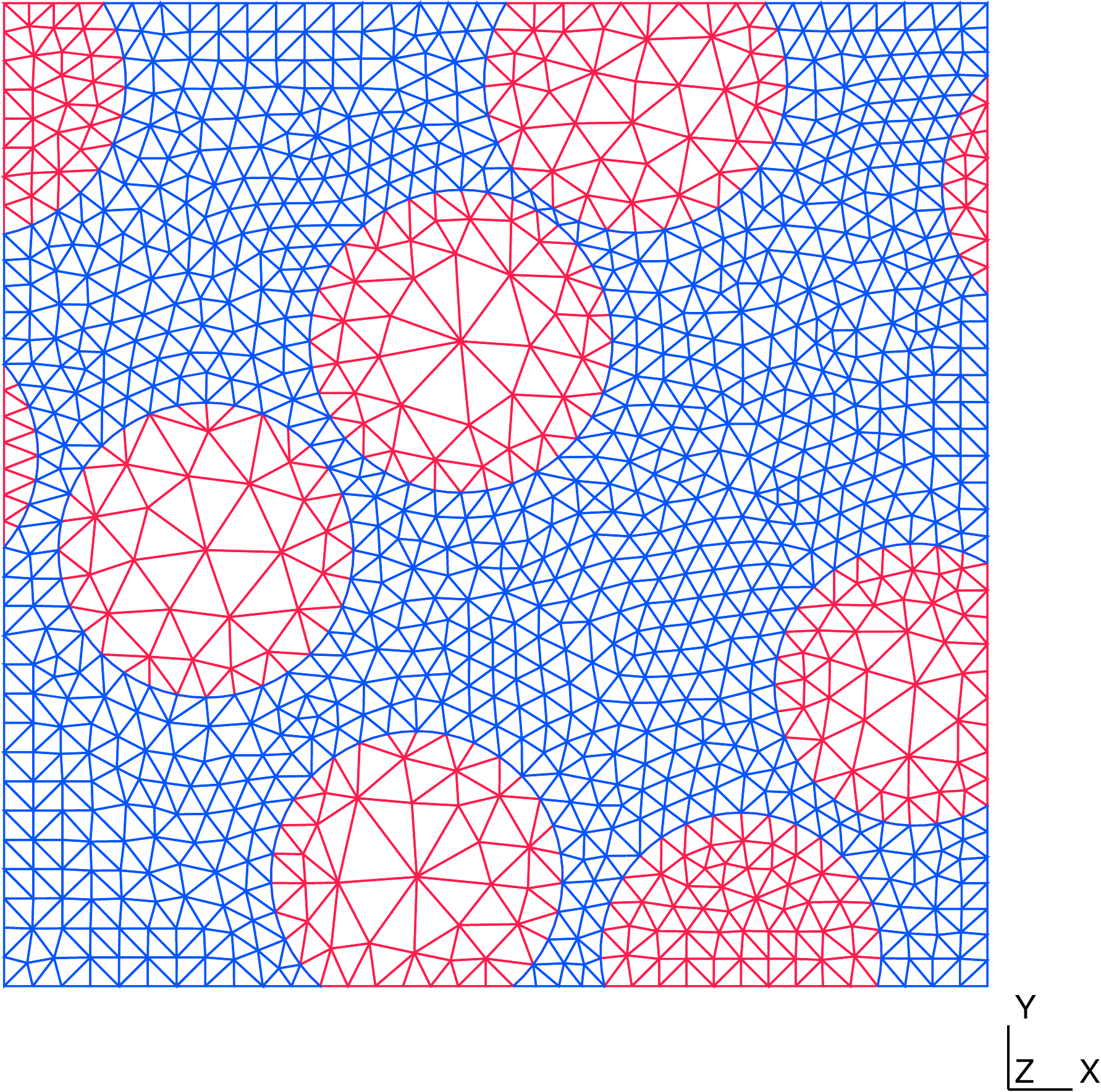}
		\caption{Finite element mesh of the micro-structure volume element of dimensions $0.02$mm$\times$$0.02$mm used to build the database.}
		\label{fig:rve}
	\end{center}
\end{figure}
The 2D RVE made of a continuous fiber reinforced elasto-plastic matrix material shown in Fig. \ref{fig:rve} is adopted for the micro-scale simulations. The volume fraction of fiber is 39.9\%.

We generate a total of $6954$ loading paths to build the training database: the training data includes $6457$ random and $497$ cyclic loading paths.
We additionally generate a total of $1551$ loading paths to build the testing database: the testing data includes $1252$ random and $299$ cyclic loading paths.

\subsubsection{Fiber}

The fibers obey a hyperelastic law based on the elastic potential
\begin{eqnarray}\label{eq:potentialFibers}
\psi_\text{fib}(\mathbf{C}) = \frac{K_\text{fib}}{2} \ln^2{J} + \frac{\mu_\text{fib}}{4} \left(\ln{\mathbf{C}}\right)^{\text{dev}}: \left(\ln{\mathbf{C}}\right)^{\text{dev}} \,,
\end{eqnarray}
where $J=\det\mathbf{F}$ is the Jacobian, $\mathbf{C}=\mathbf{F}^\text{T}\cdot\mathbf{F}$ is the right Cauchy strain tensor, $(\bullet)^\text{dev}$ denotes the deviatoric part of a second-order tensor $\bullet$, $K_\text{fib}$ and $\mu_\text{fib}$ are the material constants.
In this section the subscript ``m'' referring to the micro-scale is omitted for conciseness.
The stress tensor is deduced as
\begin{eqnarray} \label{eq:ep_piolaFibers}
\mathbf{P} =  \frac{\partial \psi_\text{fib}\left(\mathbf{F}\right)}{\partial \mathbf{F}} = K_\text{fib}\mathbf{F}^\text{-T}\text{ln}J+ \mathbf{F}^\text{-T}\cdot \left[\mu_\text{fib}\ln{\mathbf{C}}^\text{dev}\right]\,.
\end{eqnarray}

\subsubsection{Matrix}
The matrix obeys a finite strain $J_2$-elasto-plastic material model \cite{cuitino1992material}. The elastic potential energy is defined by
\begin{eqnarray}
\psi_\text{mat}(\mathbf{C}^\text{e}) = \frac{K_\text{mat}}{2} \ln^2{J} + \frac{\mu_\text{mat}}{4} \left(\ln{\mathbf{C}^\text{e}}\right)^{\text{dev}}: \left(\ln{\mathbf{C}^\text{e}}\right)^{\text{dev}} \,,
\end{eqnarray}
where $ \mathbf{C}^\text{e} = \mathbf{F}^\text{eT}\cdot\mathbf{F}^\text{e} $, and $K_\text{mat}$ and $\mu_\text{mat}$ are material constants.
In this section the subscript ``m'' referring to the micro-scale is omitted for conciseness.
The deformation gradient $ \mathbf{F} $ is decomposed into a reversible elastic part $ \mathbf{F}^\text{e} $ and an irreversible plastic part $ \mathbf{F}^\text{p} $ such that $ \mathbf{F} = \mathbf{F}^\text{e} \cdot \mathbf{F}^\text{p} $.
The first Piola--Kirchhoff stress tensor $\mathbf{P}$ thus reads
\begin{eqnarray} \label{eq:ep_piolaMatrix}
\mathbf{P} = \frac{\partial \psi_\text{mat}  \left(\mathbf{F};\mathbf{F}^\text{p}\right) }{\partial \mathbf{F}} = K_\text{mat}\mathbf{F}^\text{-T} \ln J+ \mathbf{F}^\text{e}\cdot \left[\mu_\text{mat}{\mathbf{C}^{\text{e}\,-1}}\cdot(\ln{\mathbf{C}^\text{e}})^\text{dev}\right] \cdot\mathbf{F}^\text{p\,-T}\,. 
\end{eqnarray}
The plastic part $\mathbf{F}^\text{p}$ of the deformation gradient is obtained through a $J_2$-plastic flow expressed in terms of the Kirchhoff stress. The Kirchhoff stress $\bm{\kappa}=\mathbf{P}\cdot\mathbf{F}^\text{T}$ is first computed by Eq. (\ref{eq:ep_piolaMatrix}) as
\begin{eqnarray}
\bm{\kappa} =  K_\text{mat}  \ln J \mathbf{I}+ \mathbf{F}^\text{e}\cdot \left[\mu_\text{mat}{\mathbf{C}^{\text{e}-1}}\cdot(\ln{\mathbf{C}^\text{e}})^\text{dev}\right] \cdot\mathbf{F}^\text{e\,T}\,,
\end{eqnarray}
and the equivalent von Mises stress is calculated through $\tau_\text{eq} = \sqrt{\frac{3}{2}\bm{\kappa}^\text{dev}:\bm{\kappa}^\text{dev}}$.
The von Mises stress criterion finally reads
\begin{eqnarray}\label{eq:J2yielding}
f = \tau_\text{eq} -  \tau_\text{y}^0 - R(\gamma) \le 0\,,
\end{eqnarray}
where $f$ is the yield surface, $\tau_\text{y} ^0$ is the initial yield stress, $\gamma$ is the equivalent plastic strain and $R(\gamma)= Y\left[1- \exp(-k \gamma)\right]$ is the isotropic hardening stress, with $Y$ and $ k $ being the material constants.
The evolution of $\mathbf{F}^\text{p}$ is determined by the normal plastic flow, as
\begin{eqnarray}
\dot{\mathbf{F}}^\text{p} = \dot{\gamma}\mathbf{N}\cdot\mathbf{F}^\text{p}\,,
\end{eqnarray}
where $\mathbf{N}$ is the plastic normal, see \cite{cuitino1992material} for more details. 

\secondreviewer{Finally, the material properties used in this work correspond to the ones used in \cite{WU2020113234,VANDUNG2021} and are reported in Table \ref{tab:materialProperties}.}

\begin{table}[h]  
	\caption{Material properties for fiber and matrix.}  	\label{tab:materialProperties}
	\centering  
	\begin{tabular}{cc|ccccc} 
		\hline\hline   
		\multicolumn{2}{c}{Fiber} &\multicolumn{5}{c}{Matrix}  	\\  
		\hline   
		$K_\text{fib}$ [GPa] &$\mu_\text{fib}$ [GPa] &$K_\text{mat}$ [GPa] & $\mu_\text{mat}$ [GPa]& $\sigma_\text{y}$[MPa] &$Y$ [MPa] & $k$ [-] \\  
		16.67& 12.50& 2.50 & 1.15 & 100 & 20 & 30 \\
		\hline 
	\end{tabular}  
\end{table}

\subsection{Considered state variables: Equivalent plastic strain and von Mises stress}\label{sec:stateVar}

\secondreviewer{The distributions of the equivalent plastic strains, $\gamma$, and of the von Mises stresses, $\tau_\text{eq}$, see Eq. (\ref{eq:J2yielding}), within the RVE are chosen to represent respectively a monotonic increasing state variable and a non monotonic state variable\footnotemark[1]. This allows assessing the accuracy of the surrogate models for these two different behaviors. Besides, as described in the introduction, the surrogate models can be used for posterior estimation of the structure integrity, in which case the plastic strain is relevant when predicting the damage initiation, or to obtain an estimation of the life of the composite material, in which case fatigue failure criteria based on stress tensor invariants such as the von Mises stress can be used \cite{BERREHILI20101389}, hence motivating the study of the latter.}
The dimensionality of each state variable corresponds to the number of Gauss points of the RVE finite element discretization.
Second-order triangular elements with three Gauss points are used in the direct simulations.
However, to be consistent with their visualization in the post-process, the averaged equivalent plastic strains, $\gamma$, and von Mises stresses, $\tau_\text{eq}$, of each element in the RVE are retained, \secondreviewer{although the method can readily be applied when considering the values at all the Gauss points.} 
Besides, since the fibers are purely elastic, the value of $\gamma$ is irrelevant in this phase.
Therefore, for the RVE presented in Fig. \ref{fig:rve}, we eventually have as selected state variables $\bm{Z}_{\text{M}}$ 
\begin{itemize}
	\item $\bm{\gamma} = [\gamma_1,\,\gamma_2,\,\ldots,\,\gamma_{d_\gamma}]$ with $d_\gamma = 1607$;
	\item $\bm{\tau}_\text{eq} = [{\tau_\text{eq}}_1,\,{\tau_\text{eq}}_2,\,\ldots,\,{\tau_\text{eq}}_{d_\tau}]$ with $d_\tau = 2237$.
\end{itemize} 

The training and test data are thus collected from the direct finite element simulations on the RVE in a time sequential form, with
\begin{itemize}
	\item
	input \secondreviewer{$\bm{x}_t=\left\lbrace {\mathbf{E}_\text{M}}_1,\,{\mathbf{E}_\text{M}}_2,\,\ldots,\,{\mathbf{E}_\text{M}}_t\right\rbrace$}, and
	\item
	 \secondreviewer{observation} output  $\bm{y}_t$ the union of
	\secondreviewer{$\left\lbrace \bm{\gamma}_1,\,\bm{\gamma}_2,\,\ldots,\,\bm{\gamma}_t\right\rbrace$ and $\left\lbrace {\bm{\tau}_\text{eq}}_1,\,{\bm{\tau}_\text{eq}}_2,\,\ldots,\,{\bm{\tau}_\text{eq}}_t\right\rbrace$}.	
\end{itemize}

\subsection[PCA order reductions on state variables]{PCA order reductions on $\bm{\gamma}$ and $\bm{\tau}_\text{eq}$}\label{sec:PCAInternalVar}

The linear order reduction process PCA described in Section \ref{sec:PCA} is applied on $\bm{\gamma}$ and $\bm{\tau}_\text{eq}$.
Since each loading path has about hundreds to thousands of steps and since we have more than 6000 loading paths in the training database, there are nearly $1,000,000$ samples of $\bm{\gamma}$ and $\bm{\tau}_\text{eq}$ available for the PCA process.
To be practical, only $1\%$ of the samples are randomly picked to perform the PCA.

\begin{figure}[!htb]
	\centering
	\subfigure[]{\includegraphics[scale=0.42]{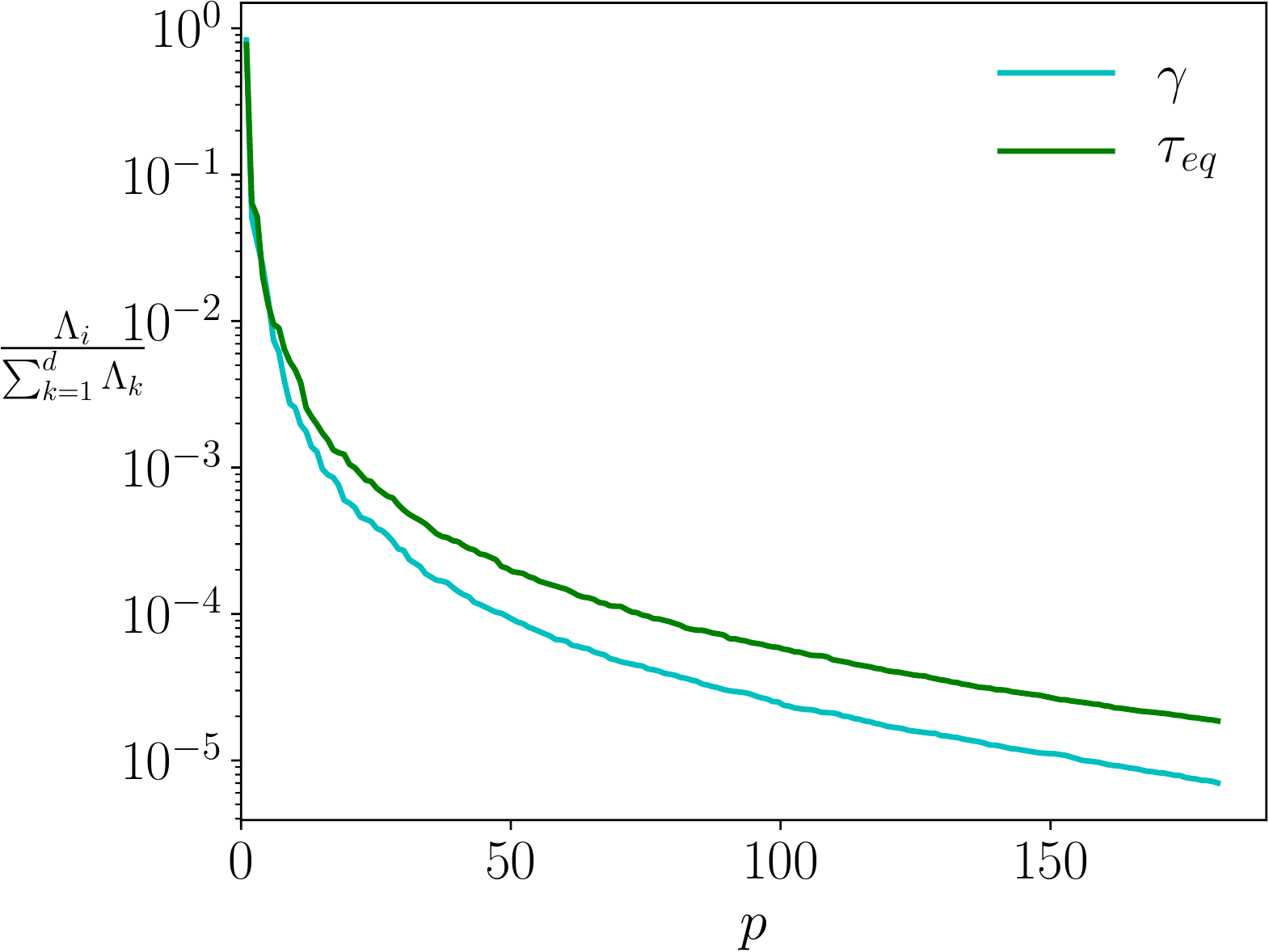}\label{fig:Eign_ratio}}\quad
	\subfigure[]{\includegraphics[scale=0.42]{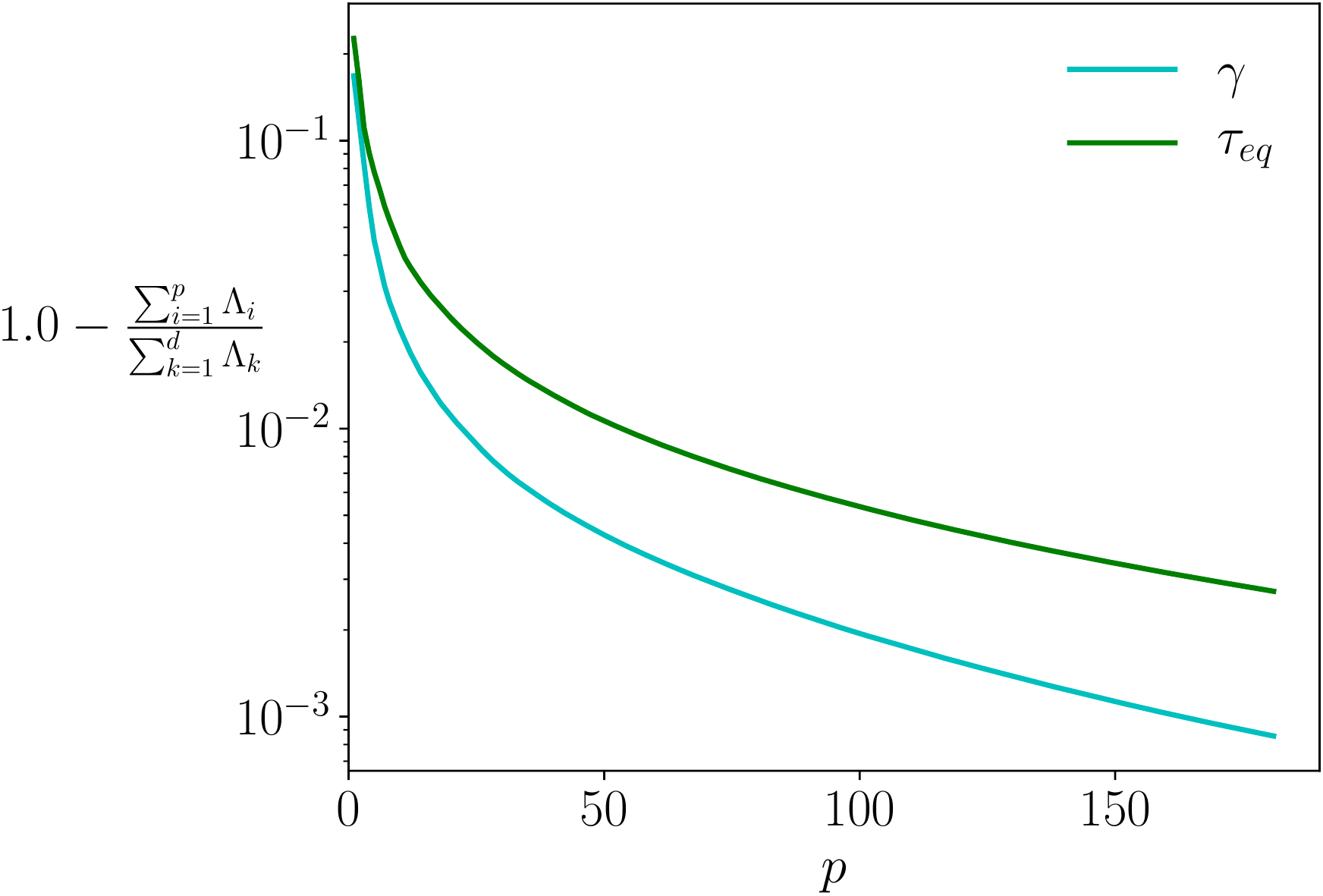}\label{fig:PCA_residual}}
	\caption{Application of PCA on the composite RVE state variables: (a) Coefficients of eigen-values; and (b) The residual fractional eigen-value, for the equivalent plastic strain $\bm{\gamma}$ and the von Mises stress $\bm{\tau}_\text{eq}$.}\label{fig:PCA_coefficient}
\end{figure}

The coefficients of eigen-value for the principal components and the residual fractional eigen-value, see Eq. (\ref{eq:reduce_accuracy}), are plotted as function of the reduced dimensionality $p$ in Fig. \ref{fig:PCA_coefficient} for both the equivalent plastic strain and the von Mises stress. 
\secondreviewer{From Fig. \ref{fig:Eign_ratio} it can be seen that the coefficients of eigen-value decrease fast for the first tens of principal components and then the decrease becomes slower. This is due to the realistic pattern of the micro-structure and the fact that it is subjected to various loading paths generated by random walk processes. As a result, many possible deformation modes exist and cannot be accurately represented by a couple of eigen-modes.}
Fig. \ref{fig:PCA_residual} provides a relative error measure of the reconstructed data from reduced dimensionality when conducting an inverse PCA.
If a rather low relative error is demanded, the reduced dimensionality $p$ will have to be high enough, especially for $\bm{\tau}_\text{eq}$.

\subsection{Mean Square Error (MSE) and relative error measure}

The error of NNWs is normally evaluated by the Mean Square Error (MSE) (\ref{eq:MSE}).
Since MSE is an absolute error defined on normalized data, its value depends on not only the accuracy of the NNWs, but also on the applied normalization.
Therefore, in our application, it is more meaningful to use a relative error instead of the MSE when we compare the errors of different \secondreviewer{surrogate models}, see Section \ref{sec:Surrogate}, in particular because some surrogate models use a PCA and others not.
Indeed, when considering the surrogate model II, see Section \ref{sec:RNN_PCA}, a PCA is applied on the observations to train the RNN, see Fig. \ref{fig:Surrogate}. Therefore, the outputs of RNNs, $\underline{\bm{\xi}}$, correspond to normalized reduced dimensionality data of PCA, each of them having a different weight on the data of original dimensionality during the reconstruction of the state variables ${\bm{Z}}_\text{M}$.
It would thus make no sense to compare the MSE of the RNN of the surrogate model II with the MSE of the RNN of the surrogate model I described in Section \ref{sec:RNN_direct} and illustrated in Fig. \ref{fig:SurrogateDR} which has as output directly the normalized state variables $\underline{\bm{Z}}_\text{M}$ obtained through a normalization by feature on the state variables $\bm{Z}_\text{M}$.

However, it is not straightforward to convert MSE to a relative error as presented in Fig. \ref{fig:PCA_residual}.
Therefore, the normalized state variables, $\underline{\bm{Z}}_\text{M}$, are chosen as the reference or target output for error evaluation of the three proposed \secondreviewer{surrogate models} developed in Section \ref{sec:Surrogate}.
In the cases of the surrogate models II and III, the RNNs predictions are first converted through inverse PCA to their original dimensionality (the dimensionality of the state variables), and the MSE loss is thus computed between the surrogate predictions and the normalized reference data (or observations).
The MSE of the PCA is defined as the error between the normalized original data and their reconstructions following the inverse PCA.
Therefore the relative error and MSE of PCA are associated through the reduced dimensionality $p$, which can serve as an intuitive connection to quantify the error of the surrogate model. 

\subsection[Surrogate model of the equivalent plastic strain distribution]{Surrogate model of the equivalent plastic strain distribution $\bm{\gamma}$} \label{sec:App_EpStrn}

The training dataset includes $6457$ random and $497$ cyclic loading paths.
All the training sequences are pre-trimmed with $\gamma_\text{crit}=6.0$, see Section \ref{sec:NNW_train}.
Then, a group of training sequences of length 800 is obtained by either padding or trimming the original ones.
A second group of training sequences of length 1200 is obtained by either padding or trimming the original sequences longer than 800.
The training mini-batches, see Section \ref{sec:minibatch}, are drawn from the groups of the two different lengths (some being thus repeated) randomly. 

In order to compare the proposed surrogate models and to study the effects of models hyper-parameters, such as the dimensionality of the hidden variables, a series of surrogate models have been trained.
Following Section \ref{sec:Surrogate}, the processes without \secondreviewer{dimensionality reduction}, with  PCA \secondreviewer{dimensionality reduction} and with PCA \secondreviewer{dimensionality reduction} and dimensionality break down are denoted as ``Surrogate I'', ``Surrogate II'' and ``Surrogate III'', respectively.
\secondreviewer{The ``Leaky ReLU'' activation function $f(\underline{\chi})=\text{max}\left(0,\,\underline{\chi}\right)+\frac{\text{min}\left(0,\,\underline{\chi}\right)}{100}$ is chosen in the feed-forward NNWs}, with no activation function applied on the input layer of $\text{NNW}_\text{I}$ and on the output layer of $\text{NNW}_\text{O}$. The initial value of hidden variables, $h_0=-1.0$, is used for the first input of a sequence.
One GRU layer is used in all the RNNs, \secondreviewer{see the details of the GRU layer in \ref{app:gru}.}

\begin{table}
	\centering
	\begin{threeparttable}		
		\caption{Hyper-parameters of RNNs to reconstruct the equivalent plastic strain distribution $\bm{\gamma}$; \firstreviewer{The notation $(n_0,\,\ldots,\,n_i,\,\ldots,\,n_N)$ holds for the number of input nodes $n_0$, of nodes $n_i$ in hidden layer $i$ and of output nodes $n_N$ --with the output nodes of a neural network holding as input nodes of the subsequent one and not being duplicated.}}\label{tab:RNN_NNW_EP}  
		\begin{tabular}{c|c|c|c} 
			\hline\hline   
			Model & $\text{NNW}_\text{I}$&hidden variables dimensionality \secondreviewer{$n_{\text{h}}$} &$\text{NNW}_\text{O}$\\ 
			\hline\hline 
			&(3, 70)&100&(800, 1607)\\
			Surrogate I&(3, 70)&200&(800, 1607)\\
			&(3, 70)&400&(800, 1607)\\
			\hline 		
			&(3, 70)&100&(800, 180)\\
			Surrogate II&(3, 70)&200&(800, 180)\\
			&(3, 70)&400&(800, 180)\\
			\hline 
			&(3, 70)&100&$(100,\,10)^{**}$\\
			Surrogate III$^{*}$&(3, 70)&200&$(100,\,10)^{**}$\\
			&(3, 70)&400&$(100,\,10)^{***}$\\
			\hline\hline		
		\end{tabular}  
		\begin{tablenotes}
		  \small
            \item {${}^{*}$  We have $Q=18$ $\text{RNN}_q$, each one having the structure reported in the columns}; 
			\item {${}^{**}$ \firstreviewer{$Q=18$, and only $\text{RNN}_q$ with $q=1,\,2,\,\ldots,\,9,\,10$ are trained and the $\text{RNN}$ step is not applied for $q=11,\,12,\ldots\,,18$}};
			\item {${}^{***}$ $Q=18$, and $\text{RNN}_q$ with $q=1,\,2,\,\ldots,\,18$ are trained}.
		\end{tablenotes}
	\end{threeparttable}
\end{table}

The considered hyper-parameters of the series of RNNs for the three presented surrogate models are detailed in Table \ref{tab:RNN_NNW_EP}.
We note that for the surrogate model III, described in Section \ref{sec:RNN_PCAsplit} and illustrated in Fig. \ref{fig:SplitRNN}, not all the $\text{RNN}_q$ have to be trained. Indeed when the number of hidden variables is limited, only the first few coefficients of the principal components can be accurately represented and training the remaining RNNs does not improve the accuracy as it will be shown \firstreviewer{in the next Section \ref{sec:hiddendimensioneffect}}. \firstreviewer{In that case the $\text{RNN}$ is not applied on the output of the groups $q=11,\,12,\ldots\,,18$ for a number of hidden variables $n_{\text{h}}=100$ or $n_{\text{h}}=200$.}

\firstreviewer{\subsubsection{The effect of the hidden variables dimensionality and surrogate models comparison}\label{sec:hiddendimensioneffect}

\begin{figure}[htb] 
\centering
\subfigure[]{\includegraphics[scale=0.4]{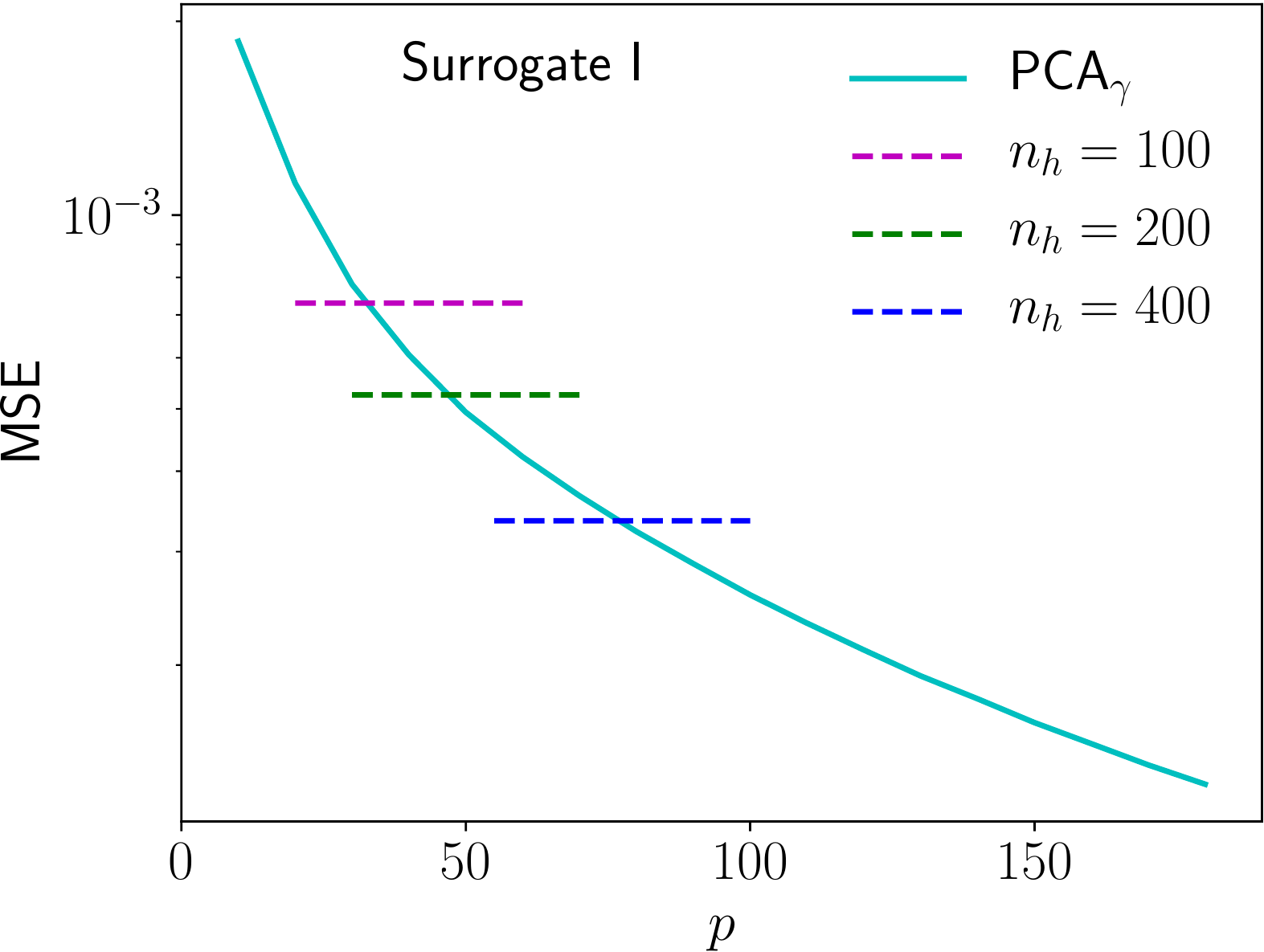}\label{fig:MSE_S1}}\quad
\subfigure[]{\includegraphics[scale=0.4]{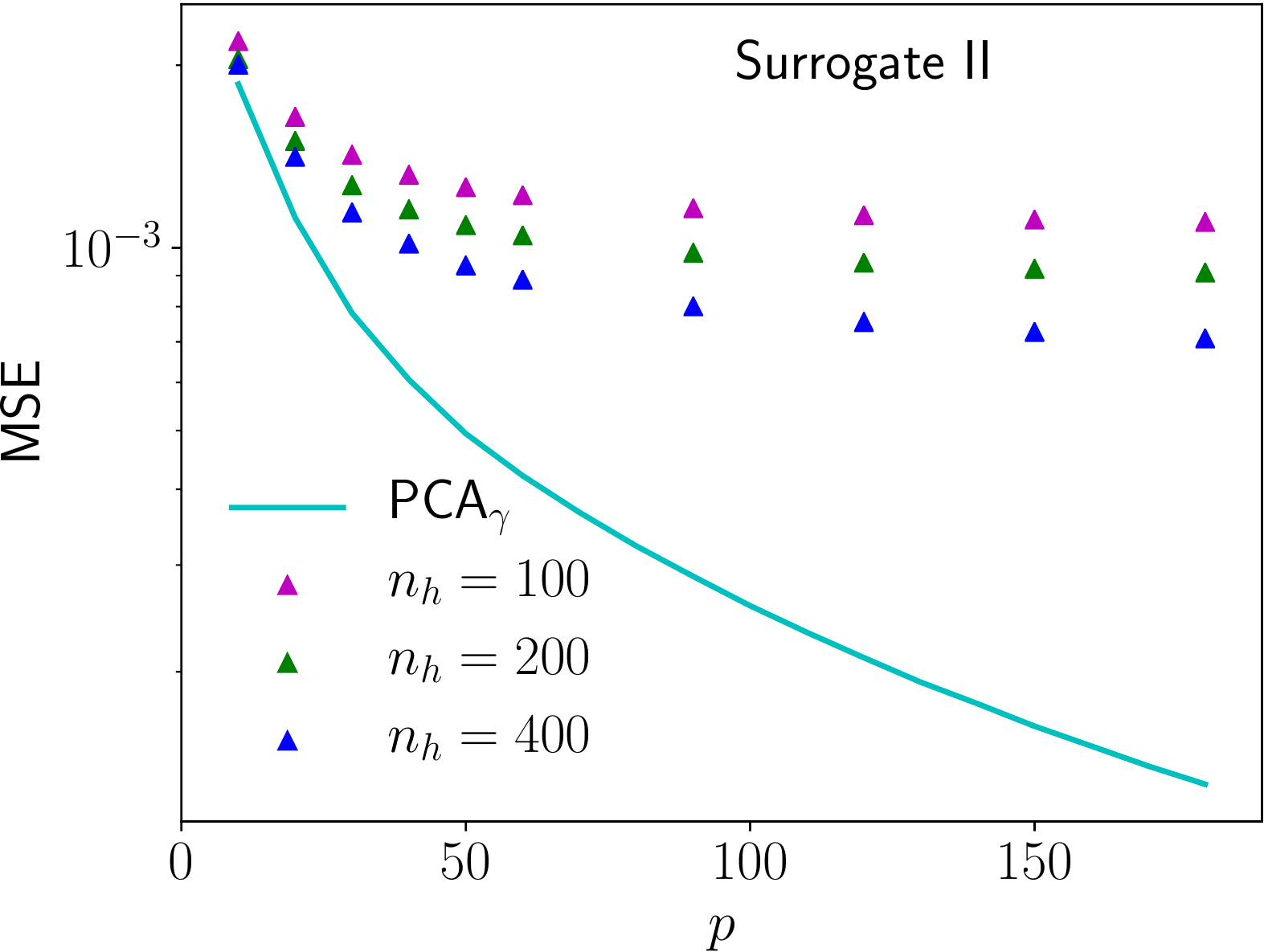}\label{fig:MSE_S2}}\\
\subfigure[]{\includegraphics[scale=0.4]{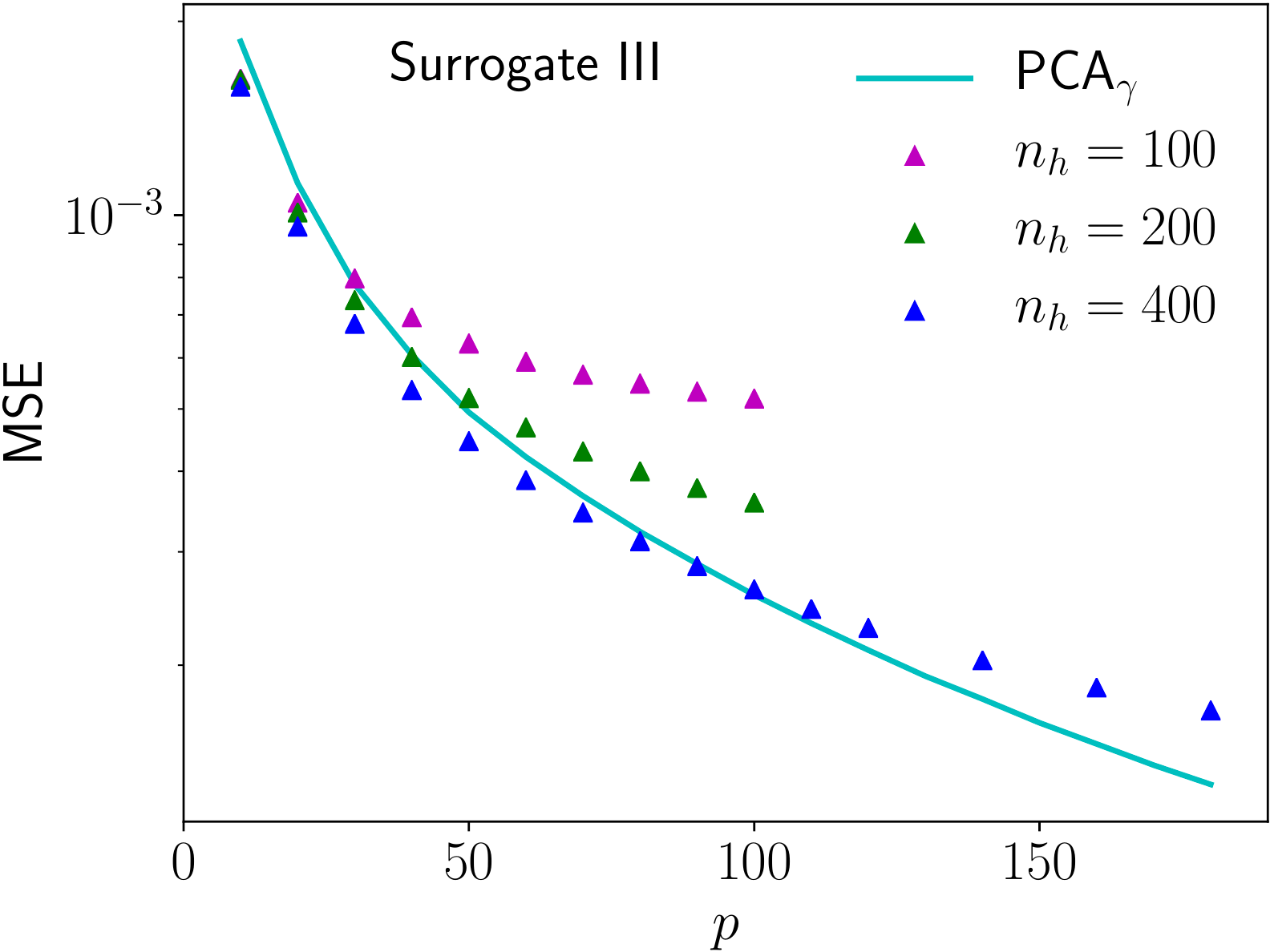}\label{fig:MSE_S3}}\quad
\subfigure[]{\includegraphics[scale=0.4]{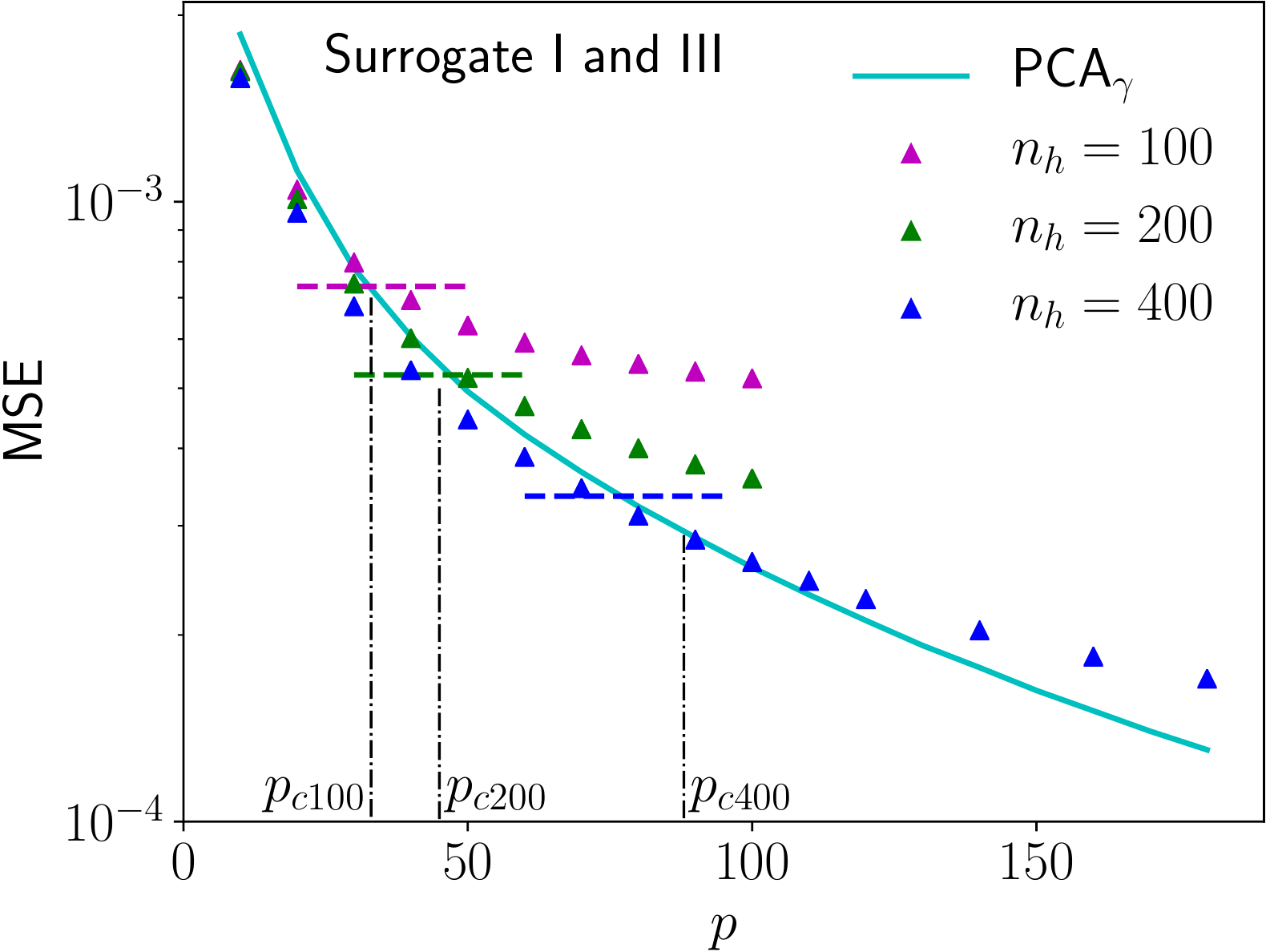}\label{fig:MSE_S1_S3}}
\caption{The MSEs of the equivalent plastic strain field, $\bm{\gamma}$, obtained by the three surrogate models on the training data. The MSE obtained with a PCA \secondreviewer{dimensionality reduction} is also provided in Fig. \ref{fig:Ep_Error} as a reference.}\label{fig:Ep_Error}
\end{figure}

The MSEs of the proposed Surrogates are evaluated with all the available training data, and plotted in Fig.\ref{fig:Ep_Error}.
The MSE obtained with a PCA \secondreviewer{dimensionality reduction} is also provided in Fig.\ref{fig:Ep_Error} as a reference.}

For the Surrogates II and III, the maximum value of $p=180$ is used during the PCA \secondreviewer{dimensionality reduction}.
Since the RNNs of these surrogate models predict the coefficients of the principal components, all the results of reduced dimensionality lower than $180$ can be extracted.
For Surrogate III with \secondreviewer{$n_{\text{h}} = 100$} and $200$, only a part of RNNs are trained to show the effect of the hidden variables number on the MSE.
In general, Fig. \ref{fig:Ep_Error} shows that the MSE decreases with an increasing number of hidden variables for all the three surrogate models.

For a constant number of hidden variables in the GRUs, Surrogate I gives a much lower error than Surrogate II, see Figs. \ref{fig:MSE_S1} and \ref{fig:MSE_S2}.
This shows that the non-linear \secondreviewer{dimensionality reduction} arising with the Surrogate I during the training is superior to the linear one of the PCA.
As a result, in order to reach the same error, more hidden variables will be required with Surrogate II than with Surrogate I.
The poor results of Surrogate II can be explained by the use of PCA in combination with the GRU.
PCA reduces the \secondreviewer{full dimensional variables} into the coefficients of the principal components, and each of the coefficients has a different weight in the original variable.
When RNNs are trained with these coefficients as  \secondreviewer{observation} output, all the elements of the output share the same hidden variables and are treated equally.
As a result, although the coefficients of the few first principal components are more important, their importance cannot be discriminated during the training of RNNs, except if an extra treatment is applied to put an emphasize on some elements of the output, which is the case with the \secondreviewer{dimensionality break down} of Surrogate III.
Indeed, Fig. \ref{fig:MSE_S3} shows that Surrogate III has the lowest MSE among the three models for the same number of hidden variables.
As explained, with Surrogates I and II, more outputs share the same hidden variables than with Surrogate III.
However, since each group of outputs of Surrogate III has its own hidden variables, in total, more hidden variables are used with Surrogate III.

Since the RNNs of Surrogates II and III are used to predict the coefficients of principal components, the lower limit of Surrogates II and III MSE is the reference MSE from PCA; this limit would be reached if the evolution of the principal components coefficients were perfectly predicted.
Although in Fig. \ref{fig:MSE_S3}, the synergistic effect of the errors from PCA and RNNs leads for low $p$ to a MSE slightly lower than that of PCA, the MSEs of Surrogate III agree well with that of PCA at low value of $p$, and diverge from that of PCA when $p$ is beyond a value ``$p_\text{c}$''; this ``$p_\text{c}$'' value increases along with the number of hidden variables.
On the one hand, we can say that more principal components coefficients can be better predicted by the RNNs of Surrogate III because of the increase of hidden variables number.
On the other hand, when using Surrogate III, Fig. \ref{fig:MSE_S3} shows that when the reduced dimensionality $p$ increases, in order to take benefit from the accuracy gain in the PCA, the number of hidden variables should increase to capture the evolution of the coefficients of the principal components.
\firstreviewer{Therefore, we can conclude that the use of PCA in a surrogate model does not reduce drastically the number of hidden variables in the RNNs required to reach a given accuracy.}
Having said that, we note that for $p>p_\text{c}$, MSE of Surrogate III keeps decreasing with the increase of $p$ although the error is larger than that of the PCA.
This means that, although the RNNs predictions are not accurate for the coefficients of principal components higher than $p_\text{c}$, the evolution trends of those coefficients are still predicted.
Besides, the PCA can reduce the scale of the output feed-forward networks, $\text{NNW}_\text{O}$, which can be important for problems with \secondreviewer{high dimensional output}.
According to the number of learnable parameters in Eqs. (\ref{eq:NP_NNW}) and (\ref{eq:NP_GRU}), the PCA \secondreviewer{dimensionality reduction} will become less effective for complex physical processes, in which a large number of hidden variables are needed. 

Finally, in Fig. \ref{fig:MSE_S1_S3}, the dashed lines corresponding to the MSEs of Surrogate I with respectively 100, 200 and 400 hidden variables cross the PCA reference line at $p\approx p_\text{c}$. \firstreviewer{Therefore it appears that the Surrogates I and III have a similar accuracy when using the same number of hidden variables. So the evolution of the coefficients of the principal components that can be accurately represented with Surrogate III depends strongly on the number of hidden variables.}

Although there are a few RNNs needed to be trained for Surrogate III, their training is still computational efficient when the training scheme in Fig. \ref{fig:flowchart} is used as compared to the training of Surrogate I when targeting a comparable MSE.
With one GPU of 14 Gbit memory, the training of Surrogate III with \secondreviewer{$n_{\text{h}}=400$} for each $\text{RNN}_q$, see Table \ref{tab:RNN_NNW_EP}, takes around 48 hours, while the training of Surrogate I  with \secondreviewer{$n_{\text{h}}=400$} takes around 108 hours.
Besides, Surrogate III is also more flexible than the other two processes: its RNNs, the independent $\text{RNN}_q$, can have different structures, in terms of hidden variables number and also number of outputs.
Finally Surrogate II is not recommended, since for a comparable accuracy, the required number of hidden variables will increase drastically.

\subsubsection{Principal modes of PCA and physical information: the connection between the physical state and the RNN hidden variables}

\begin{figure}[!htb] 
	\centering
	\subfigure[]{\includegraphics[scale=0.26]{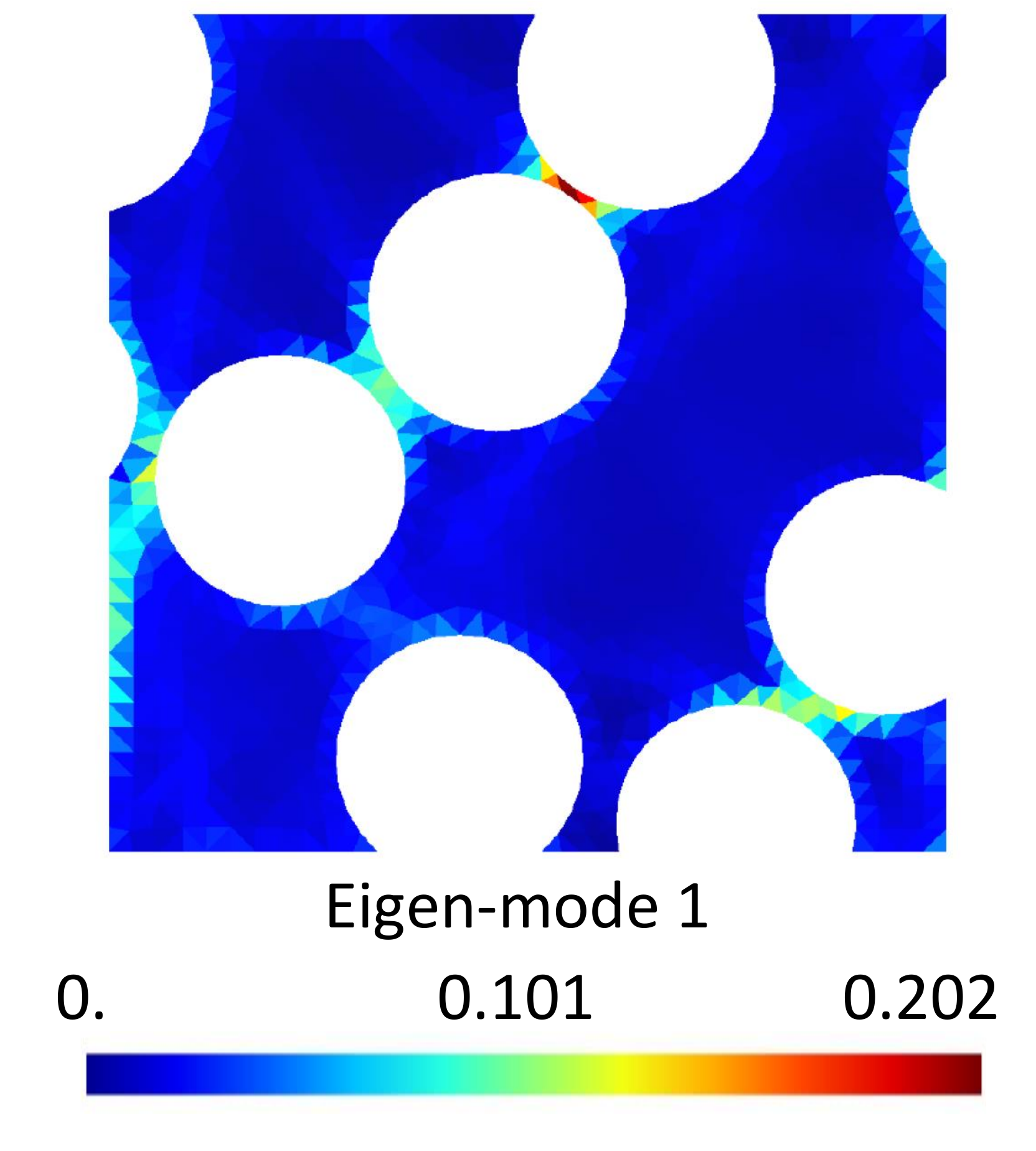}\label{fig:mode1}}\,
	\subfigure[]{\includegraphics[scale=0.26]{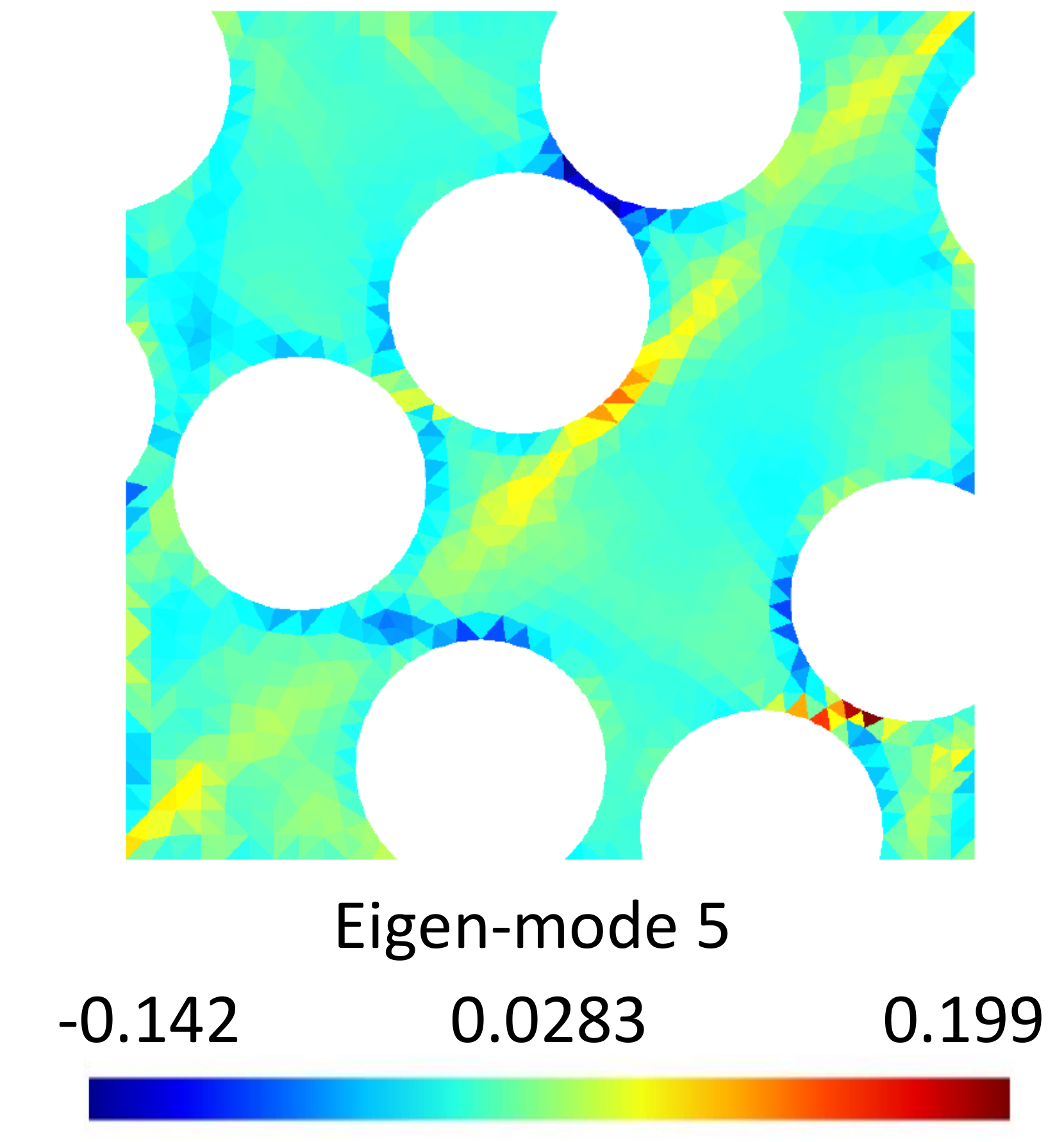}\label{fig:mode5}}\,
	\subfigure[]{\includegraphics[scale=0.26]{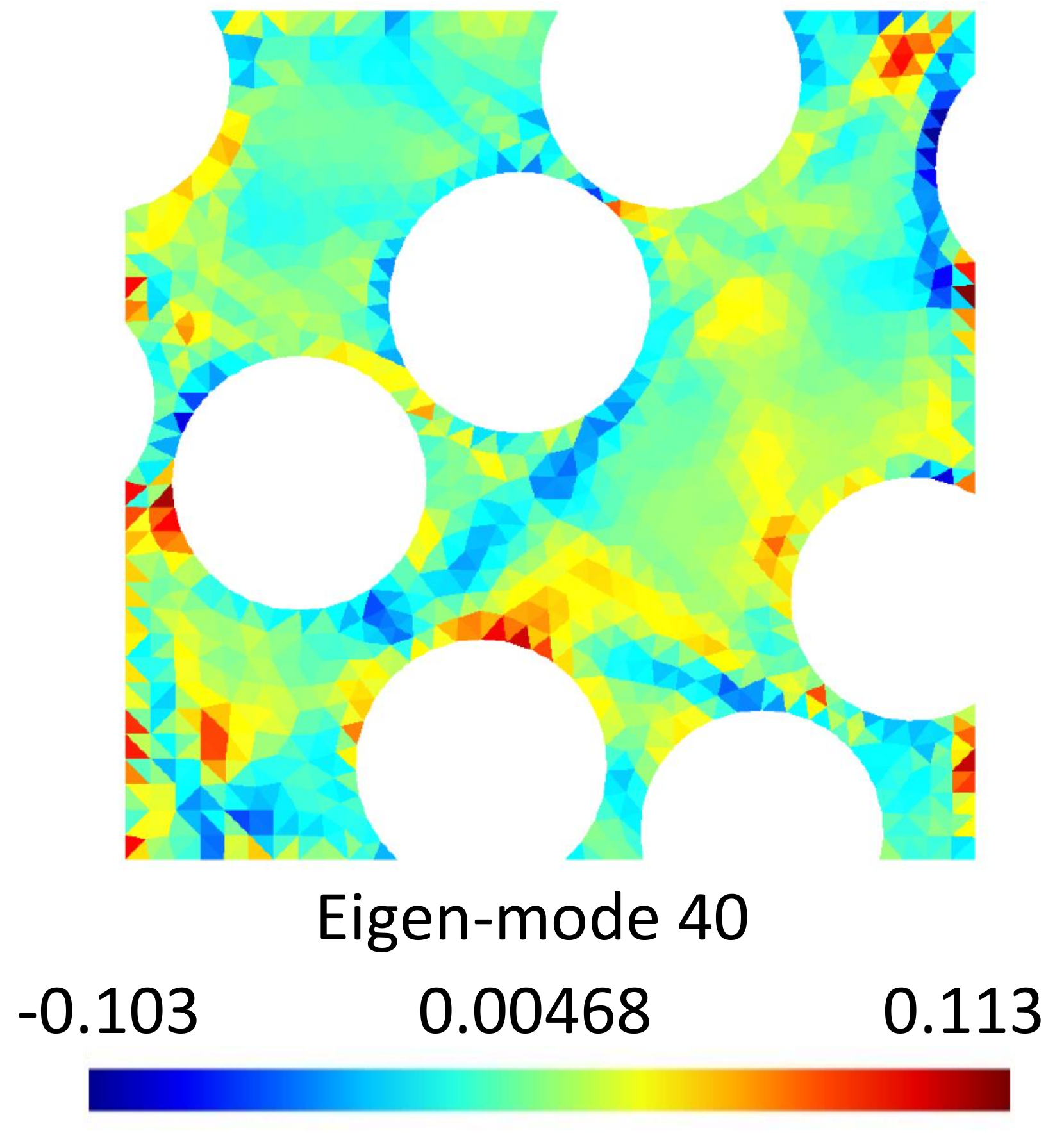}\label{fig:mode40}}\\
	\subfigure[]{\includegraphics[scale=0.26]{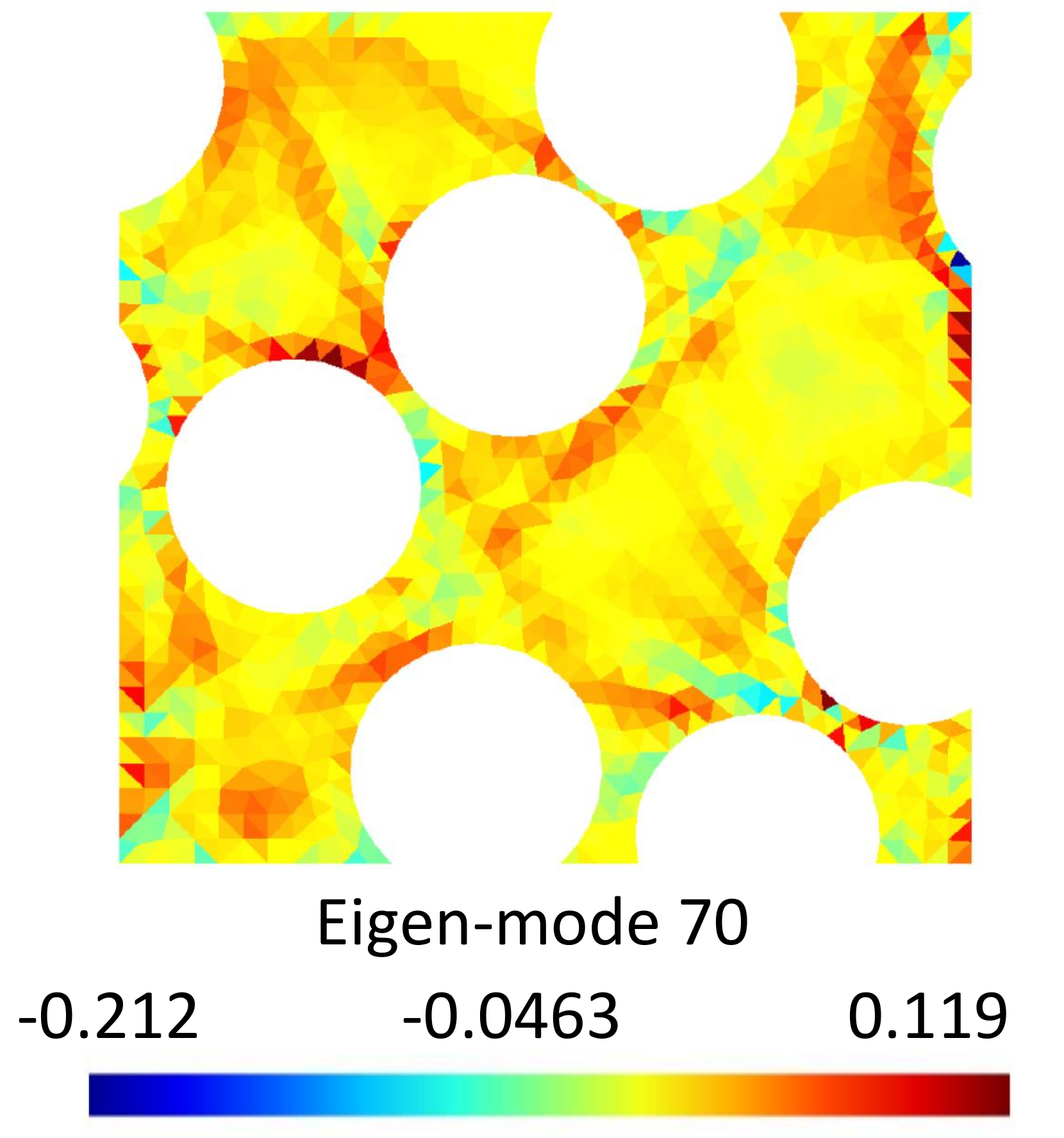}\label{fig:mode70}}\,
	\subfigure[]{\includegraphics[scale=0.26]{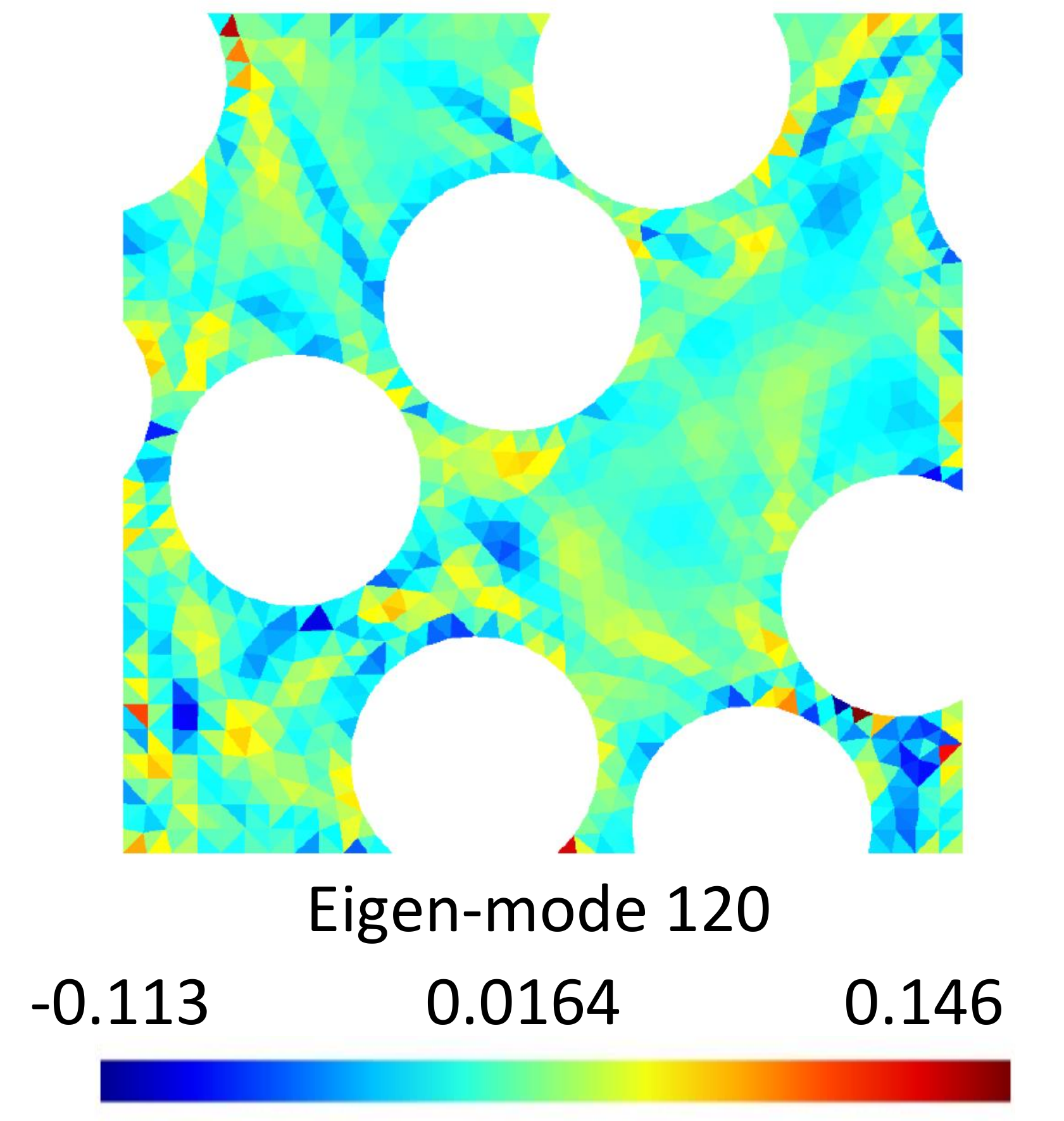}\label{fig:mode120}}\,
	\subfigure[]{\includegraphics[scale=0.26]{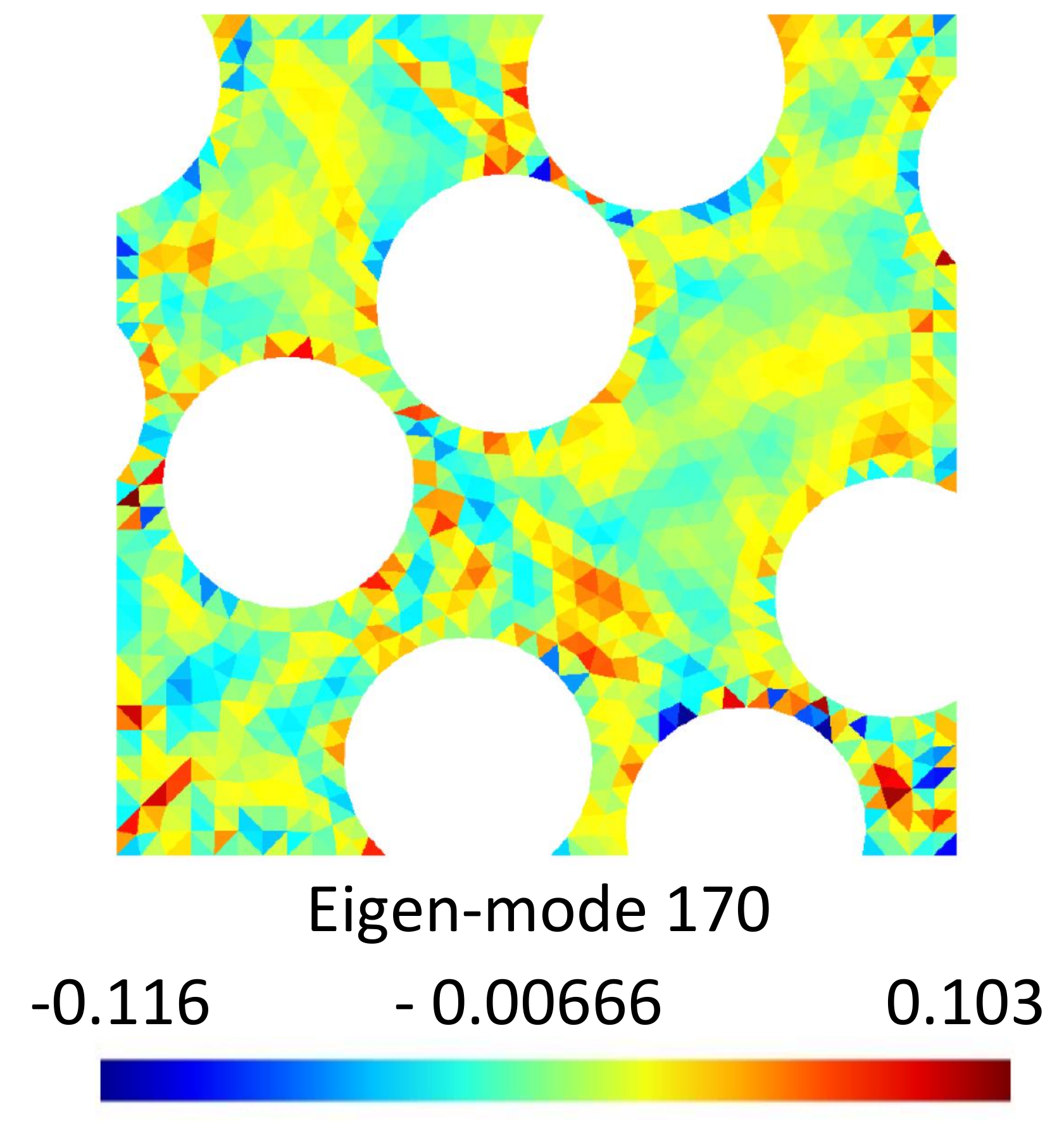}\label{fig:mode170}
}	\caption{The eigen-modes of the equivalent plastic strain field, $\bm{\gamma}$, obtained from PCA.}\label{fig:PCA_mode}
\end{figure}

A further study on the principal components of output data reveals the connection between the hidden variables and the physical information.
The principal components, $\underline{\bm{v}}_i,\,(i=1,\,\ldots,p)$, obtained from PCA can be visualized in a finite element discretization and are called principal modes in their visualization forms.
A few principal modes are presented in Fig. \ref{fig:PCA_mode}, where the index "$i$" of a mode corresponds to the principal component "$\underline{\bm{v}}_i$".

In Fig. \ref{fig:PCA_mode}, the first few modes show quite regular patterns which can be clearly divided into a few regions, while a pattern fragmentation can be seen in the higher modes, such as modes 120 and 170.
The regular patterns contain more homogenized physical information than the fragmented patterns, which have much more detailed physical information.
Therefore, to trace the evolution of the coefficients of low modes, such as modes 1 and 5, less hidden variables are required in the RNNs than for the higher modes.
\firstreviewer{This can be explained by the fact that if we want to replace these modes by some variables, such as using a simple \secondreviewer{dimensionality reduction} method--clustering, it is expected that the higher modes need much more variables than the lower modes.}

\subsubsection[Predictions of the equivalent plastic strain distribution on the testing data]{Predictions of $\bm{\gamma}$ on the testing data}\label{sec:ptesting}

\begin{table}
\centering
\begin{threeparttable}		
	\caption{MSE of the three surrogate models on testing data; all the RNNs use \secondreviewer{$n_{\text{h}}=400$}, and $h_0$=-1.0.}\label{tab:RNN_EP_test}  
	\begin{tabular}{c|c|c|c} 
	\hline\hline   
	Model & $\text{NNW}_\text{I}$ &$\text{NNW}_\text{O}$&MSE\\ 
	\hline\hline 
	Surrogate I&(3, 70)&(800, 1607)&0.00066\\
	\hline 		
	Surrogate II&(3, 70)&(800, 180)& 0.00153\\
	\hline 
	Surrogate III&(3, 70)&$(100,\,10)^{*}$&0.00047\\
	\hline\hline		
	\end{tabular}  
	\begin{tablenotes}
	\small
	\item {${}^*$ $Q=18$, and $\text{RNN}_q$ with $q=1,2, ..,18$ are trained}.
	\end{tablenotes}
 	\end{threeparttable}
\end{table}

The accuracy of the trained surrogate models with 400 hidden variables is now assessed with the testing data, which include $1252$ random and $299$ cyclic loading paths.
The surrogate models MSEs on the testing data are presented in Table \ref{tab:RNN_EP_test}.
As expected, the MSEs on testing data are higher than on the training data.
\firstreviewer{As observed during training, for a given number of hidden variables, the least accurate model, in terms of MSE, is the Surrogate II and the most accurate one is the Surrogate III.}

\begin{figure}[!htb]
	\centering
	\subfigure[]{\includegraphics[scale=0.48]{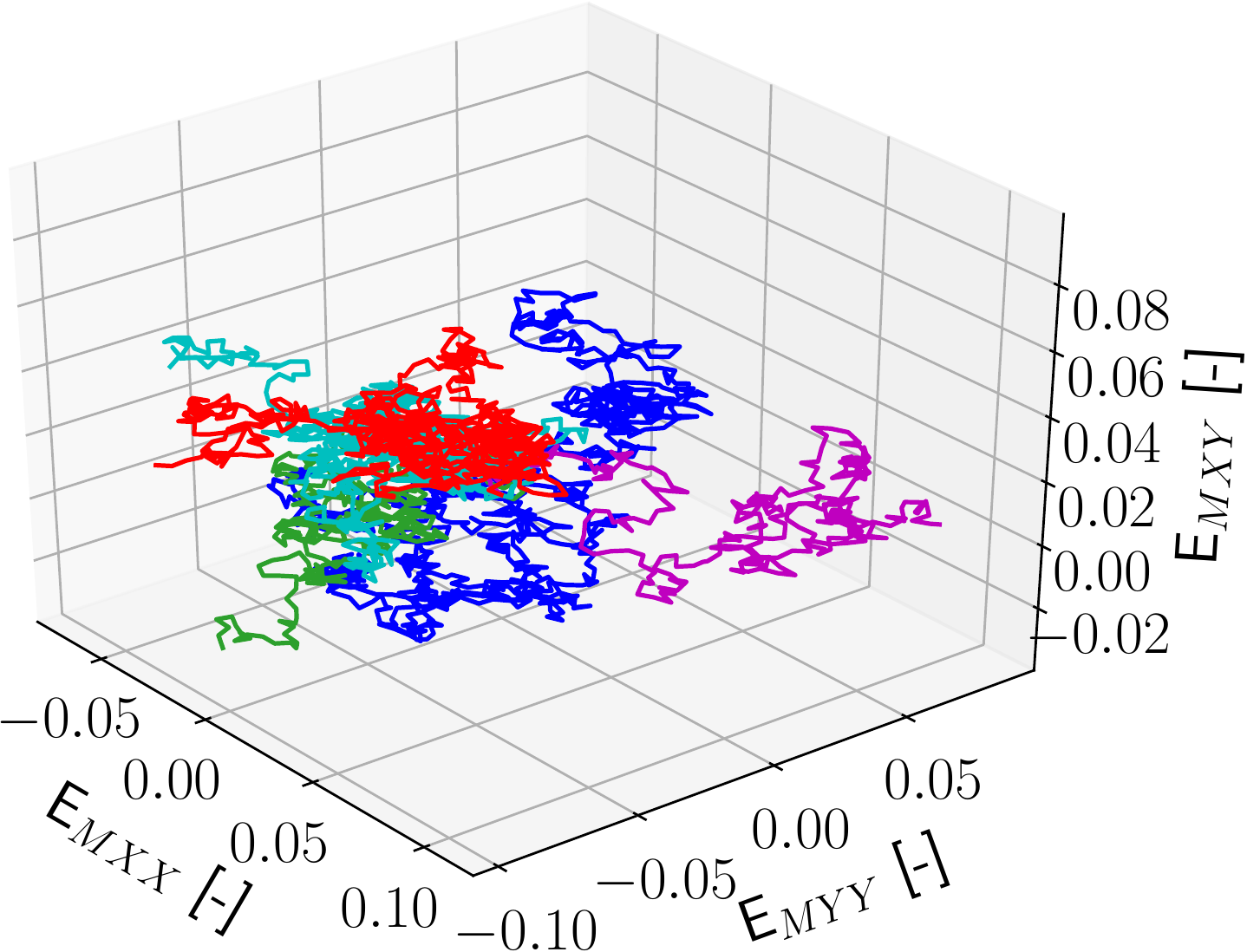}\label{fig:LPath_R}}
	\subfigure[]{\includegraphics[scale=0.48]{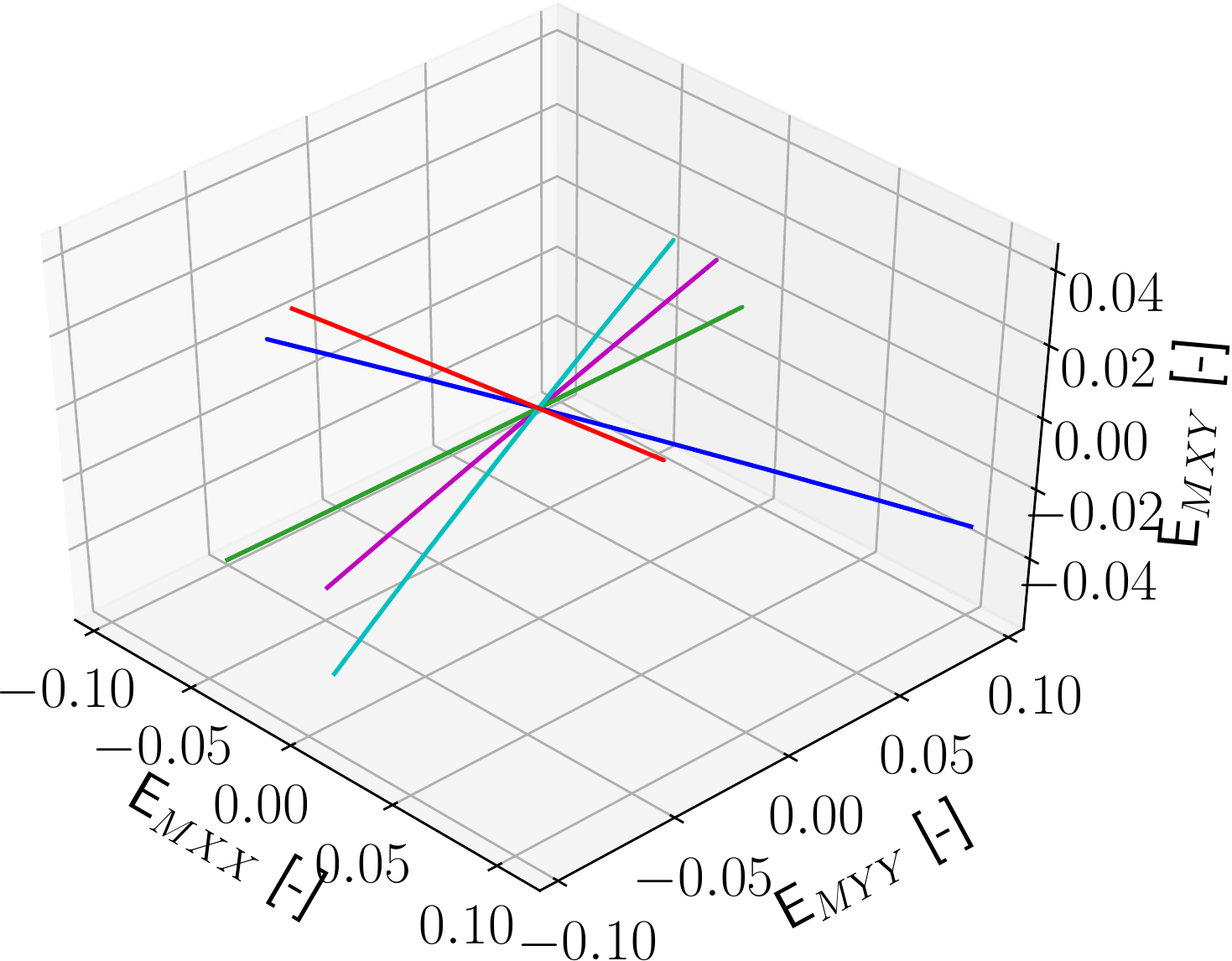}\label{fig:LPath_C}}
	\caption{Examples of loading paths of the testing data: (a) \firstreviewer{Random loading paths}, and (b) Cyclic loading paths.}\label{fig:LPath}
\end{figure}

\begin{figure}[!htb]
	\centering
	\subfigure[]{\includegraphics[scale=0.38]{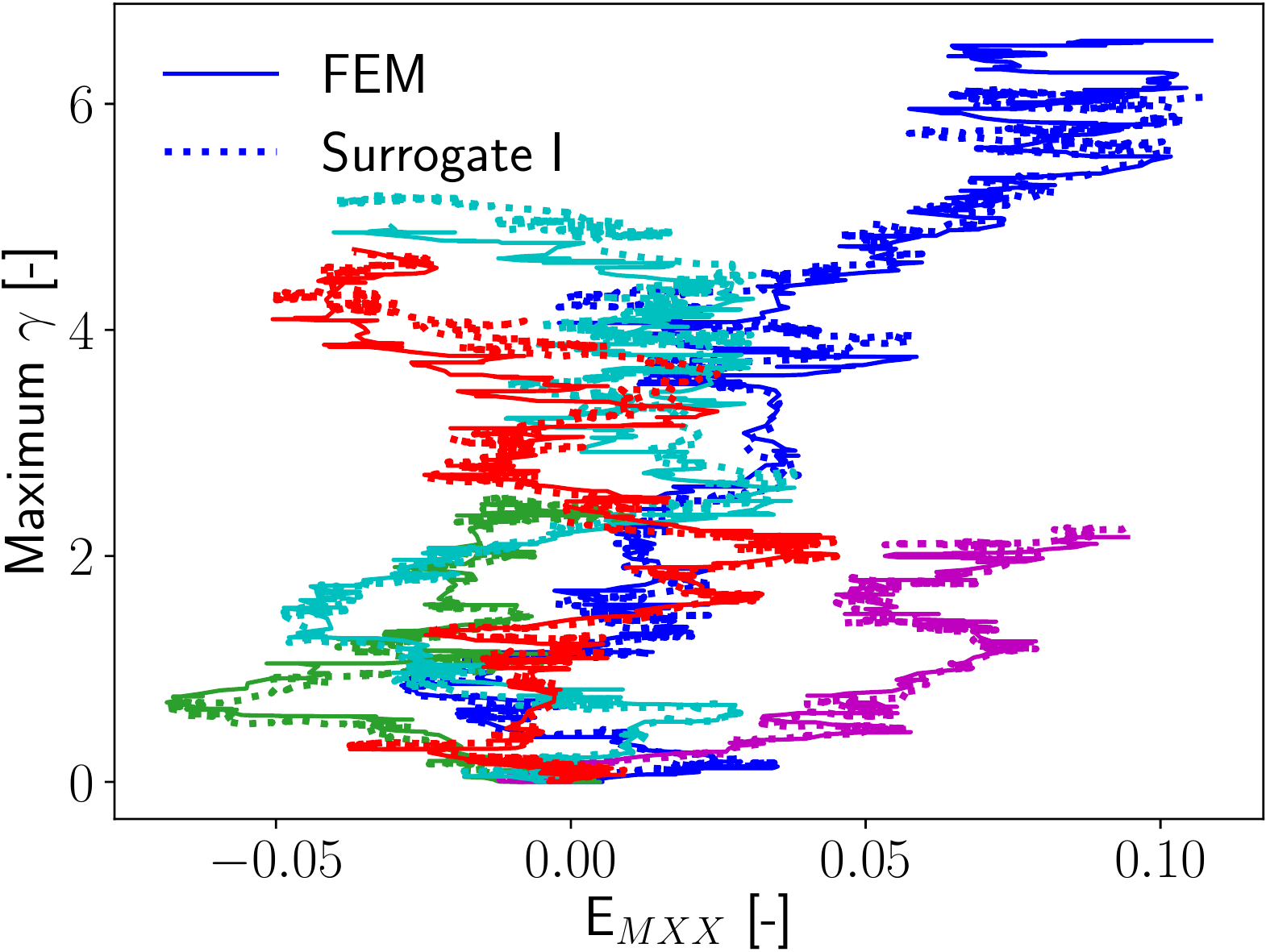}\label{fig:MaxEp_RP_I}}\,
	\subfigure[]{\includegraphics[scale=0.38]{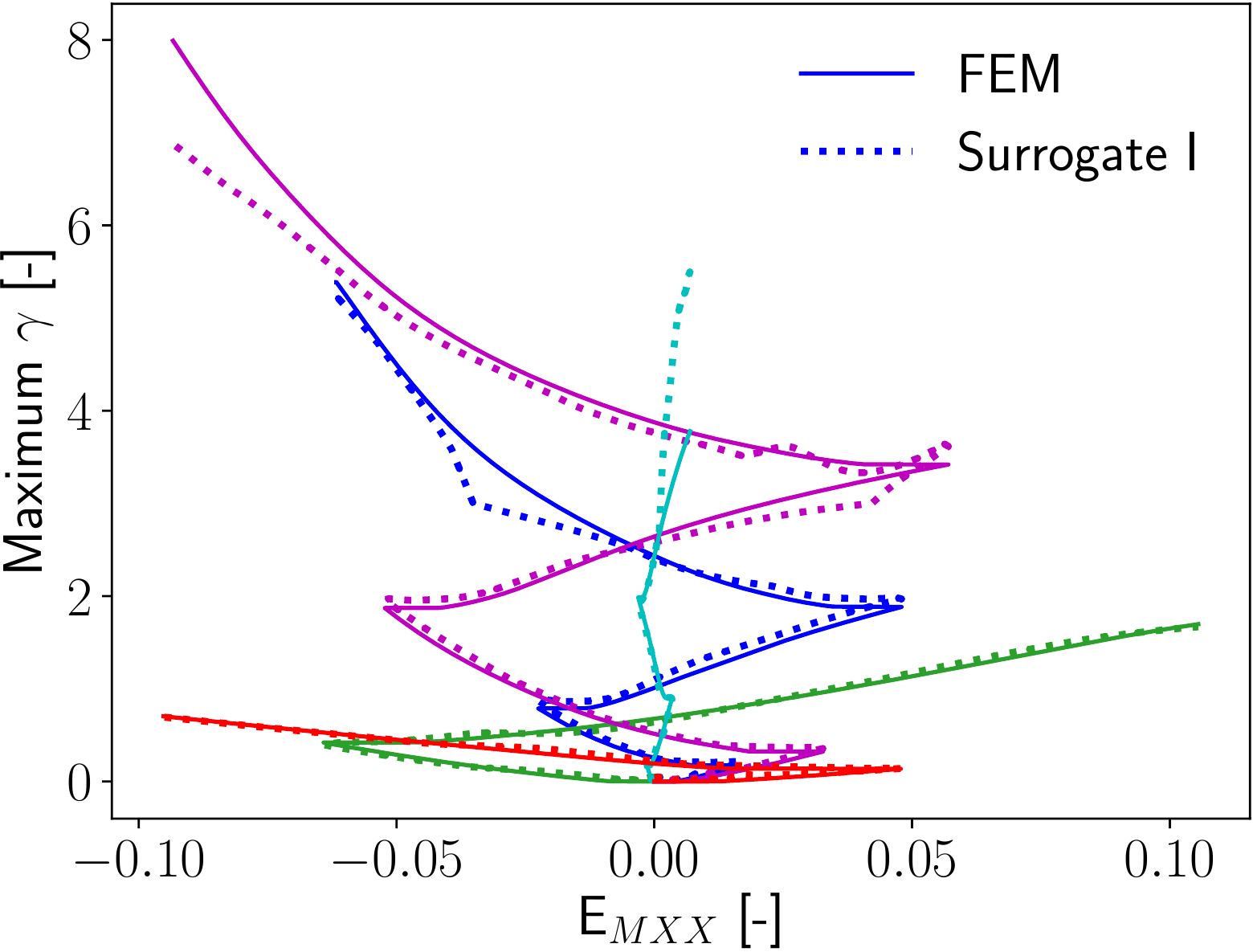}\label{fig:MaxEp_CP_I}}\\
	\subfigure[]{\includegraphics[scale=0.38]{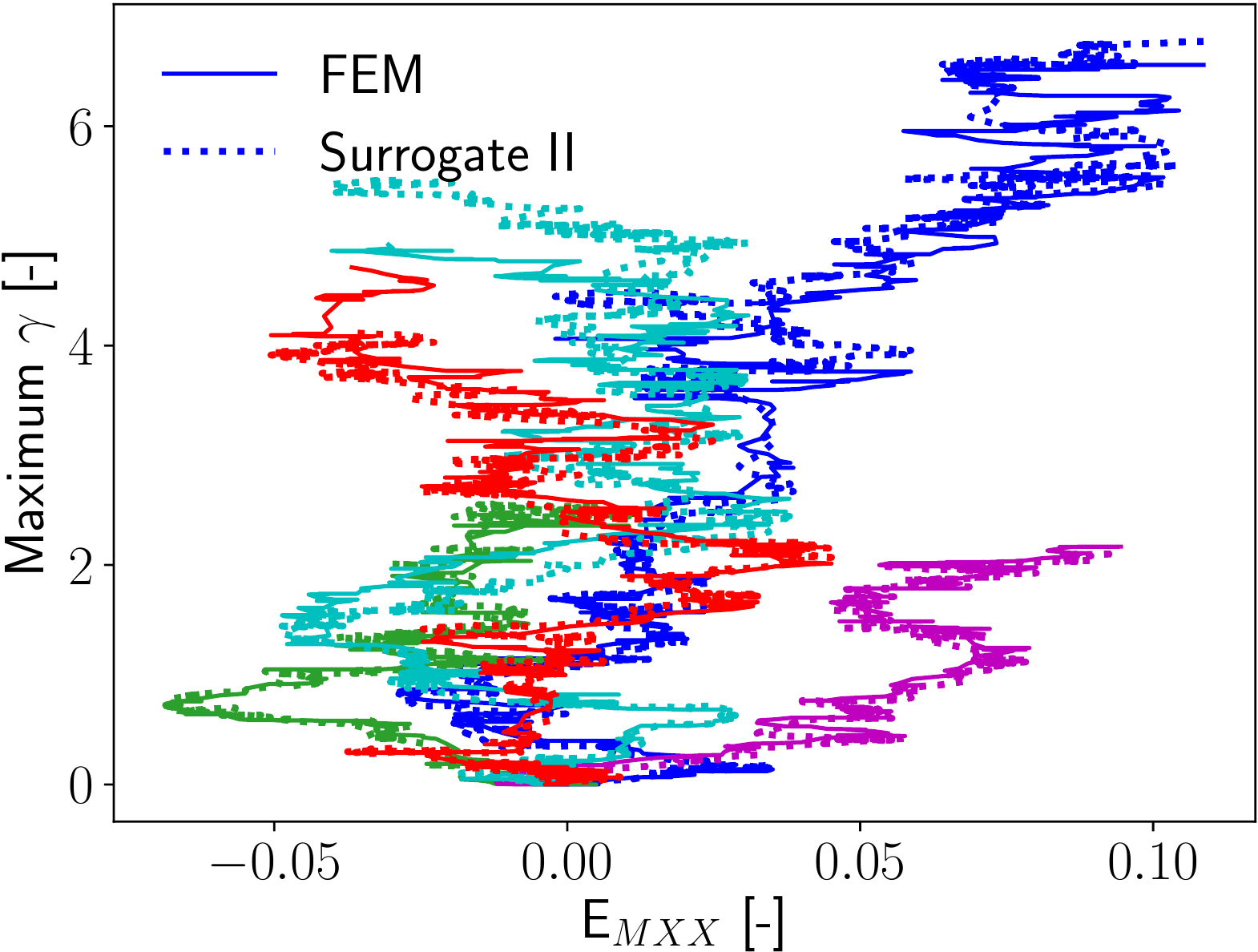}\label{fig:MaxEp_RP_II}}\,
	\subfigure[]{\includegraphics[scale=0.38]{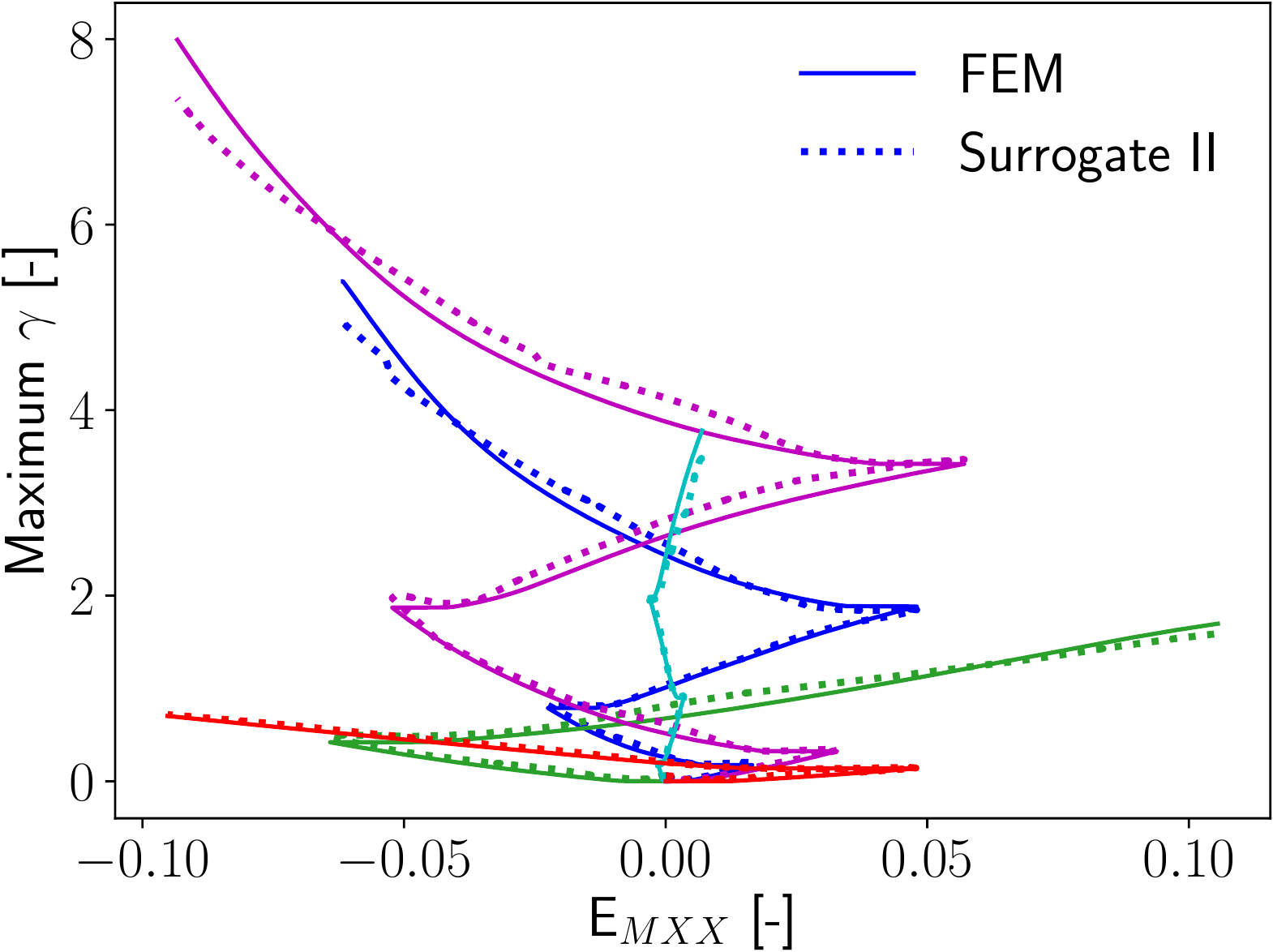}\label{fig:MaxEp_CP_II}}\\
	\subfigure[]{\includegraphics[scale=0.38]{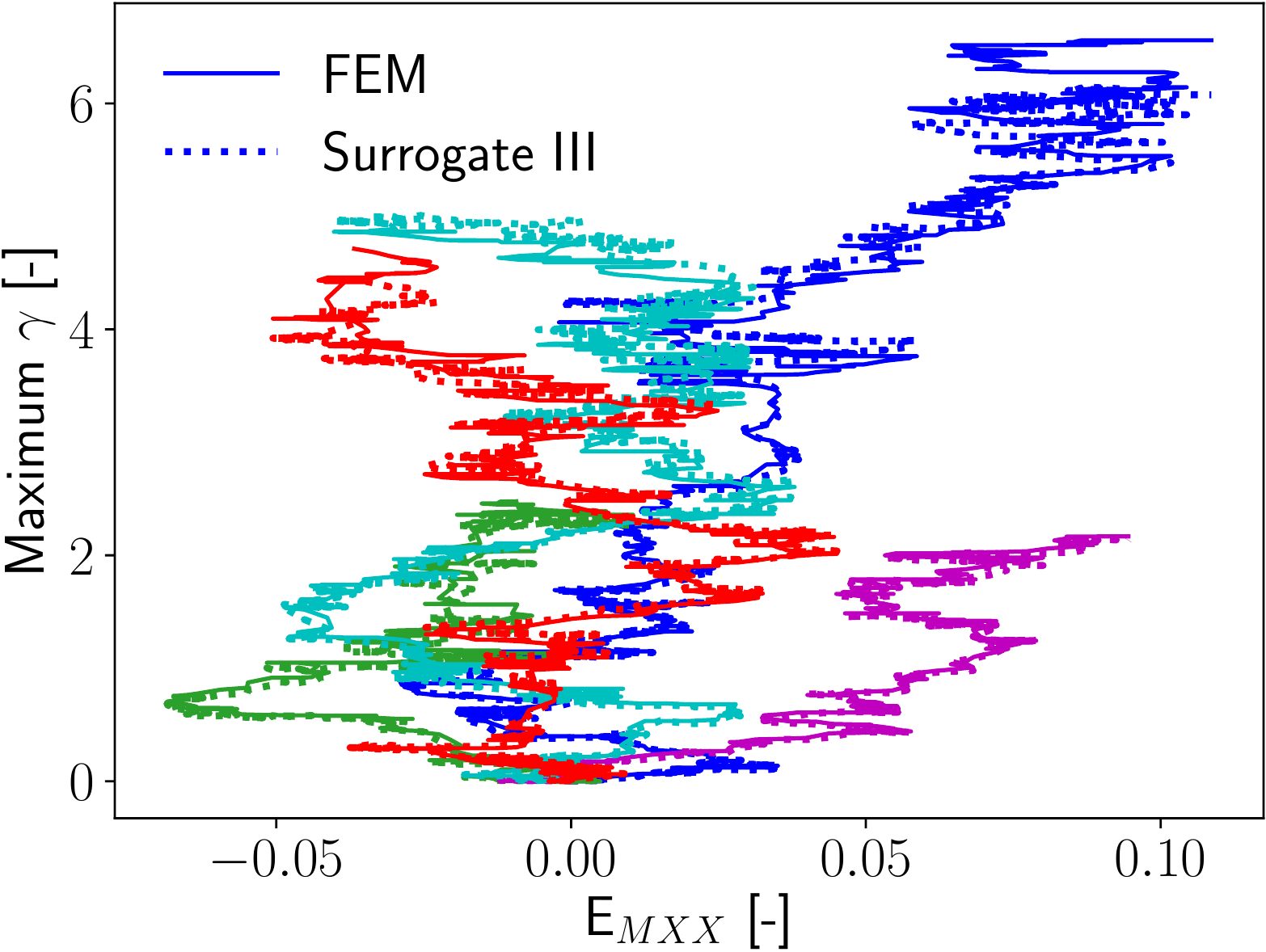}\label{fig:MaxEp_RP_III}}\,
	\subfigure[]{\includegraphics[scale=0.38]{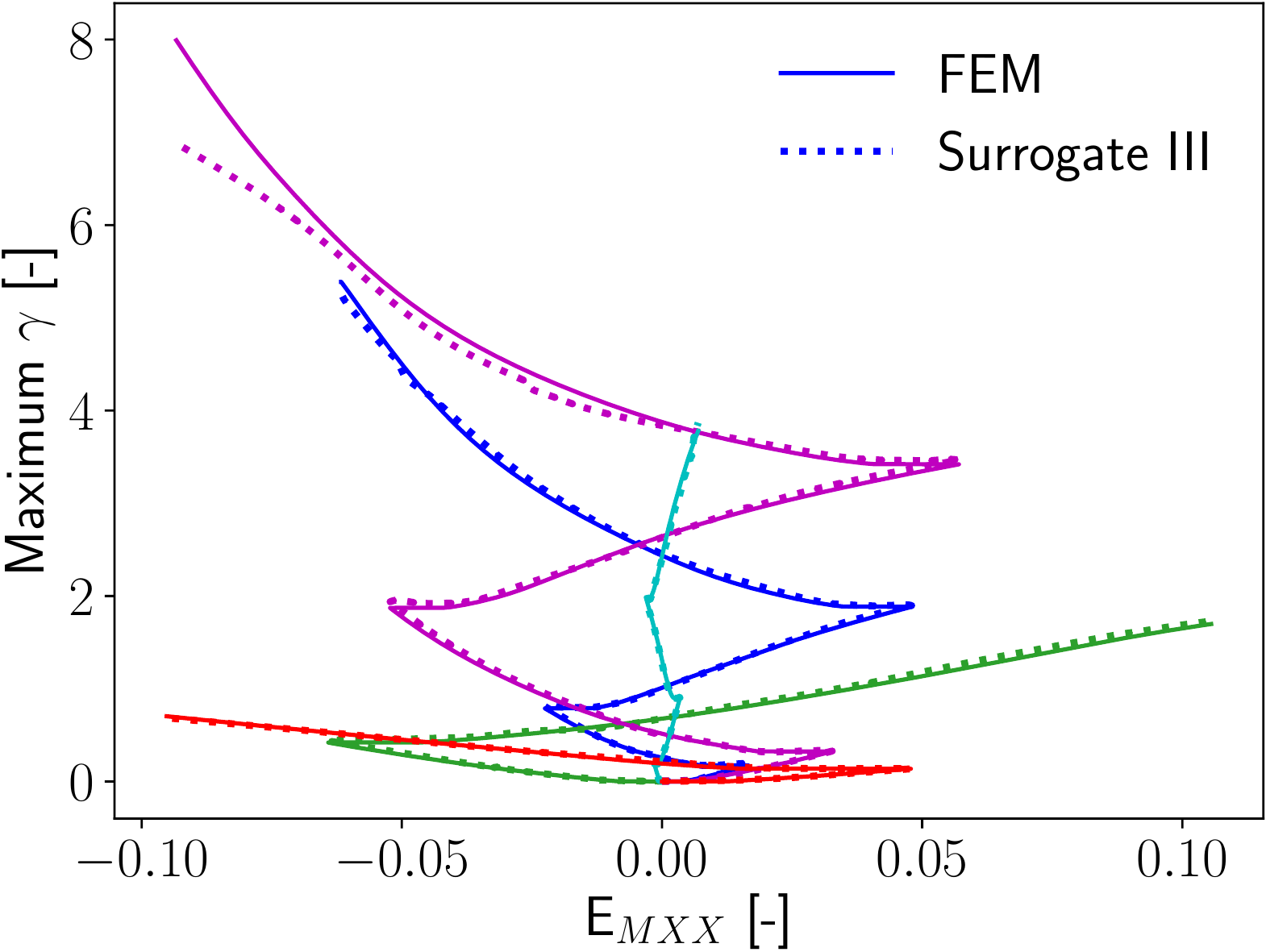}\label{fig:MaxEp_CP_III}}
	\caption{Evolution of the maximum equivalent plastic strain in the RVE at each loading step: (a), (c) and (e) For random loading paths; and (b), (d) and (f) For cyclic loading paths.}\label{fig:Max_Ep}
\end{figure}

A few randomly picked loading paths, which include five random and five cyclic paths, see Figs. \ref{fig:LPath_R} and \ref{fig:LPath_C}, are used to compare and verify the predictions of the proposed models. The evolution of the maximum equivalent plastic strains in the RVE, at each loading step, predicted by the surrogate models are plotted in Fig. \ref{fig:Max_Ep} and compared to the ones obtained with the direct finite element simulations.
Good agreements can be seen between the predictions of the surrogate models and the direct finite element results when the maximum equivalent plastic strains are moderate, \emph{e.g.} $\gamma_\text{max}<3$.
Therefore, the high MSEs of the surrogate models predictions on the testing data may be due to insufficient training data with high value of $\gamma$.

\begin{figure}[!htb]
	\centering
    \includegraphics[scale=0.5]{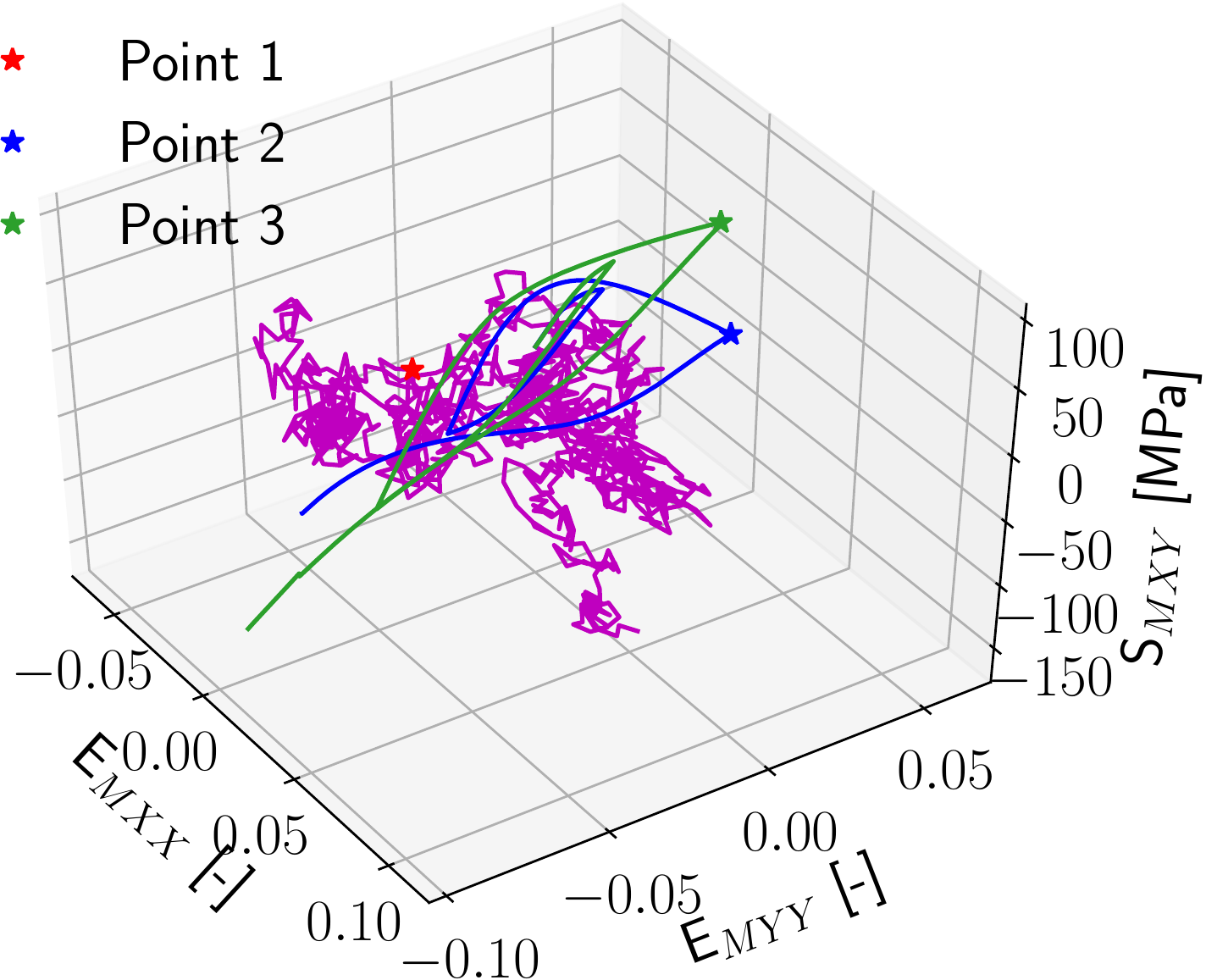}
	\caption{Three loading paths --two cyclic and one random-- of the testing data with picked loading steps, or configurations, marked with a star ``$\star$''}\label{fig:LStep}
\end{figure}

\secondreviewer{
\begin{figure}[!htb]
	\centering
	\subfigure[]{\includegraphics[scale=0.23]{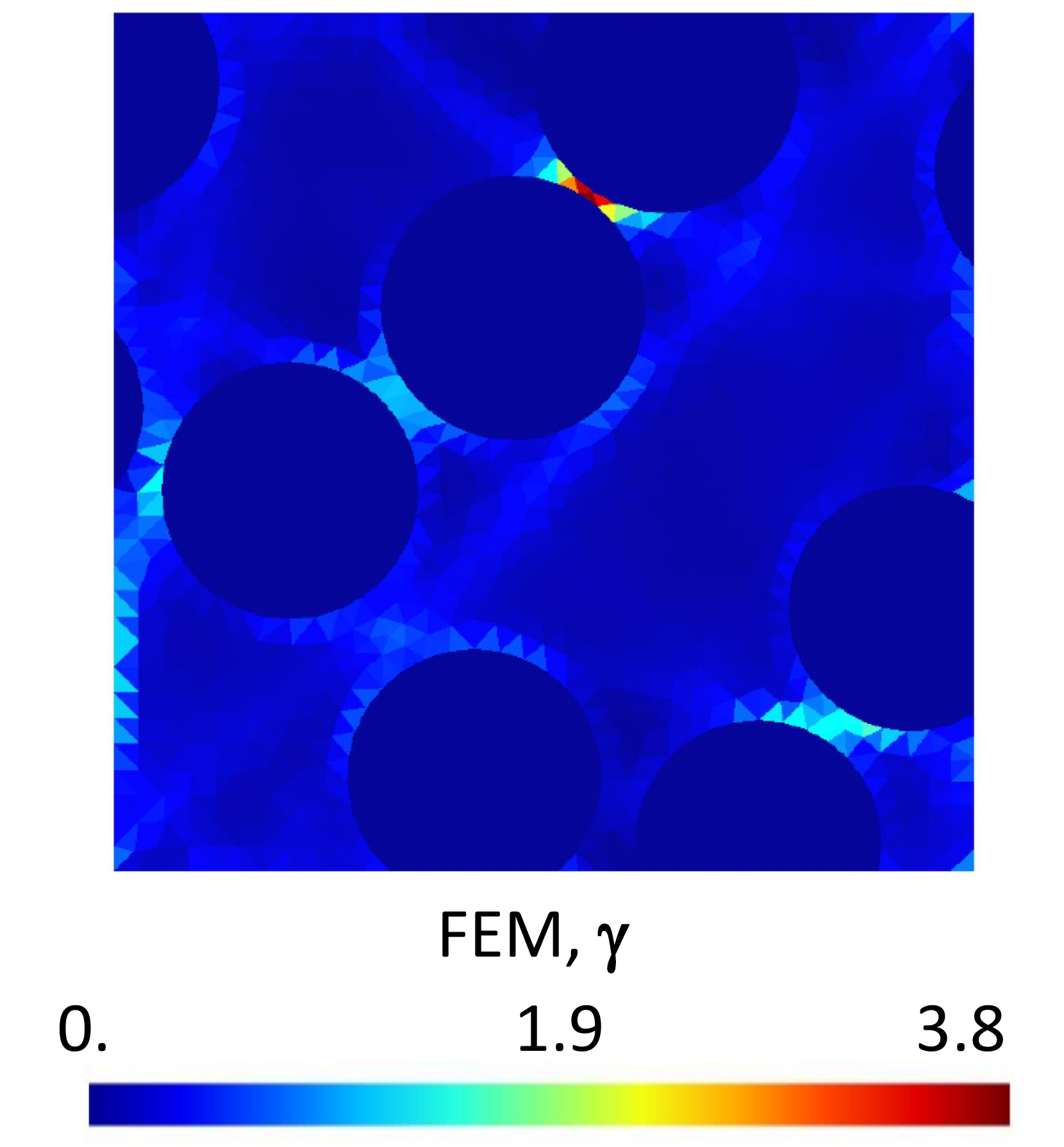}\label{fig:FE_Ep1}}\quad\quad\quad\quad\quad\quad\quad\quad\quad\quad\,
	\subfigure[]{\includegraphics[scale=0.23]{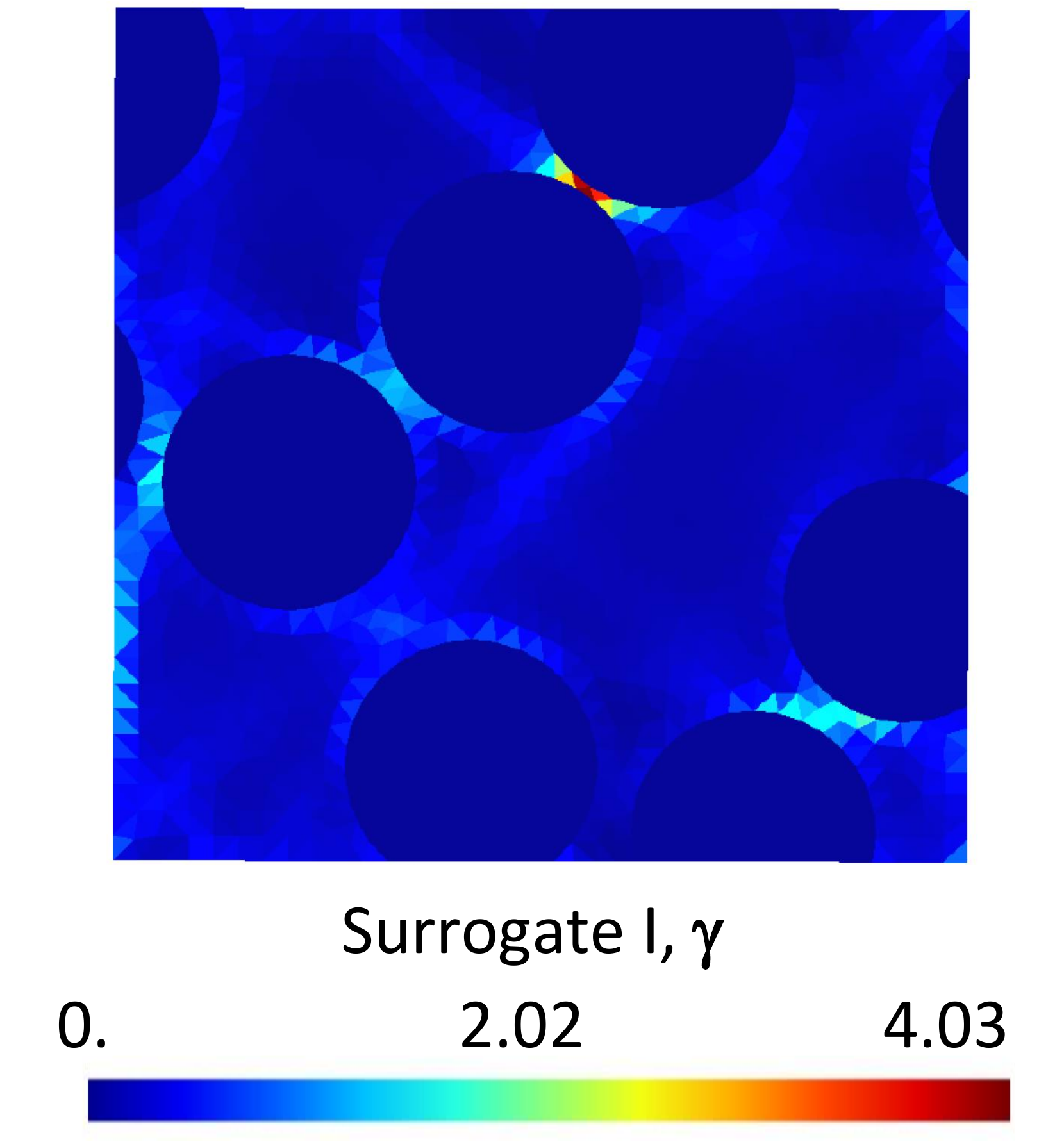}\label{fig:S1_P1}}\,
	\subfigure[]{\includegraphics[scale=0.23]{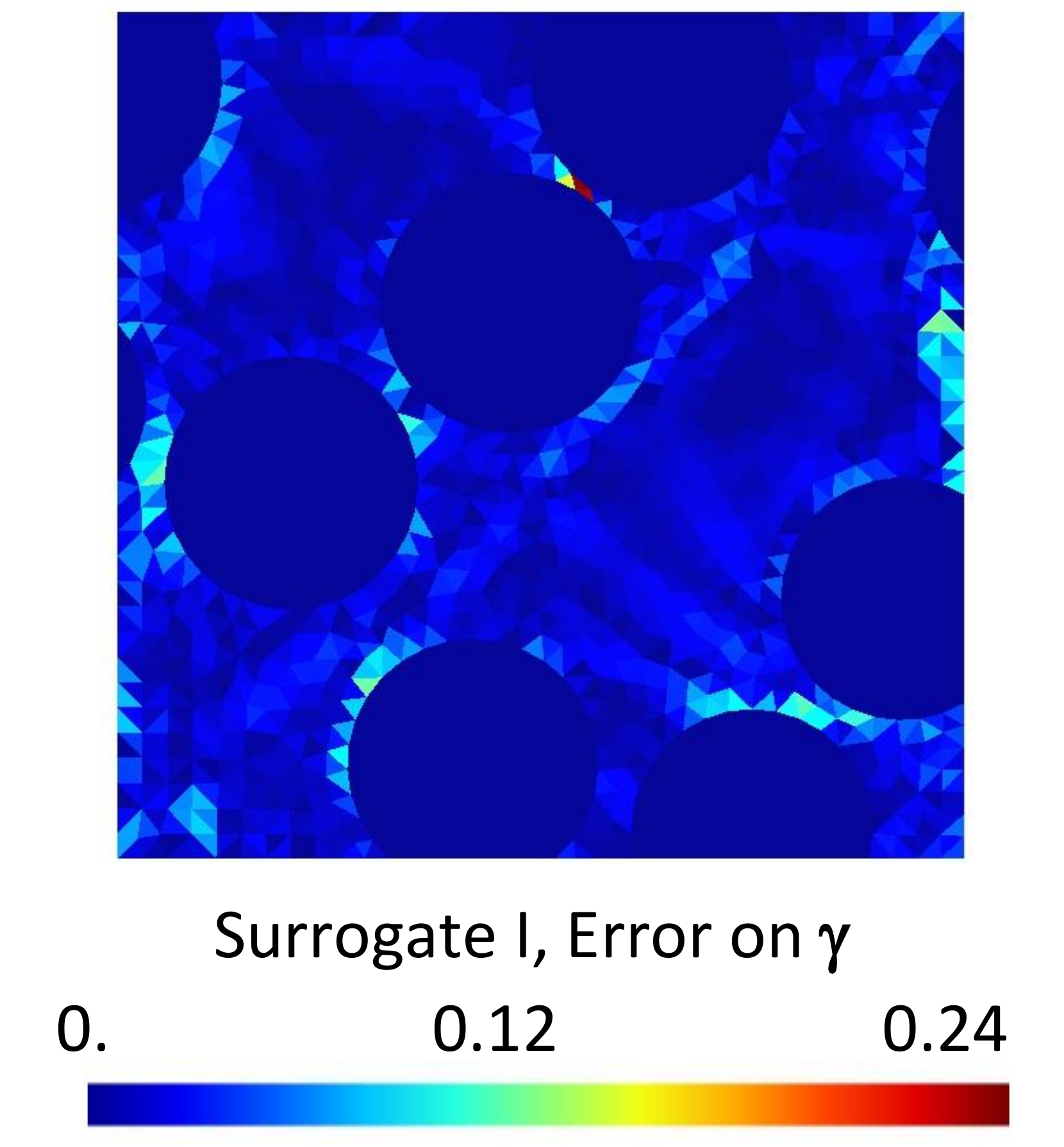}\label{fig:S1_P1_error}}\\	
	\subfigure[]{\includegraphics[scale=0.23]{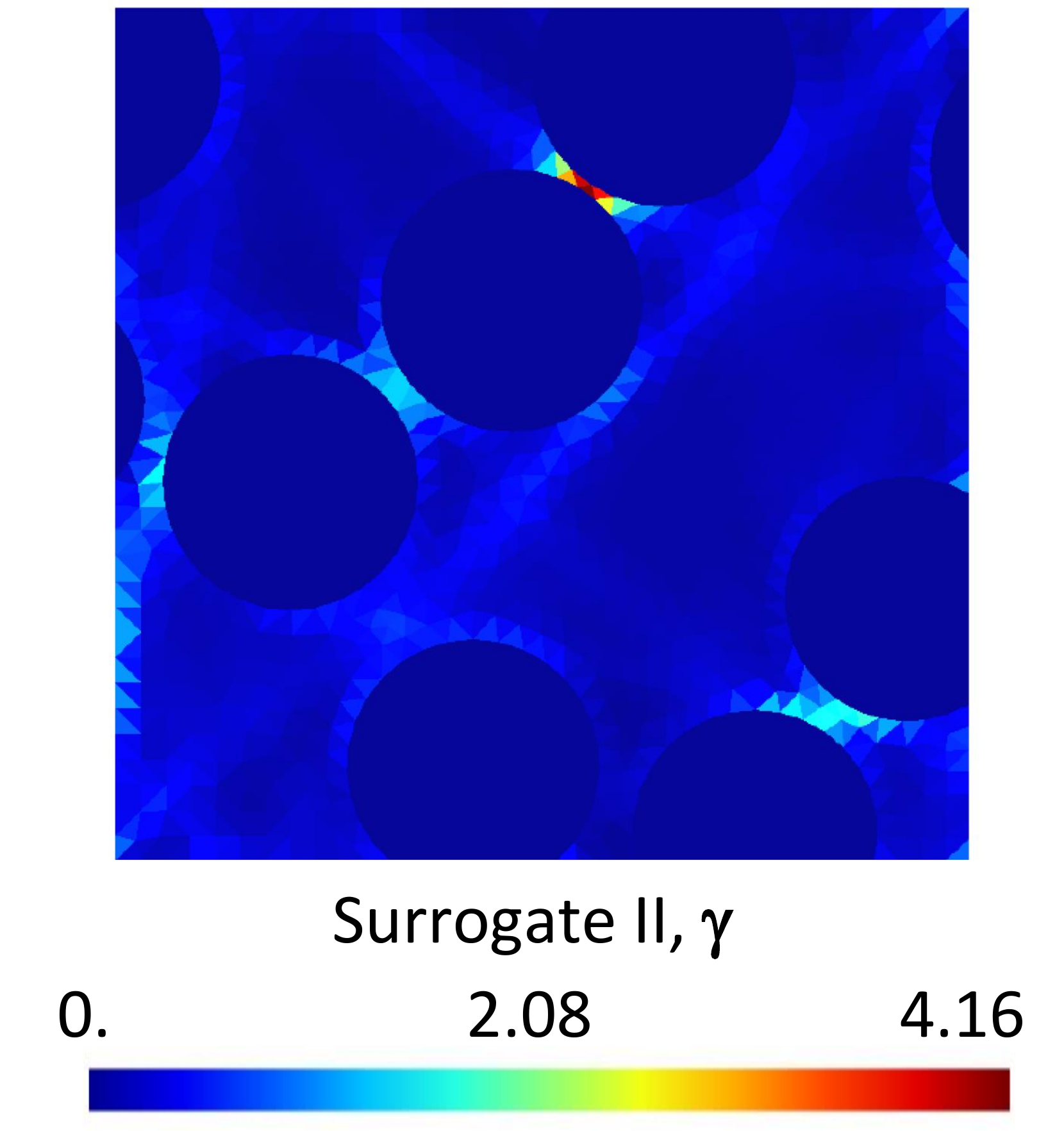}\label{fig:S2_P1}}\,
	\subfigure[]{\includegraphics[scale=0.23]{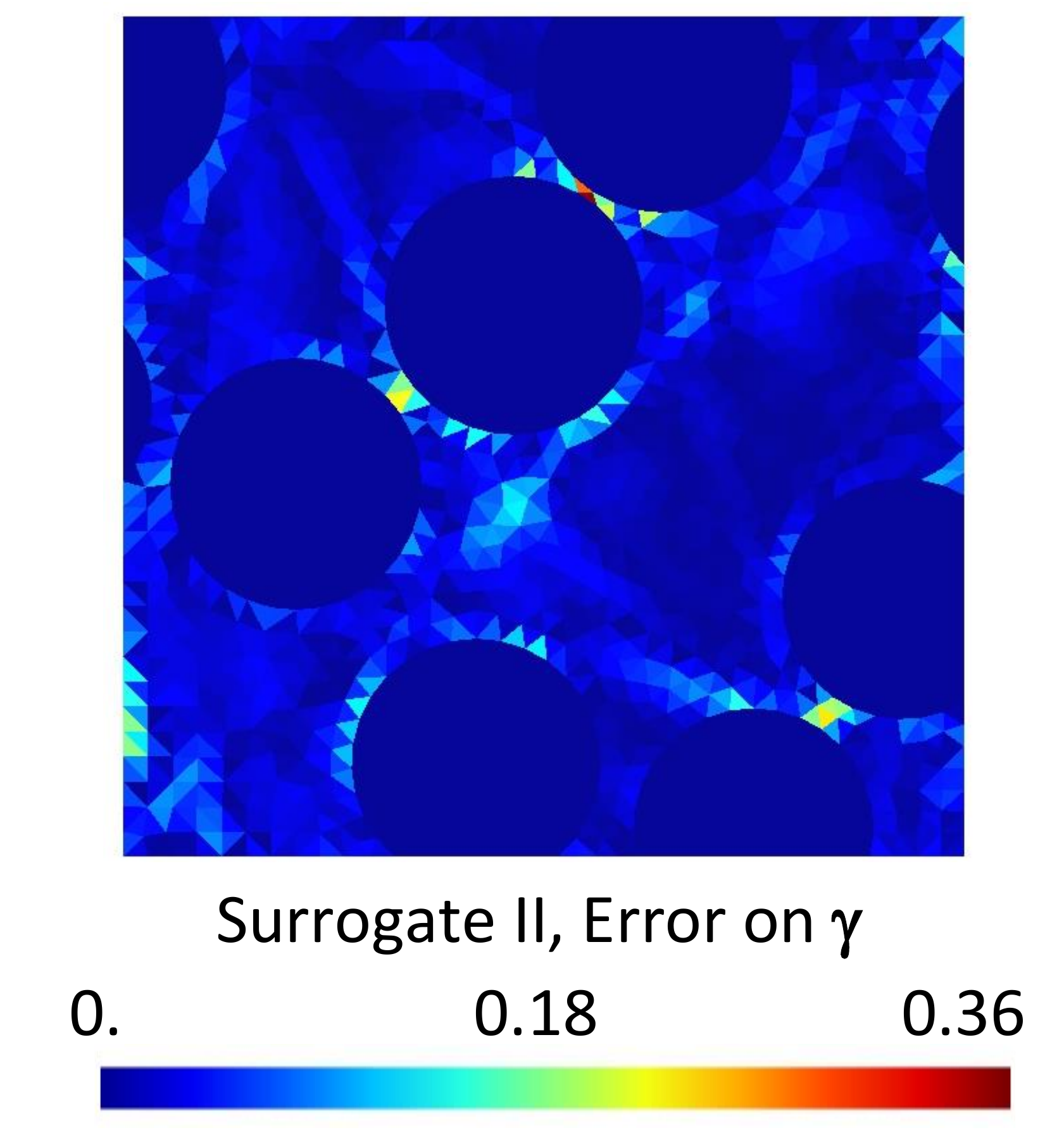}\label{fig:S2_P1_error}}\,
	\subfigure[]{\includegraphics[scale=0.23]{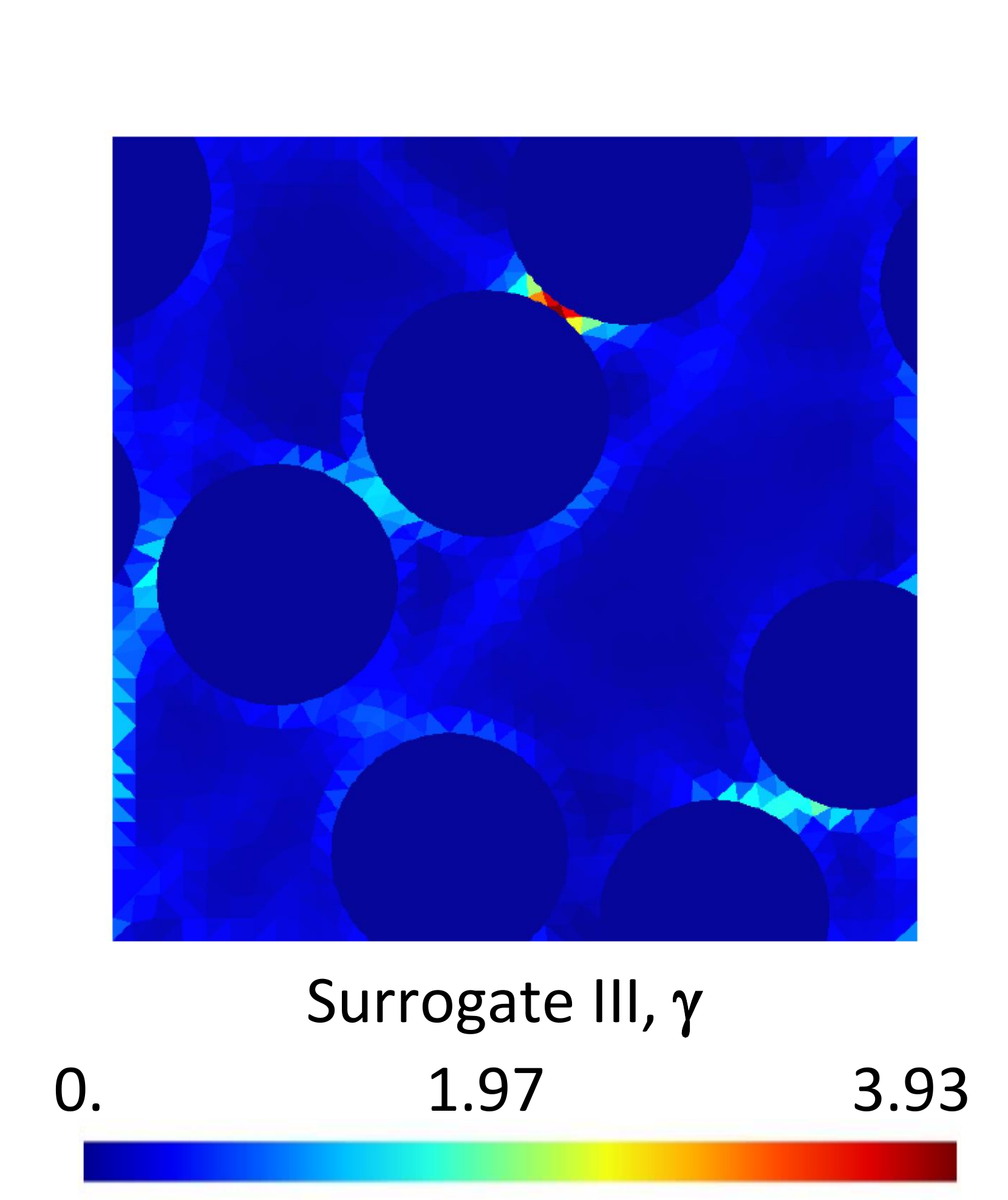}\label{fig:S3_P1}}\,
        \subfigure[]{\includegraphics[scale=0.23]{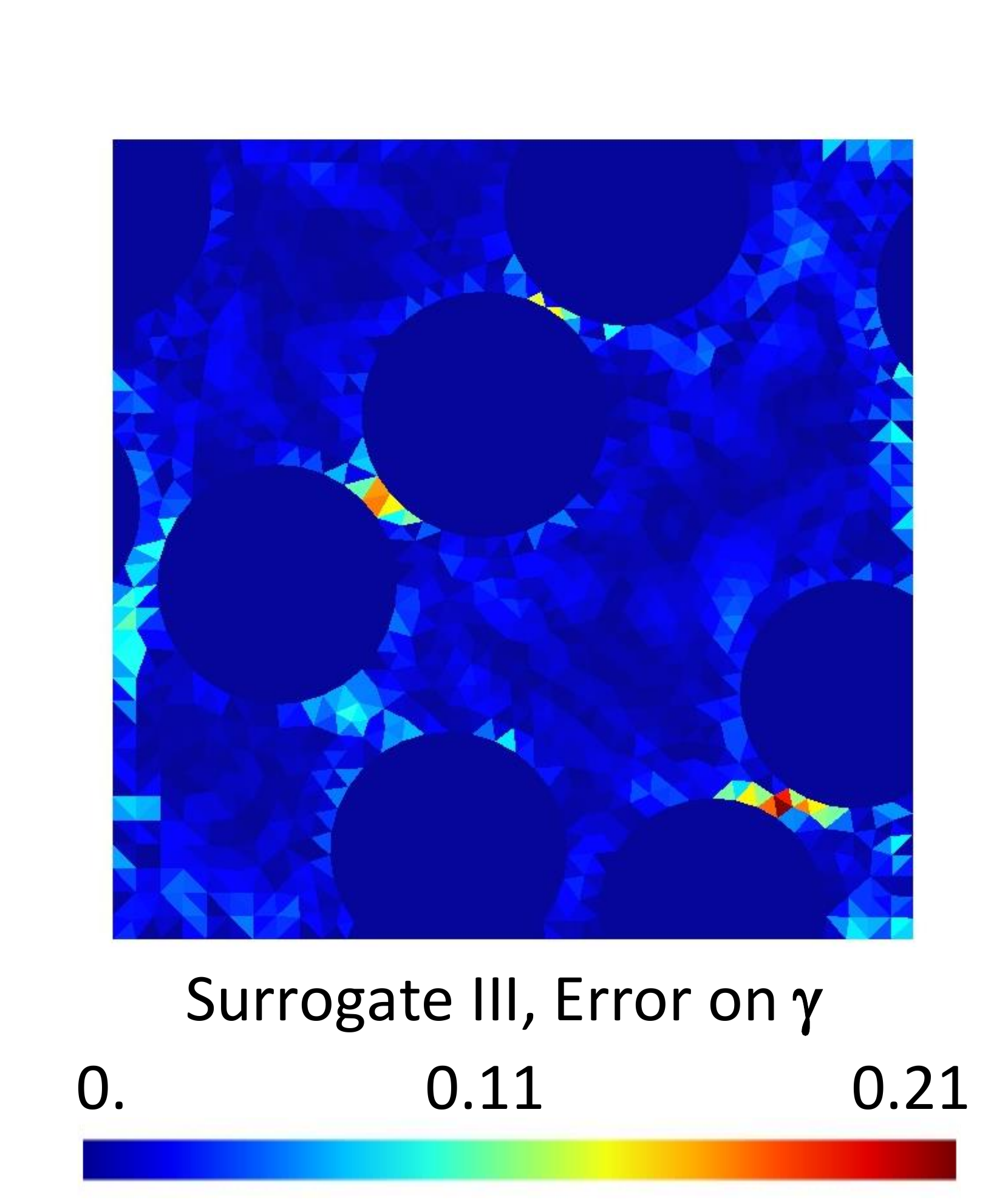}\label{fig:S3_P1_error}}\,		
	\caption{Distributions of the equivalent plastic strain in the RVE at loading point ``1'', see Fig. \ref{fig:LStep}: (a) From finite element simulations; and (b, c), (d, e) and (f, g) Respectively using Surrogates I, II and III. Errors are in terms of absolute values.}\label{fig:Ep_FieldP1}
\end{figure}

\begin{figure}[!htb]
	\centering
	\subfigure[]{\includegraphics[scale=0.23]{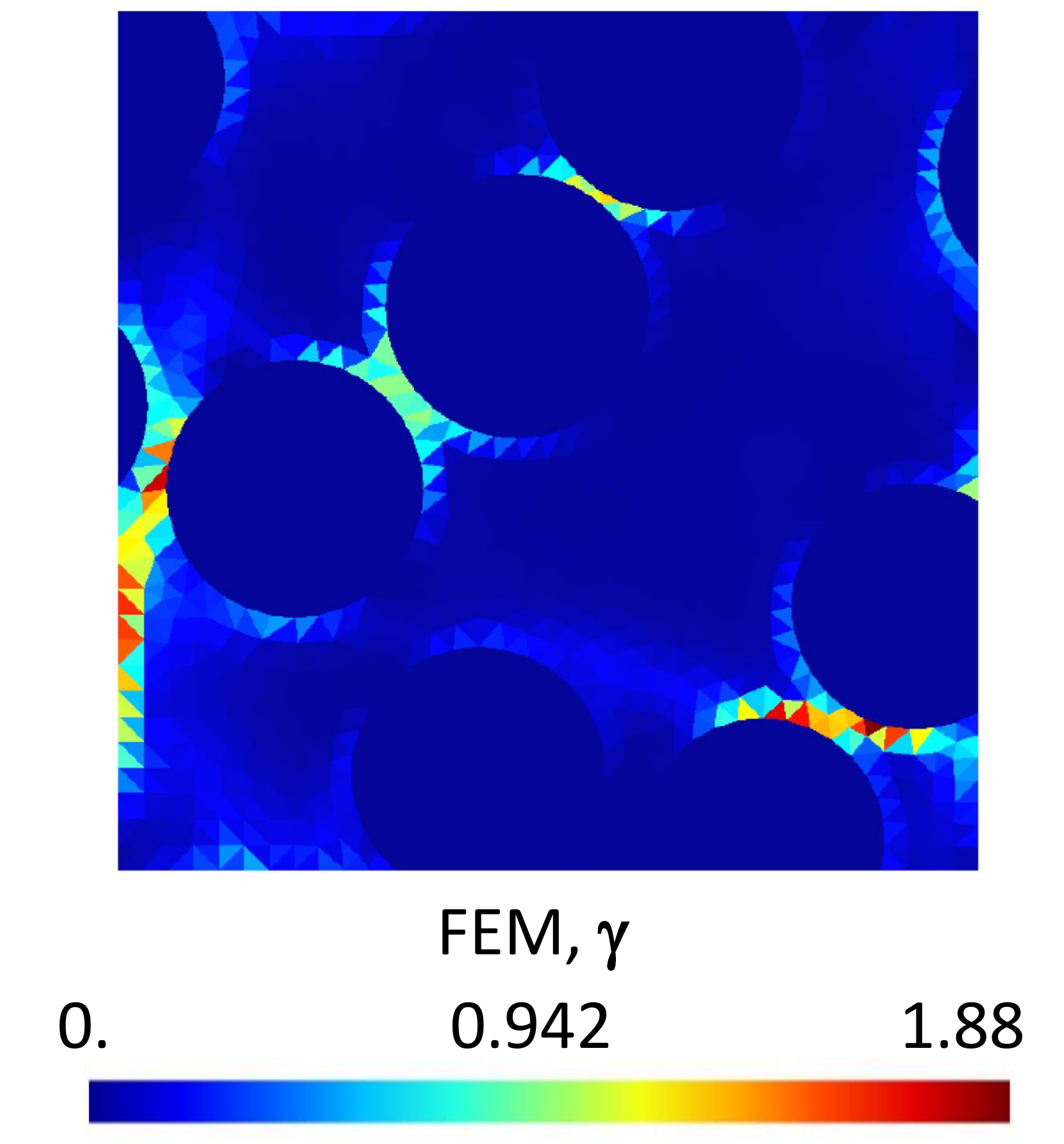}\label{fig:FE_Ep2}}\quad\quad\quad\quad\quad\quad\quad\quad\quad\quad\,
	\subfigure[]{\includegraphics[scale=0.23]{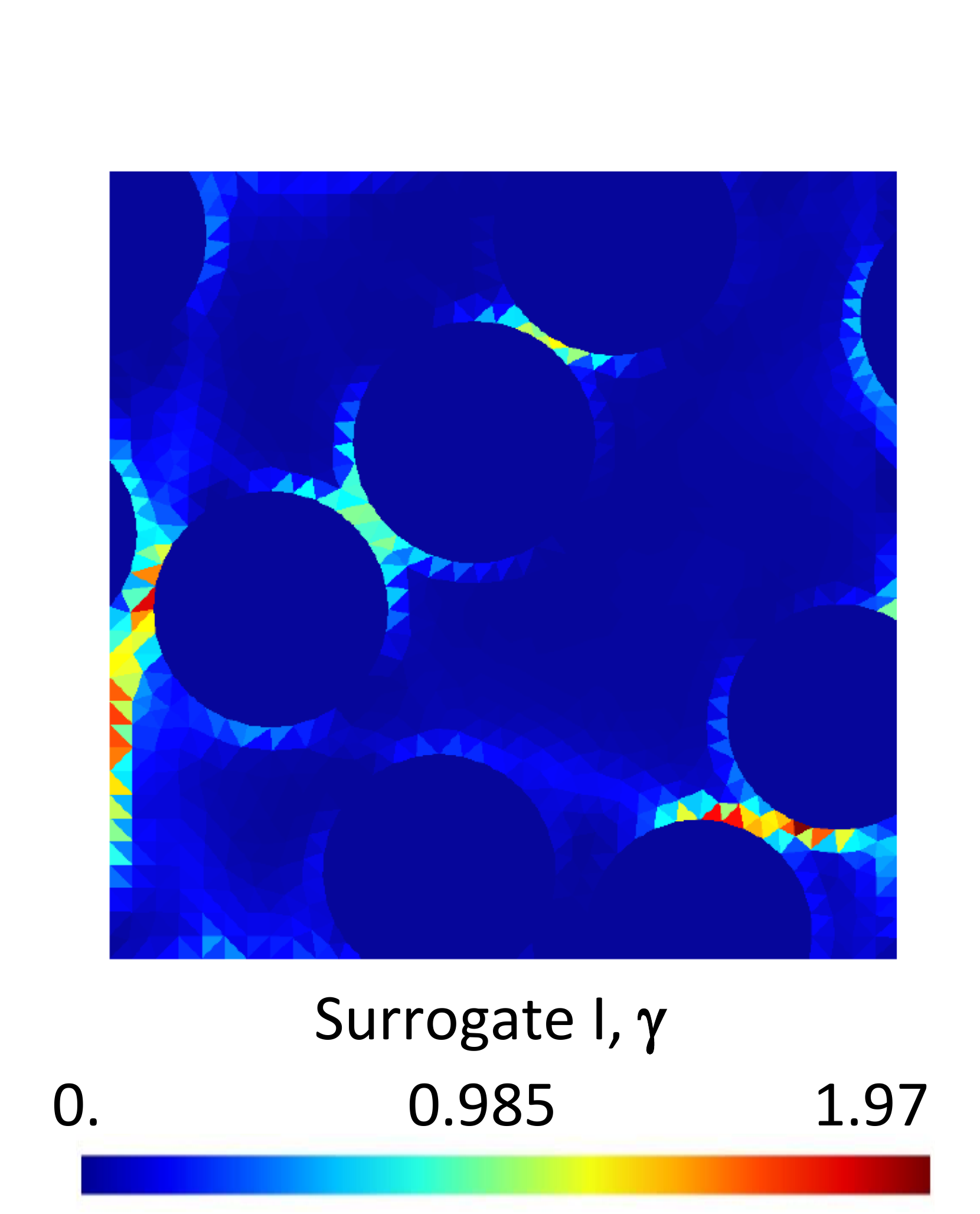}\label{fig:S1_P2}}\,
	\subfigure[]{\includegraphics[scale=0.23]{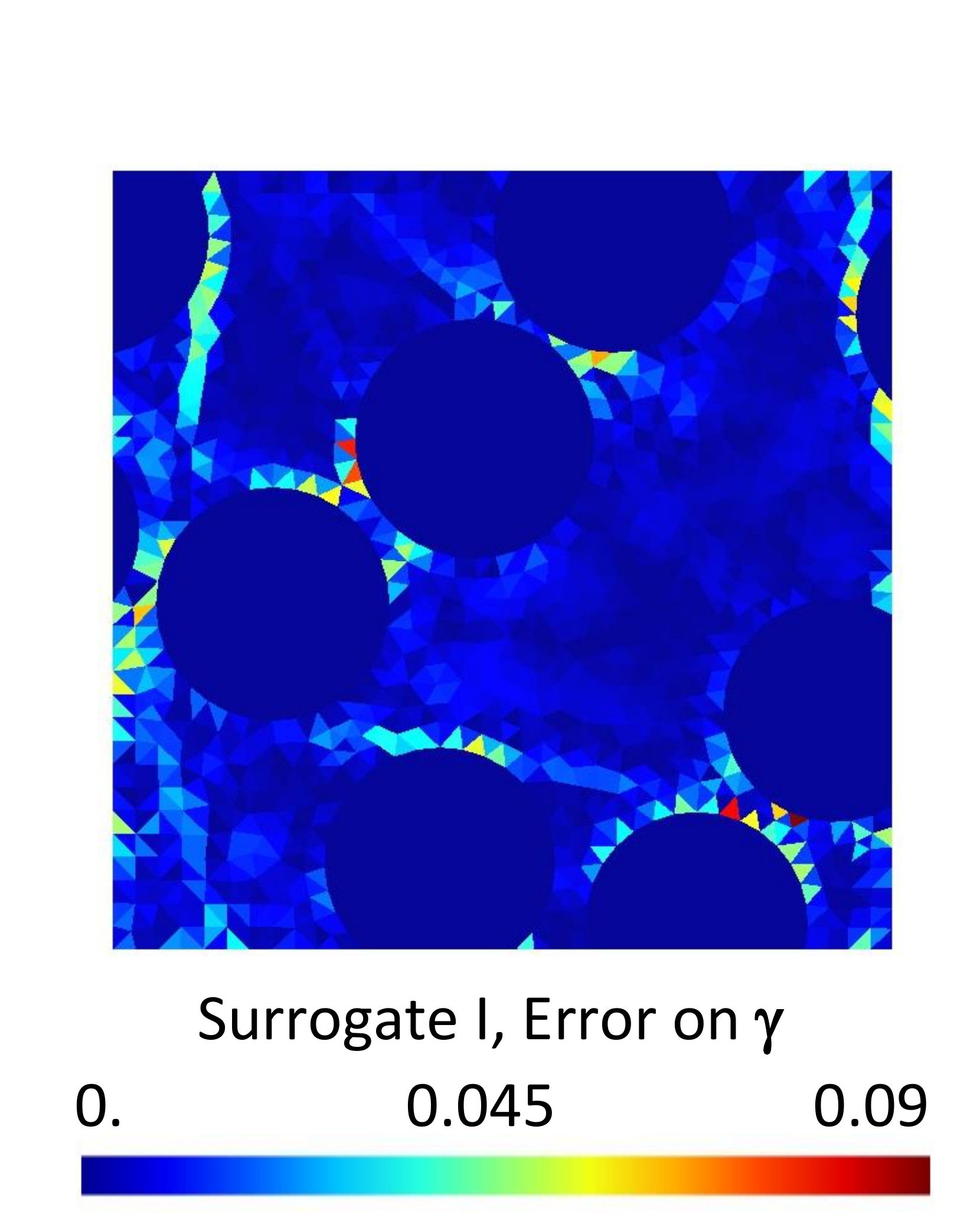}\label{fig:S1_P2_error}}\\		
	\subfigure[]{\includegraphics[scale=0.23]{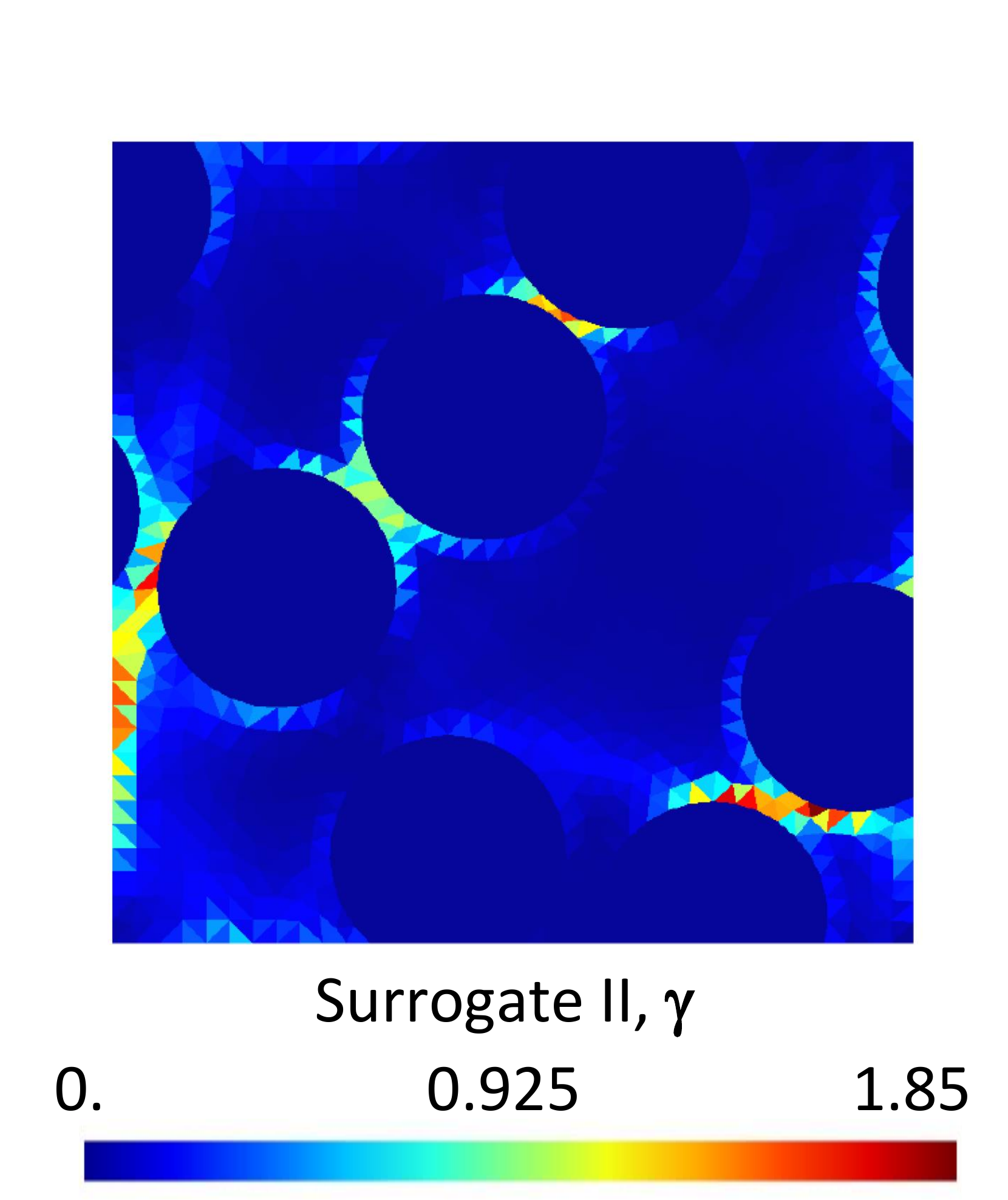}\label{fig:S2_P2}}\,
	\subfigure[]{\includegraphics[scale=0.23]{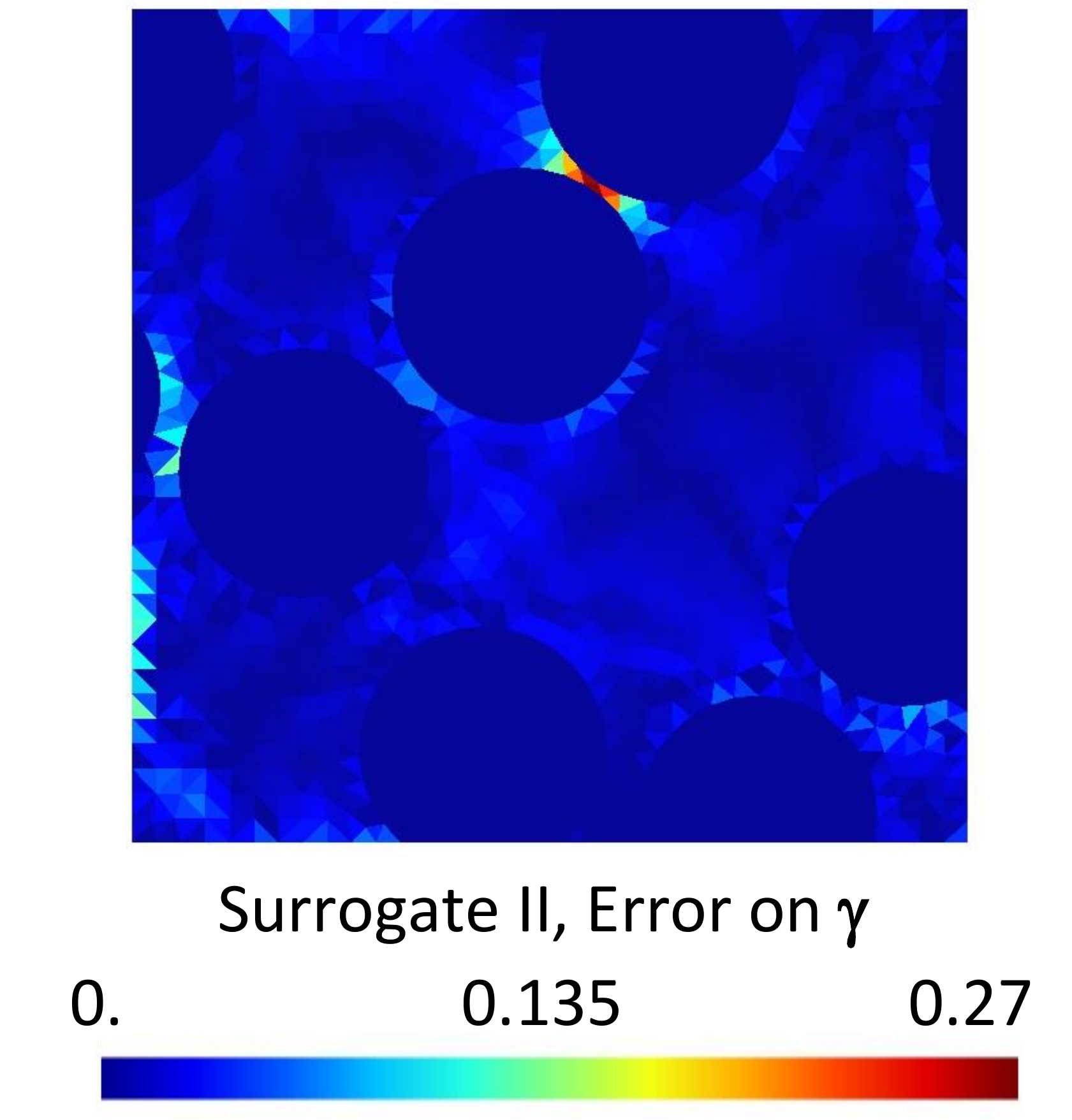}\label{fig:S2_P2_error}}\,
	\subfigure[]{\includegraphics[scale=0.23]{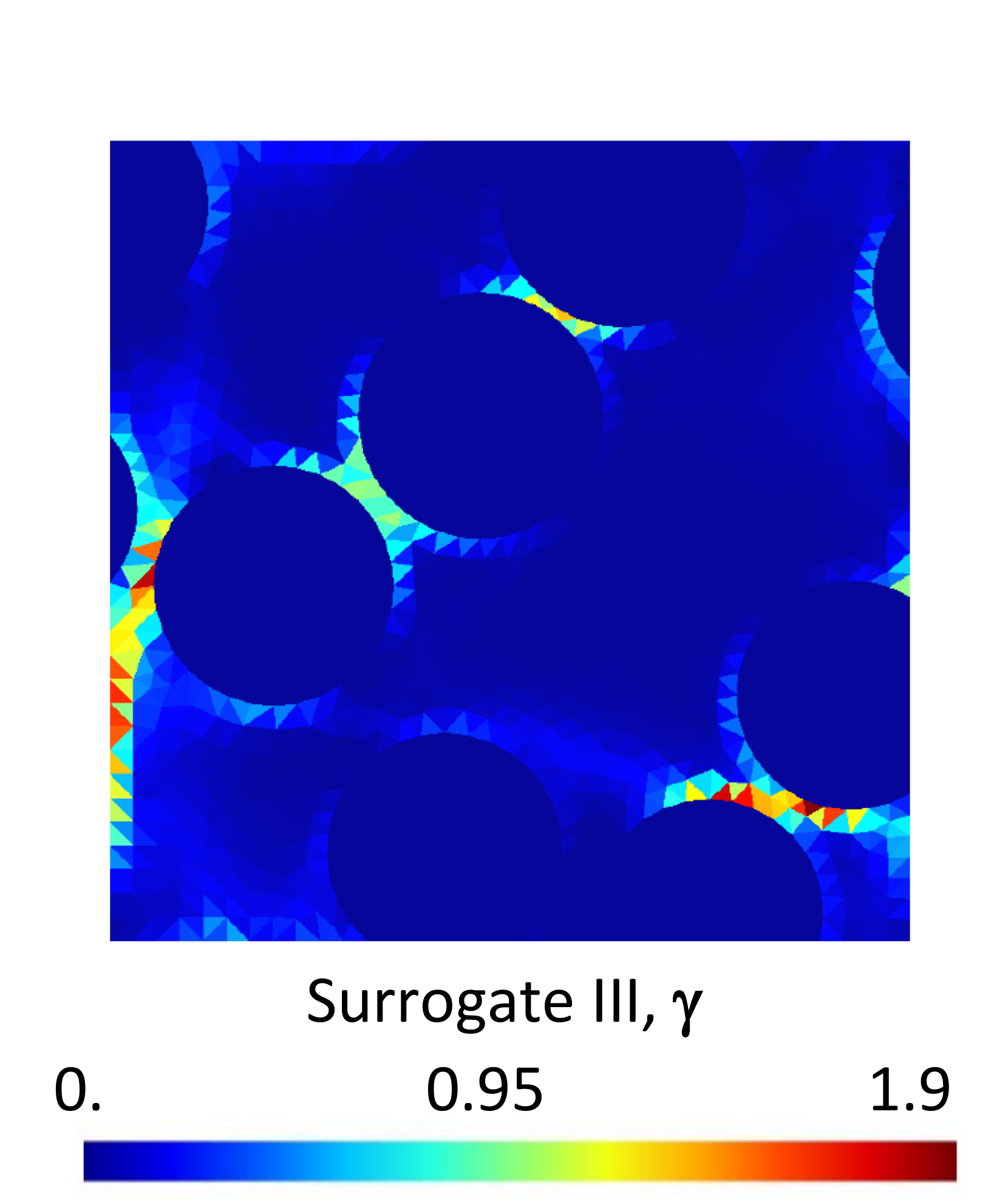}\label{fig:S3_P2}}\,
	\subfigure[]{\includegraphics[scale=0.23]{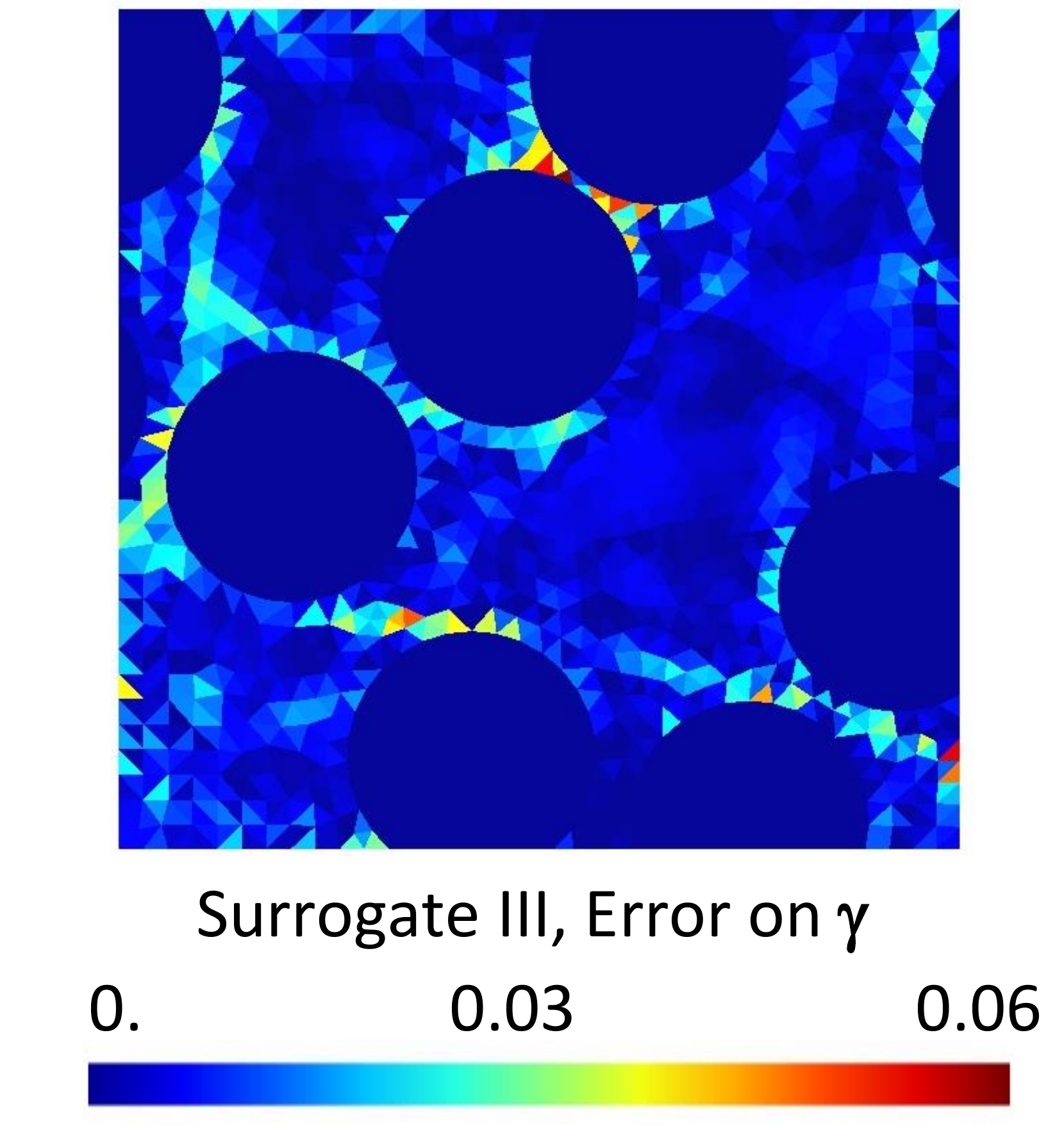}\label{fig:S3_P2_error}}
	\caption{Distributions of the equivalent plastic strain in the RVE at loading point ``2'', see Fig. \ref{fig:LStep}: (a) From finite element simulations; and (b, c), (d, e) and (f, g) Respectively using Surrogates I, II and III. Errors are in terms of absolute values.}\label{fig:Ep_FieldP2}
\end{figure}

\begin{figure}[!htb]
	\centering
	\subfigure[]{\includegraphics[scale=0.23]{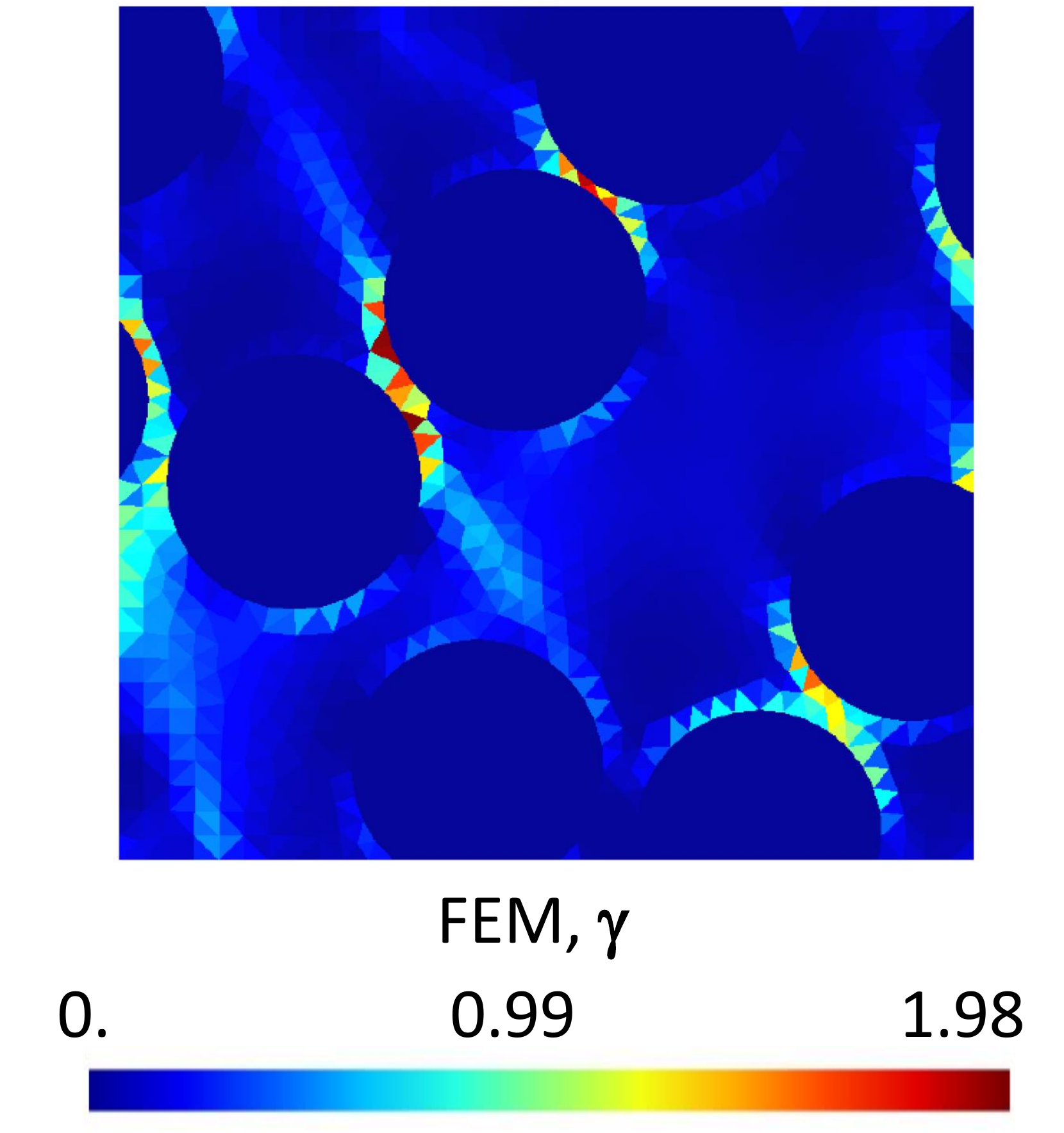}\label{fig:FE_Ep3}}\quad\quad\quad\quad\quad\quad\quad\quad\quad\quad\,
	\subfigure[]{\includegraphics[scale=0.23]{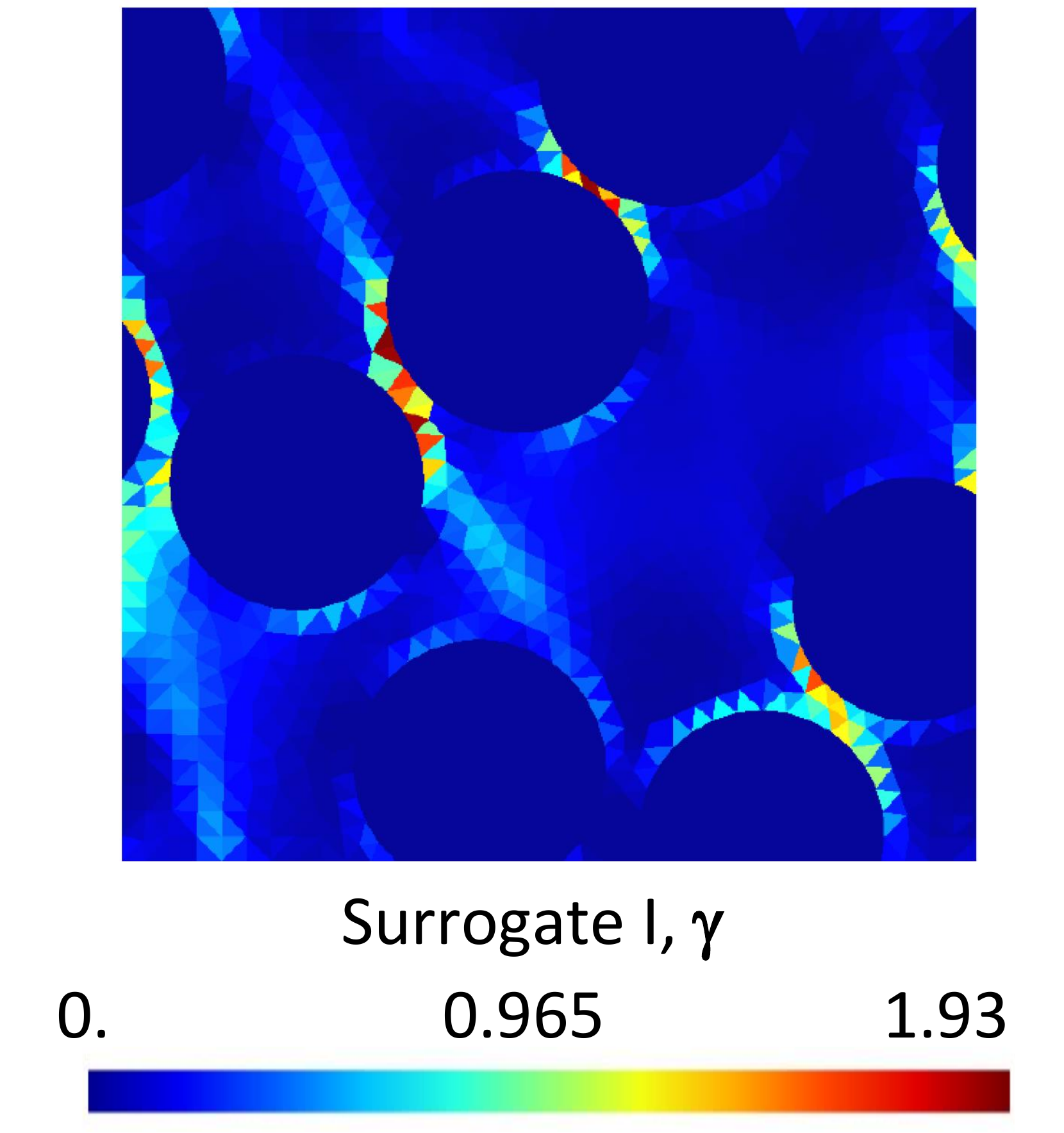}\label{fig:S1_P3}}\,
	\subfigure[]{\includegraphics[scale=0.23]{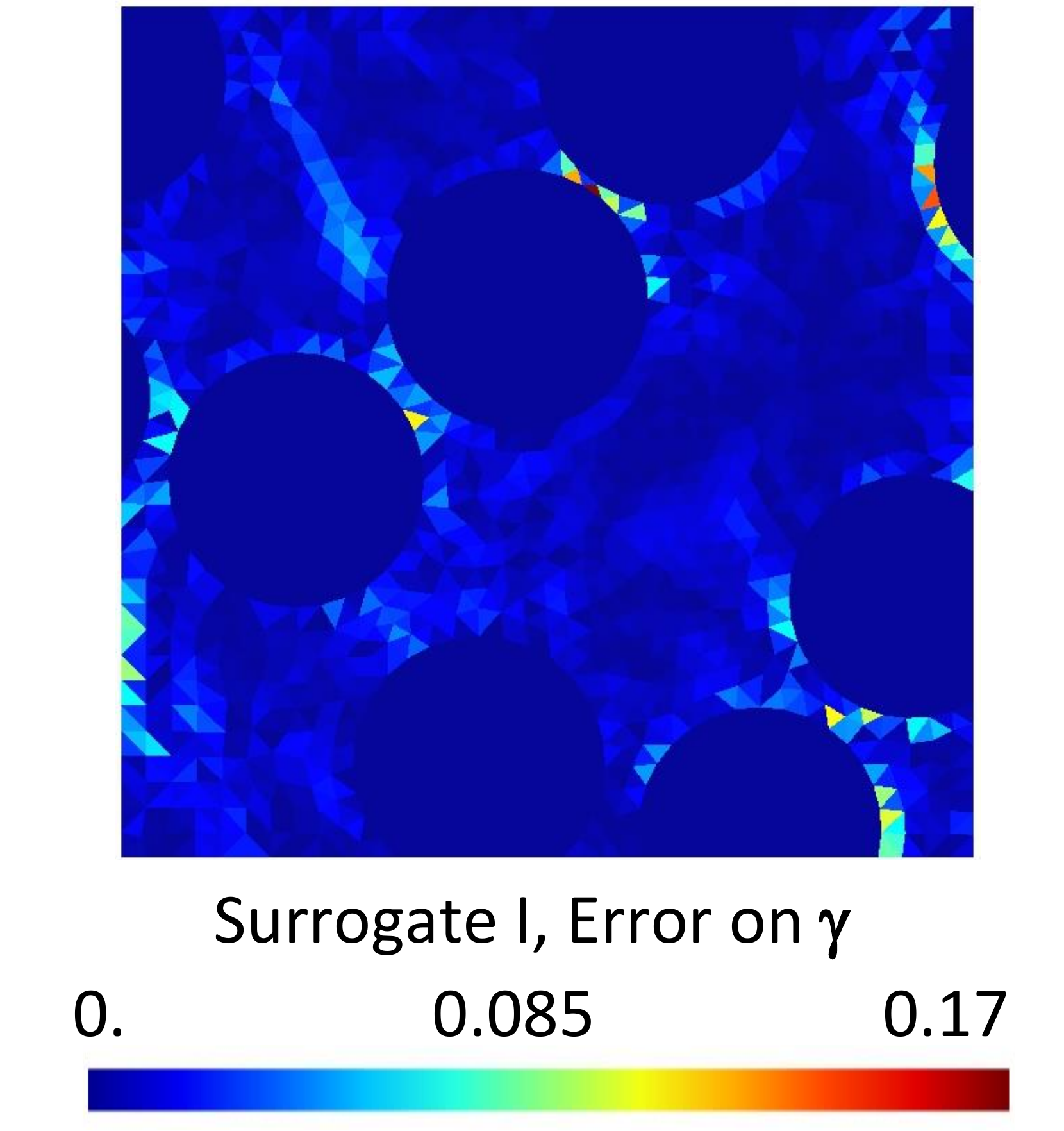}\label{fig:S1_P3_error}}\\
	\subfigure[]{\includegraphics[scale=0.23]{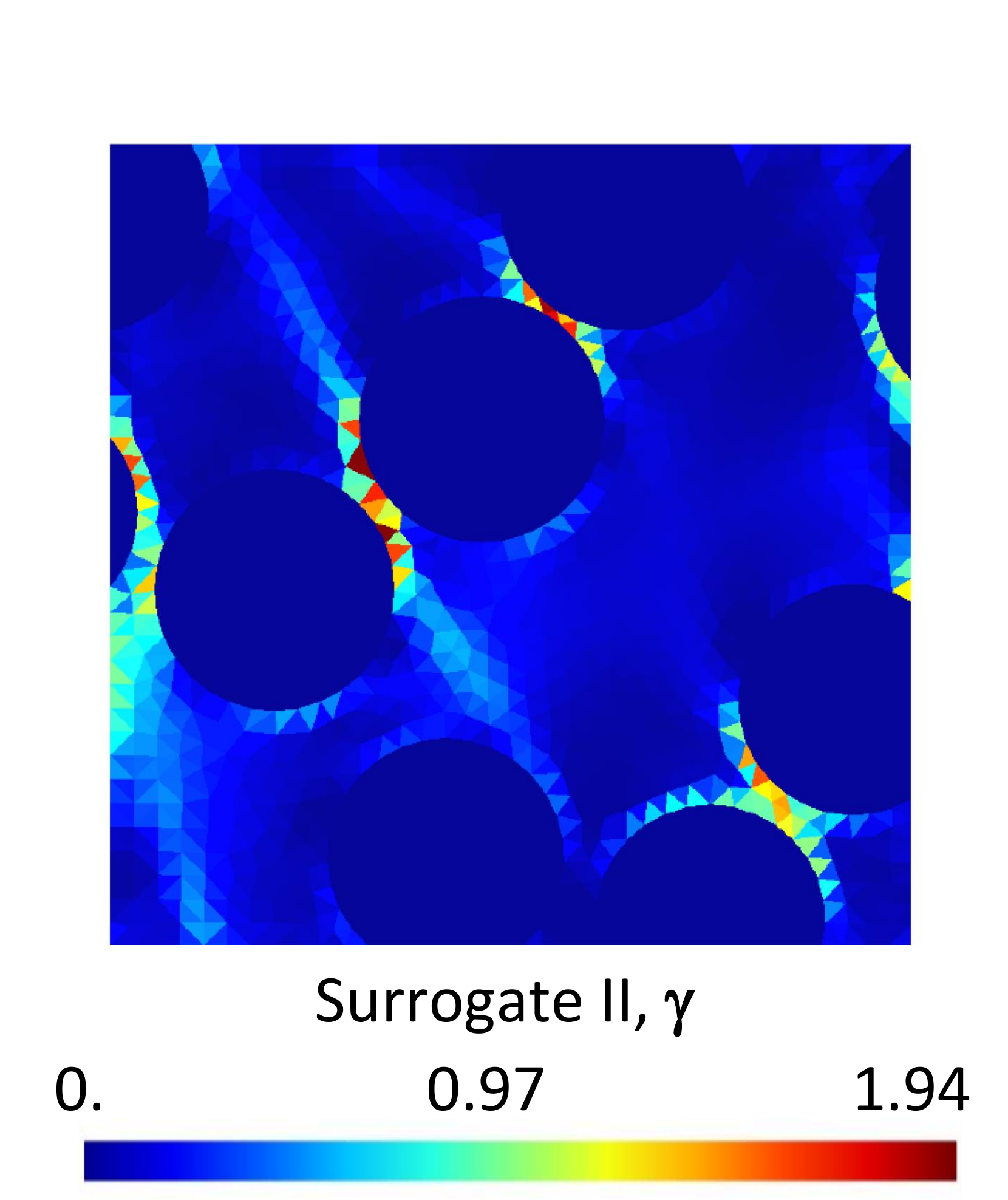}\label{fig:S2_P3}}\,
        \subfigure[]{\includegraphics[scale=0.23]{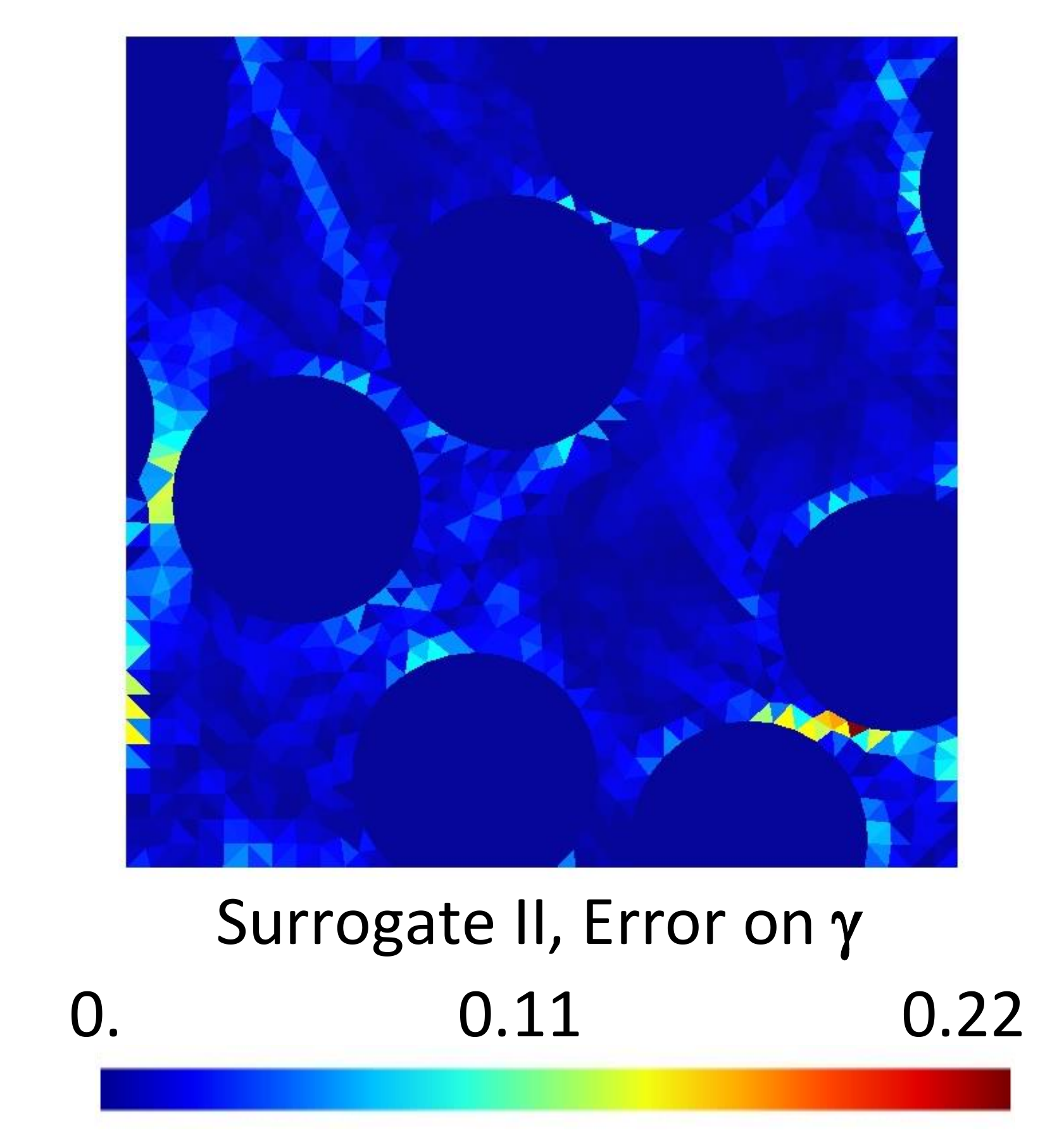}\label{fig:S2_P3_error}}\,
	\subfigure[]{\includegraphics[scale=0.23]{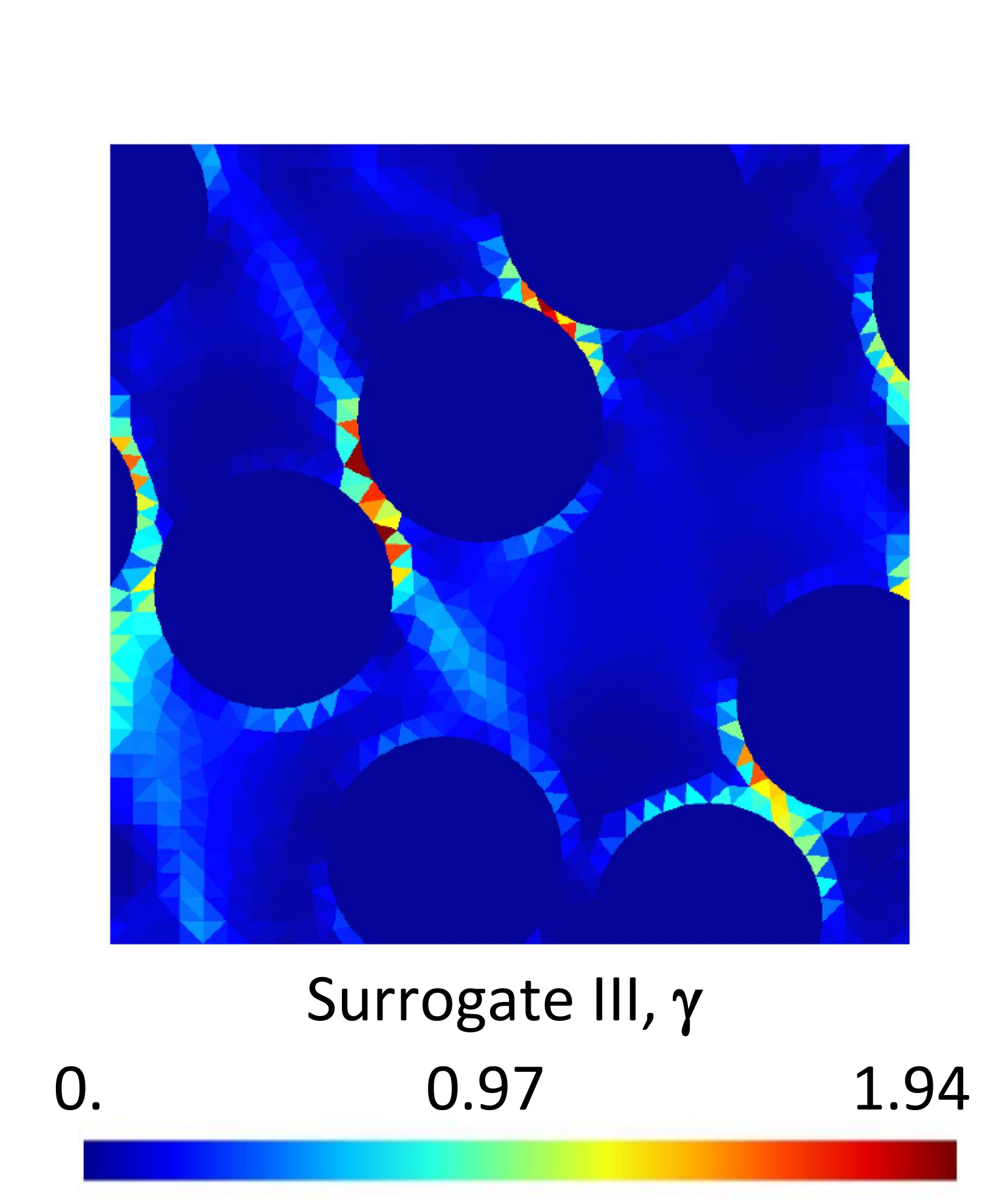}\label{fig:S3_P3}}\,
	\subfigure[]{\includegraphics[scale=0.23]{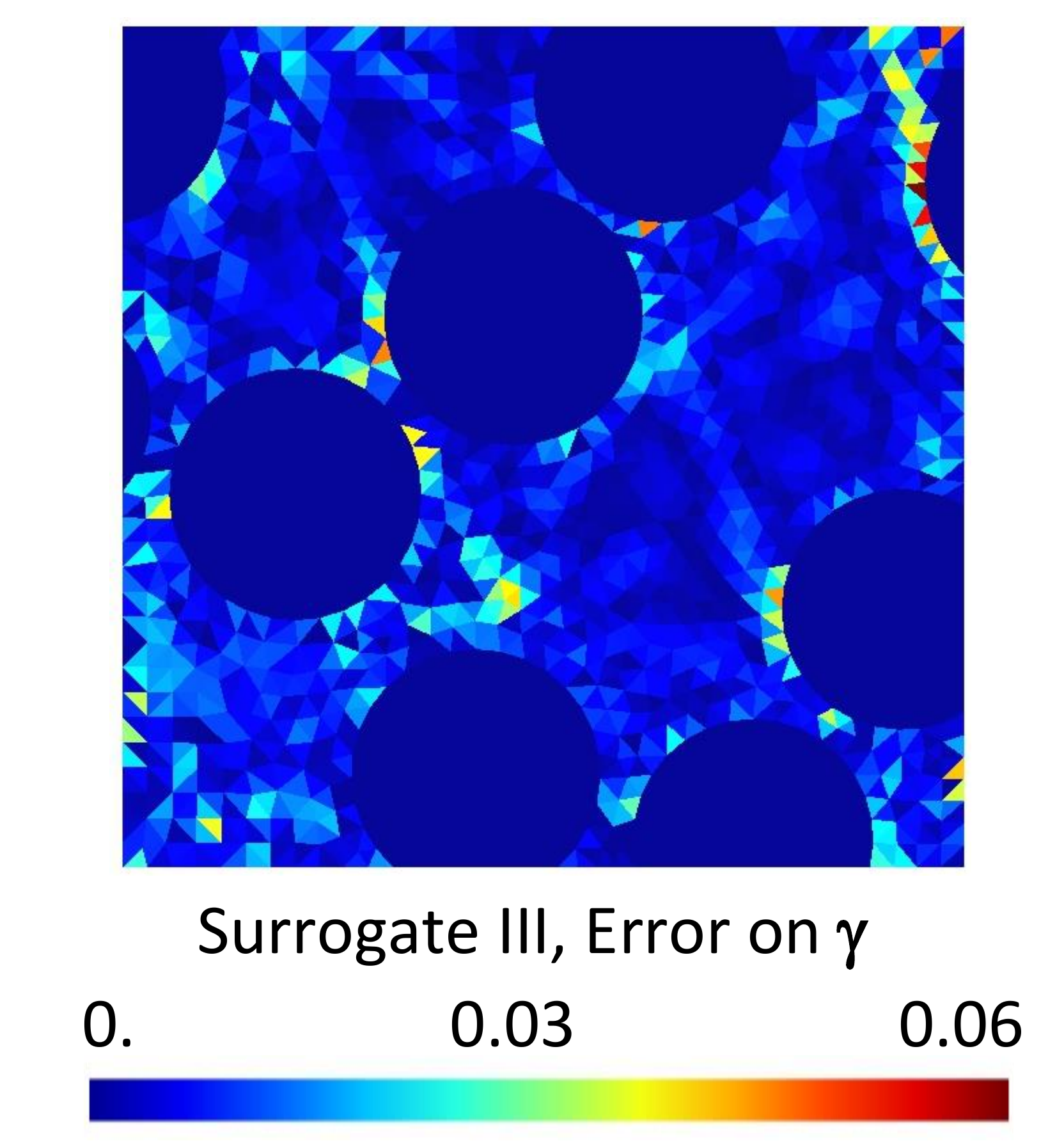}\label{fig:S3_P3_error}}	
	\caption{Distributions of the equivalent plastic strain in RVE at loading point ``3'', see Fig. \ref{fig:LStep}: (a) From finite element simulations; and (b, c), (d, e) and (f, g) Respectively using Surrogates I, II and III. Errors are in terms of absolute values.}\label{fig:Ep_FieldP3}
\end{figure}
}

The distributions of the equivalent plastic strains $\bm{\gamma}$ within the RVE are also compared at different loading steps/configurations.
Three loading steps/configurations are marked on the random and cyclic testing paths in Fig. \ref{fig:LStep}\changes{. Their} corresponding equivalent plastic strain fields $\bm{\gamma}$ within the RVE are presented in Figs. \ref{fig:Ep_FieldP1}-\ref{fig:Ep_FieldP3}. The results of the direct finite element simulations and the reconstructions obtained with the three surrogate models are illustrated.
Since the predictions of the surrogate models are not bounded and could even be negative in some elements, some elements predicted negative values of the order of 0.0001-0.001 and these negative values were set to zero in Figs. \ref{fig:Ep_FieldP1}-\ref{fig:Ep_FieldP3}.
The three considered configurations correspond to three totally different plastic strain distributions predicted by the direct finite element simulations, see Figs. \ref{fig:FE_Ep1}-\ref{fig:FE_Ep3}.
These three different patterns of the equivalent plastic strain fields $\bm{\gamma}$ are all well predicted by the three surrogate models. \secondreviewer{The error with Surrogate III remains limited to around 5\% of the maximum value. This is due to the fact that the distribution patterns are dominated by the first few principal components of the state variable field reduction, so that even the least accurate surrogate model can recover the pattern. Because of the use of a dimensionality reduction for the Surrogate II and Surrogate III, the larger errors can be found in locations which do not correspond to the maximum plastic strain, see \emph{e.g.} Figs. \ref{fig:S2_P1_error} and \ref{fig:S3_P1_error}.}

\subsection[Surrogate model of the von Mises stress distribution]{Surrogate model of the von Mises stress distribution $\bm{\tau}_\text{eq}$}\label{sec:App_TauEq}

From the analysis conducted in Section \ref{sec:App_EpStrn} it appears that there is no need to try arbitrarily different numbers of hidden variables when training the RNNs of Surrogates I or III. 
Indeed, according to the desired relative error, the required PCA reduced dimensionality $p$ can be determined from Fig. \ref{fig:PCA_residual}.
Therefore, using a \secondreviewer{dimensionality break down}, the \secondreviewer{trial and error approach} can be carried out on a RNN with only one output, ${\underline{\xi}_\tau}_{p}$, which is the coefficient of principal component ${\underline{\bm{v}}_\tau}_p$, yielding the required number of hidden variables of the RNNs.
As soon as the number of hidden variables is sufficient for this 1D RNN, it is also sufficient for all the RNNs of the Surrogates I and III.
We discard Surrogate II in this application because of its poor efficiency.

In this application, $\bm{\tau}_\text{eq}$ is considered and $0.01$ is used as the desired relative error.
According to Fig. \ref{fig:PCA_residual}, this error corresponds to $p\approx60$.
Therefore, the normalized coefficient of the principal component ${\underline{\bm{v}}_\tau}_{60}$, ${\underline{\xi}_\tau}_{60}$, is used as the unique output of a RNN when performing the \secondreviewer{trial and error approach} of its design. This \secondreviewer{trial and error approach} is summarized as follows
\begin{itemize}
	\item
	The constant $\text{NNW}_\text{I}$: (3, 70) and $\text{NNW}_\text{O}$: (30, 1) are used. The \secondreviewer{trial and error approach} starts from \secondreviewer{$n_{\text{h}}=100$};
	\item
	\firstreviewer{The RNN is trained for 5000 epochs after which we check whether the evolution of ${\underline{\xi}_\tau}_{60}$ can be qualitatively captured by this RNN;}
	\item
	If the trained RNN is not accurate enough, we increase \secondreviewer{$n_{\text{h}}$} by 100 and start a new RNN training up to reaching the required MSE.
\end{itemize}   

\begin{figure}[!htb]
	\centering
	\subfigure[]{\includegraphics[scale=0.38]{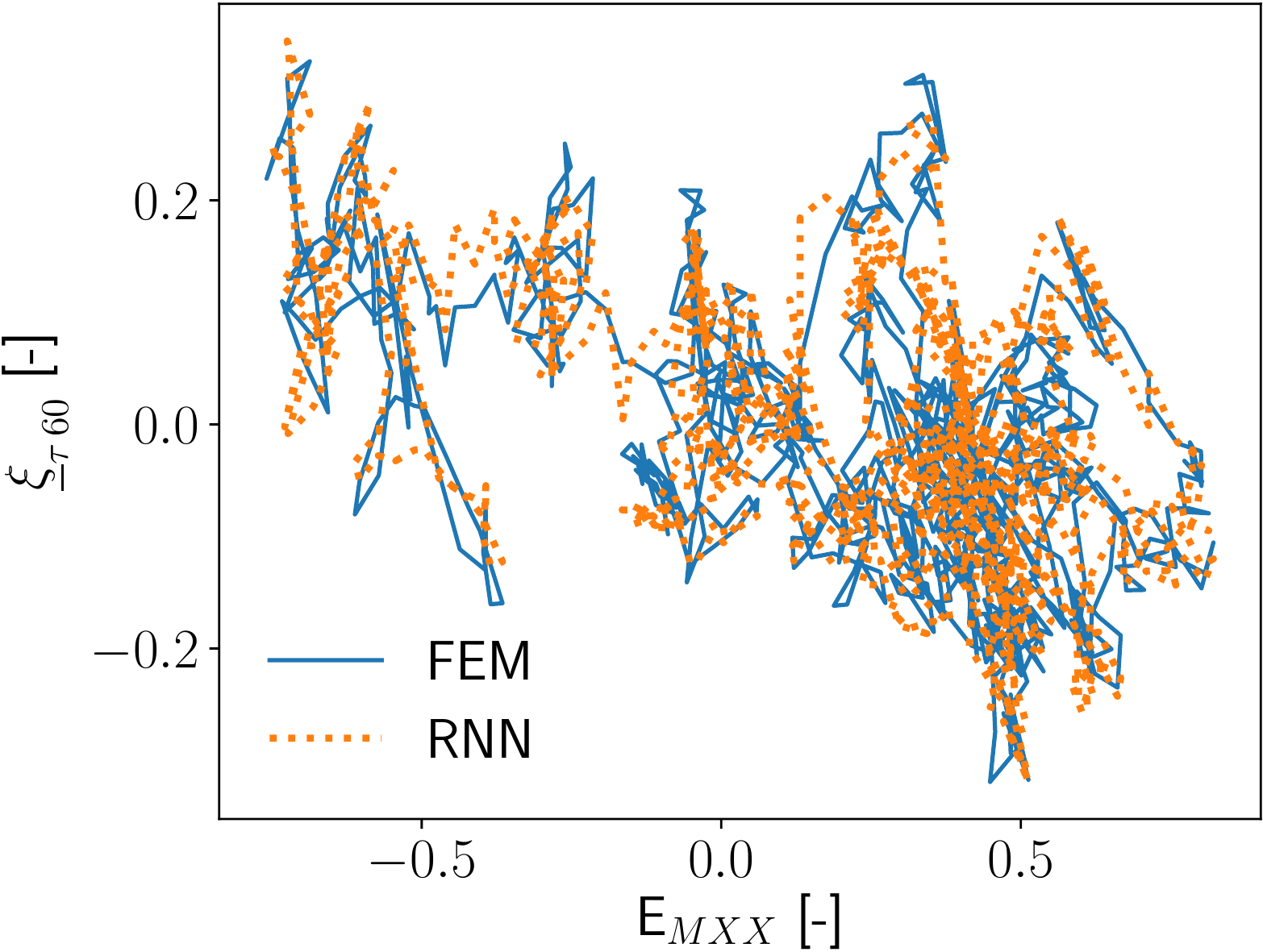}\label{fig:Xi60_1}}
	\subfigure[]{\includegraphics[scale=0.38]{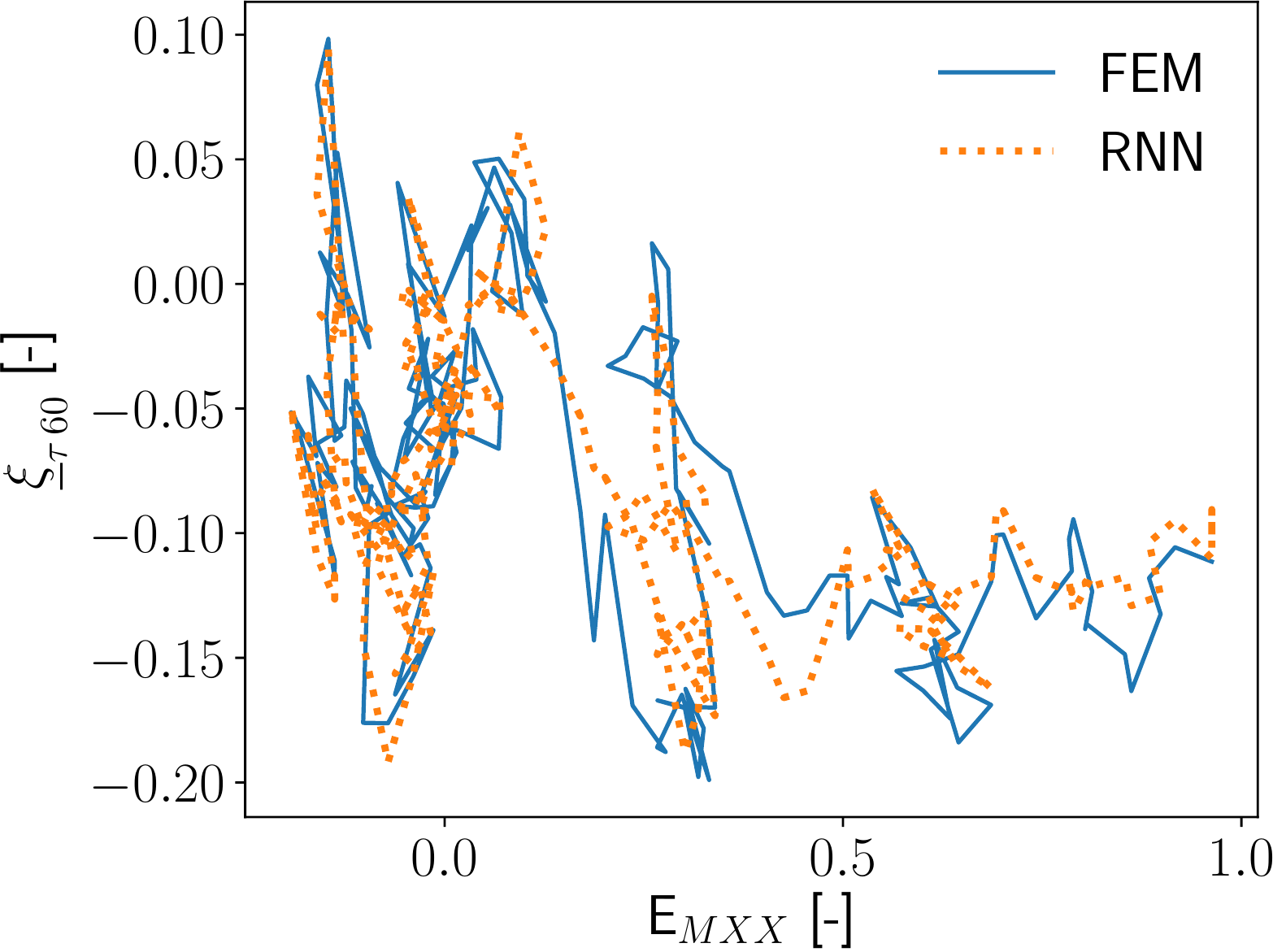}\label{fig:Xi60_2}}	
	\caption{Evolution of ${\underline{\xi}_\tau}_{60}$ reached for \secondreviewer{$n_{\text{h}}=600$} in the \secondreviewer{trial and error approach} of the surrogate model design. The training is stopped after 5000 epochs.}\label{fig:Xi60}
\end{figure}

The same loading sequences as in the case of \changes{the} equivalent plastic strain, see Section \ref{sec:App_EpStrn}, are used for training and testing in this case.    
The \secondreviewer{trial and error approach} yields that \secondreviewer{$n_{\text{h}}=600$} is needed in order to capture the evolution \changes{trends of} ${\underline{\xi}_\tau}_{60}$, and two examples are shown in Fig. \ref{fig:Xi60}.
It needs to be noticed that the RNN predictions are not really accurate with respect to the finite element results, because the training is stopped after 5000 epochs.
When considering surrogate models with PCA \secondreviewer{dimensionality reduction}, since more than 60 principal components will be used, the accuracy requirement of ${\underline{\xi}_\tau}_{60}$ could be relaxed during the trial and \changes{error approach}.

\begin{table}
	\centering
	\begin{threeparttable}		
		\caption{Hyper-parameters of RNNs of Surrogates I and III to reconstruct the von Mises stress distribution $\bm{\tau}_\text{eq}$; All RNNs use \secondreviewer{$n_{\text{h}}=600$}, and $h_0$=-1.0; \firstreviewer{The notation $(n_0,\,\ldots,\,n_i,\,\ldots,\,n_N)$ holds for the number of input nodes $n_0$, of nodes $n_i$ in hidden layer $i$ and of output nodes $n_N$ --with the output nodes of a neural network holding as input nodes of the subsequent one and not being duplicated}.}\label{tab:RNN_SVM}  
		\begin{tabular}{c|c|c|c} 
			\hline\hline   
			Model & $\text{NNW}_\text{I}$ &$\text{NNW}_\text{O}$&MSE on testing data\\ 
			\hline\hline 
			Surrogate I&(3, 70)&(1200, 2237)&0.00204\\
			\hline 
			Surrogate III&(3, 70)&$(200,\,20)^{*}$&0.00174\\
			\hline\hline		
		\end{tabular}  
		\begin{tablenotes}
			\small
			\item {${}^*$ PCA with $p=180$ and \secondreviewer{dimensionality break down} with $Q=9$ ($\text{RNN}_q$ with $q=1,2, ..,9$ are trained)}.
		\end{tablenotes}
	\end{threeparttable}
\end{table}

\begin{figure}[htb]
	\centering
	\includegraphics[scale=0.5]{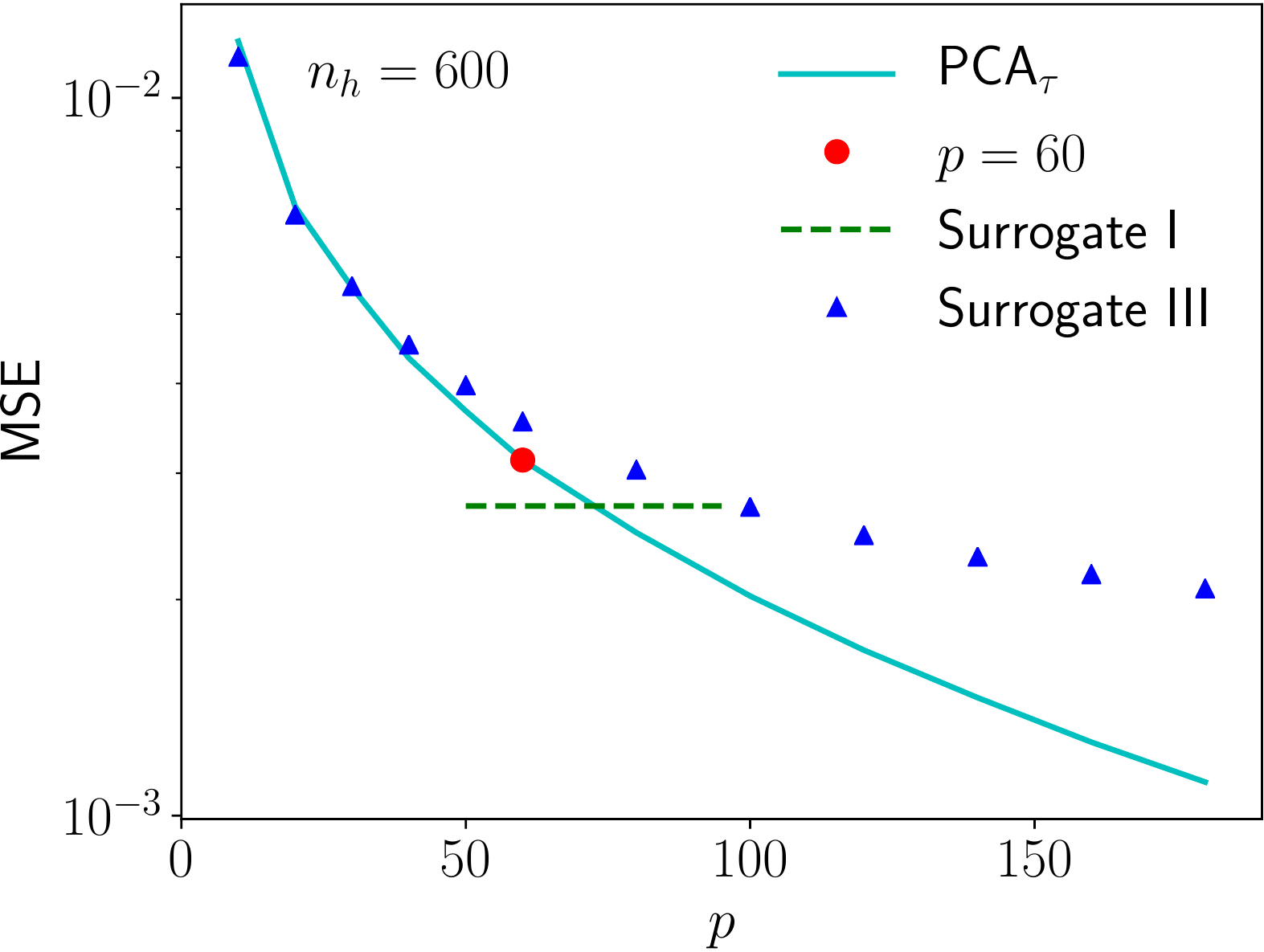}
	\caption{The MSEs of the equivalent von Mises stress field, $\bm{\tau}_\text{eq}$, obtained by the surrogate models I and III on training data. The MSE obtained with a PCA \secondreviewer{dimensionality reduction} is also provided as a reference.}\label{fig:SVM_Error}
\end{figure}

Surrogates I and III with \secondreviewer{$n_{\text{h}}=600$} are adopted to reconstruct the evolution of the equivalent von Mises stress $\bm{\tau}_\text{eq}$ in the RVE.
Their structures are presented in Table \ref{tab:RNN_SVM}.
The obtained MSEs on the training data are plotted in Fig. \ref{fig:SVM_Error}, together with the PCA MSE as a reference.
As expected, the MSEs obtained by surrogate models I and III with $p=180$ are below the MSE of PCA at $p=60$.
The MSEs of the surrogate models on the testing data are reported in Table \ref{tab:RNN_SVM} and are lower than on the training data.
Since the training data of the von Mises stress has a much wider range than for the equivalent plastic strain, large and small values are part of the  trained data, while they were exceptional in the equivalent plastic strain distribution.
Hence the difference in MSE of the equivalent plastic strain which was smaller with the training data than with the testing data is no longer observed with the von Mises stress distribution, which confirms the origin of this error pointed out in Section \ref{sec:ptesting}. 

\begin{figure}[!htb]
	\centering
	\subfigure[]{\includegraphics[scale=0.38]{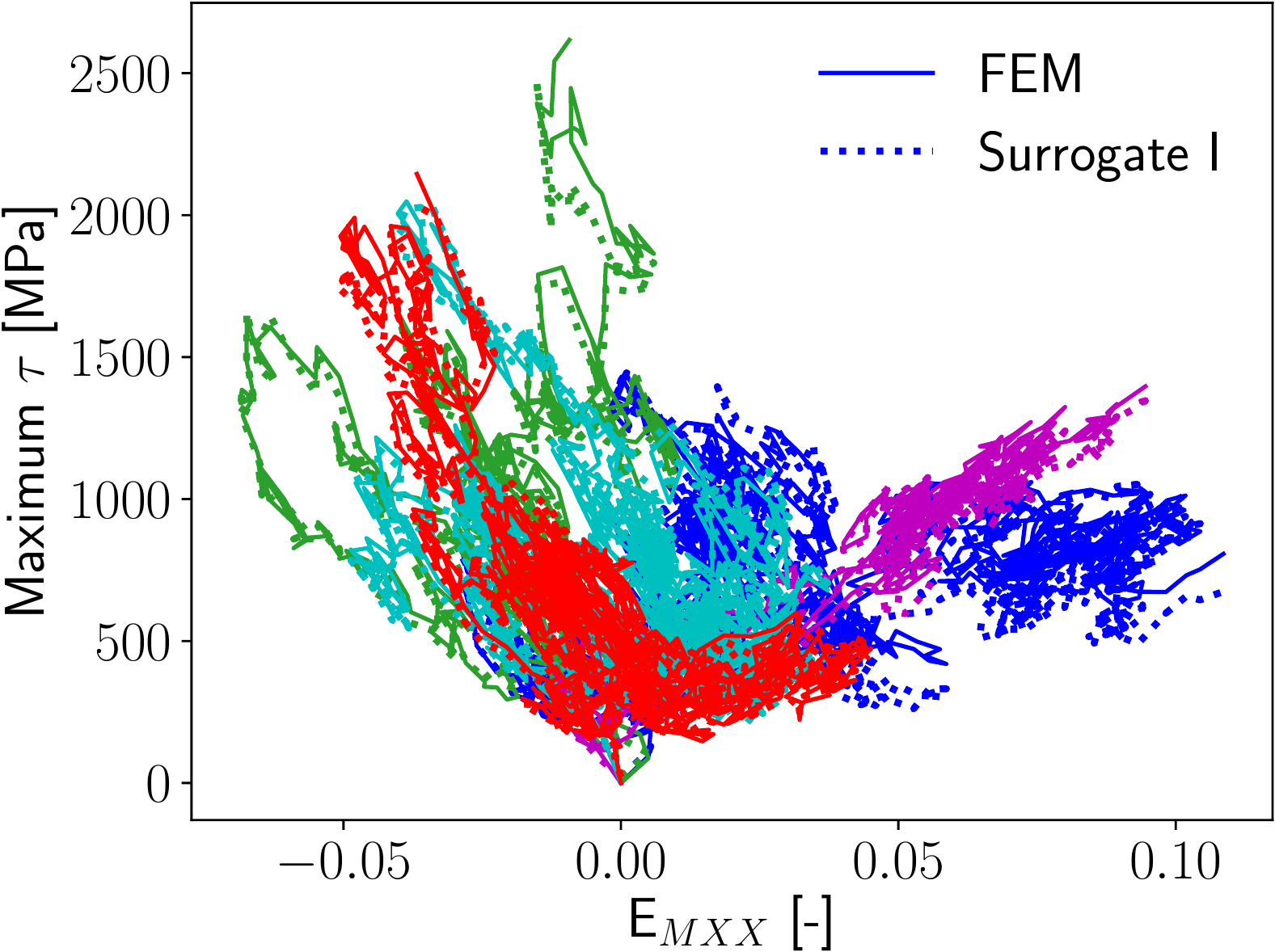}}\label{fig:MaxSVM_RP_I}\,
	\subfigure[]{\includegraphics[scale=0.38]{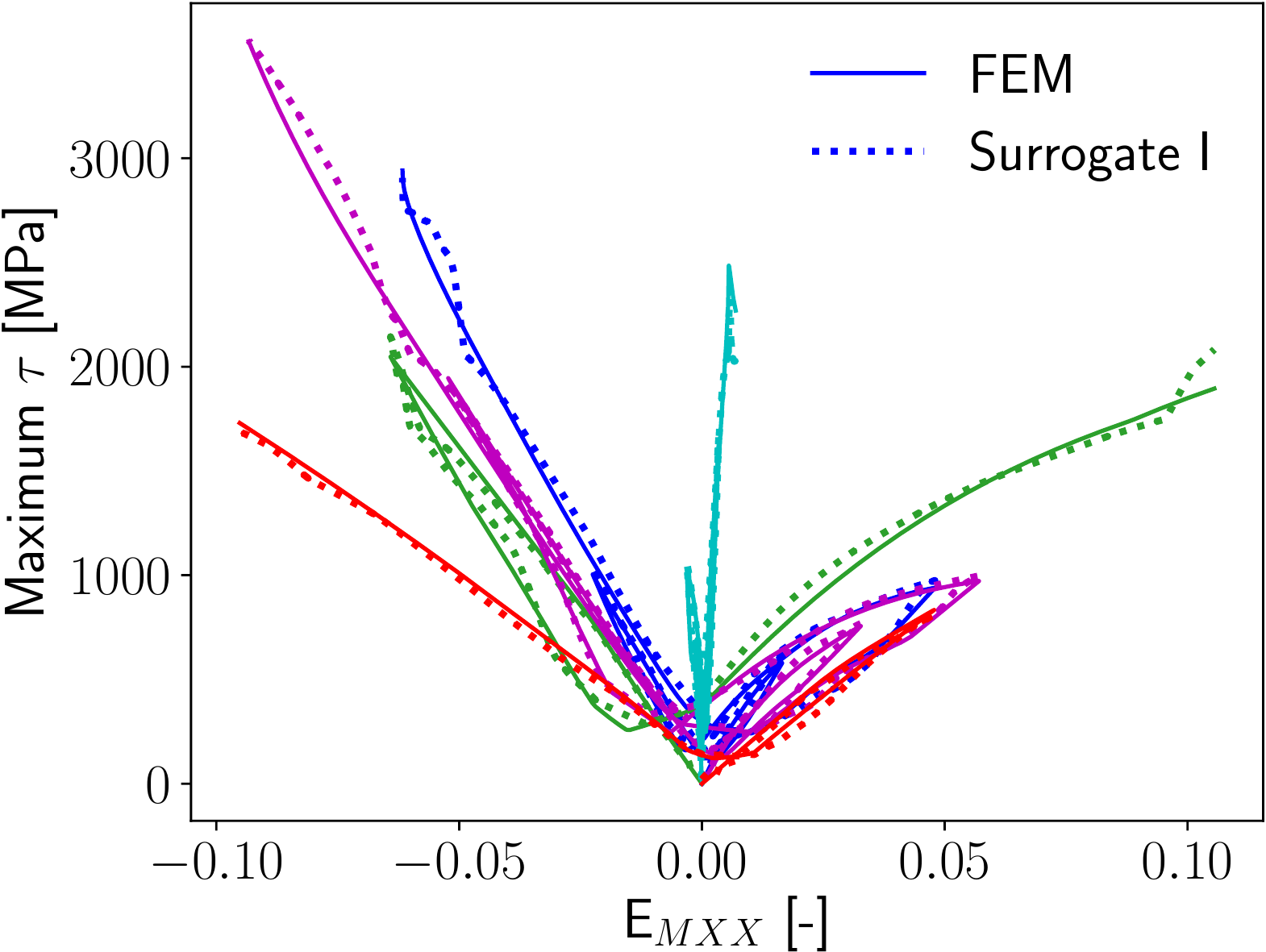}}\label{fig:MaxSVM_CP_I}\\
	\subfigure[]{\includegraphics[scale=0.38]{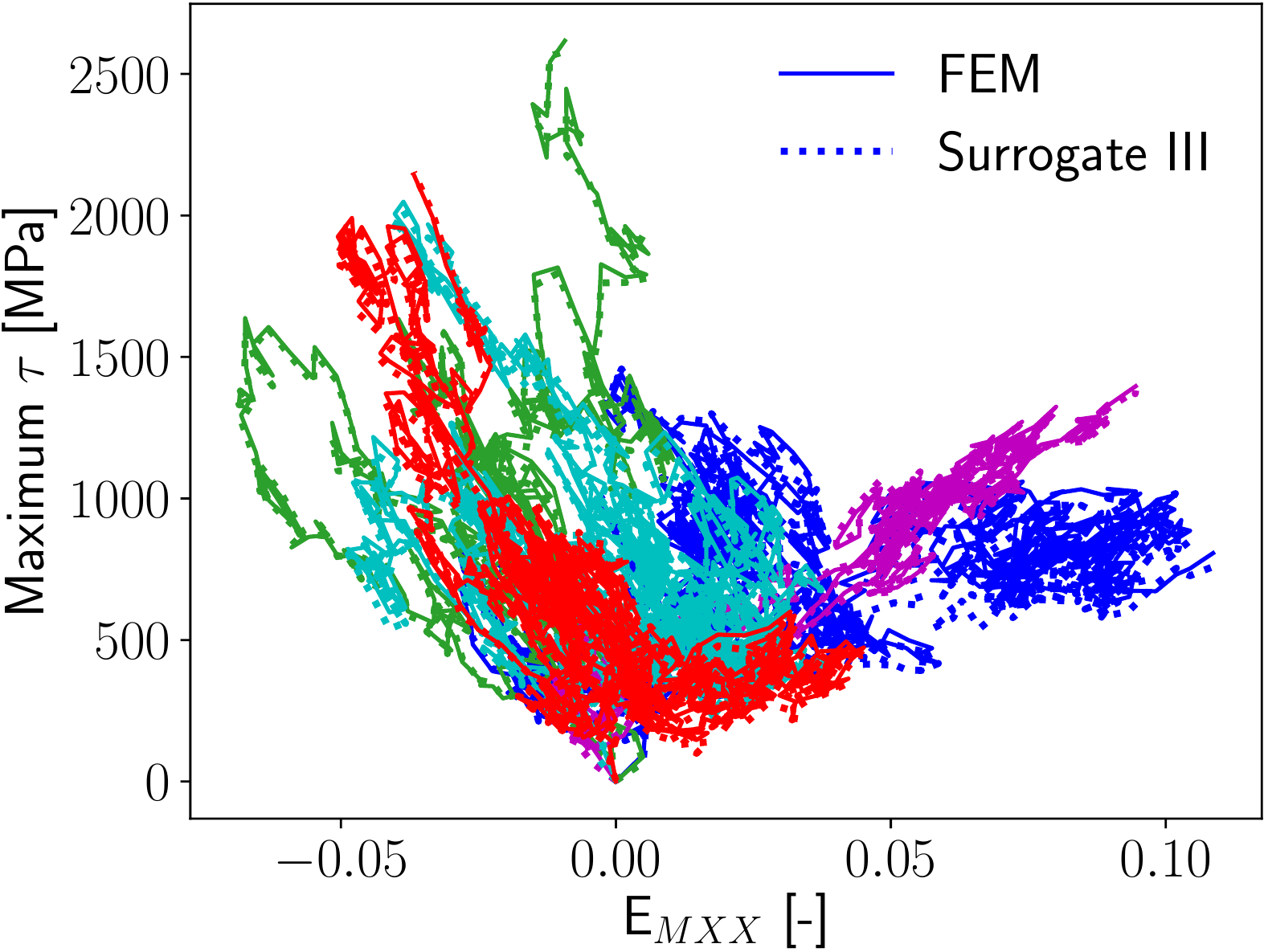}}\label{fig:MaxSVM_RP_III}\,
	\subfigure[]{\includegraphics[scale=0.38]{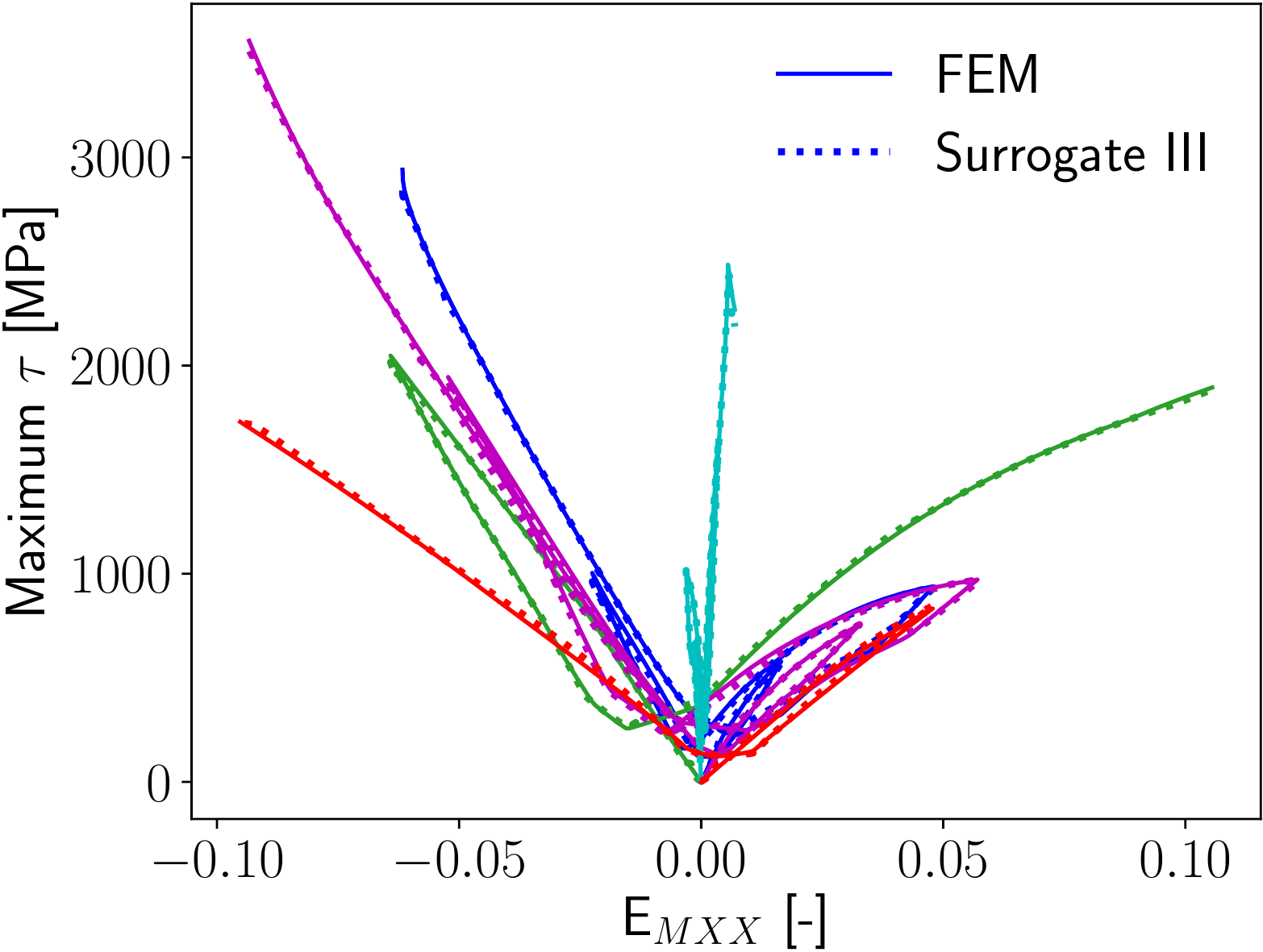}}\label{fig:MaxSVM_CP_III}
	\caption{Evolution of the maximum equivalent von Mises stress in the RVE at each loading step: (a) and (c) For random loading paths; and (b) and (d) For cyclic loading paths.}\label{fig:Max_SVM}
\end{figure}

\secondreviewer{
\begin{figure}[!htb]
	\centering
	\subfigure[]{\includegraphics[scale=0.23]{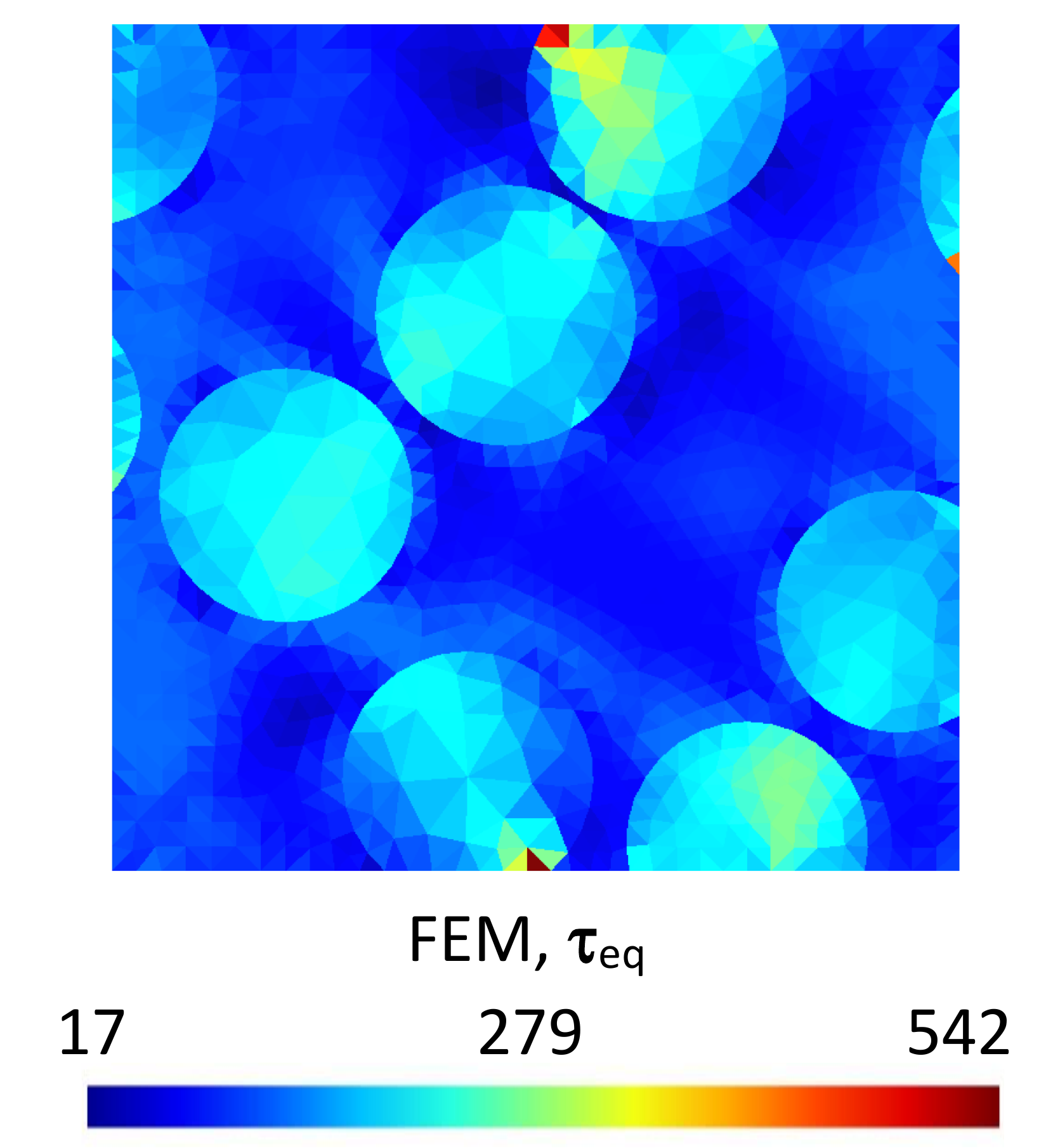}\label{fig:FE_SVM1}}\\
	\subfigure[]{\includegraphics[scale=0.23]{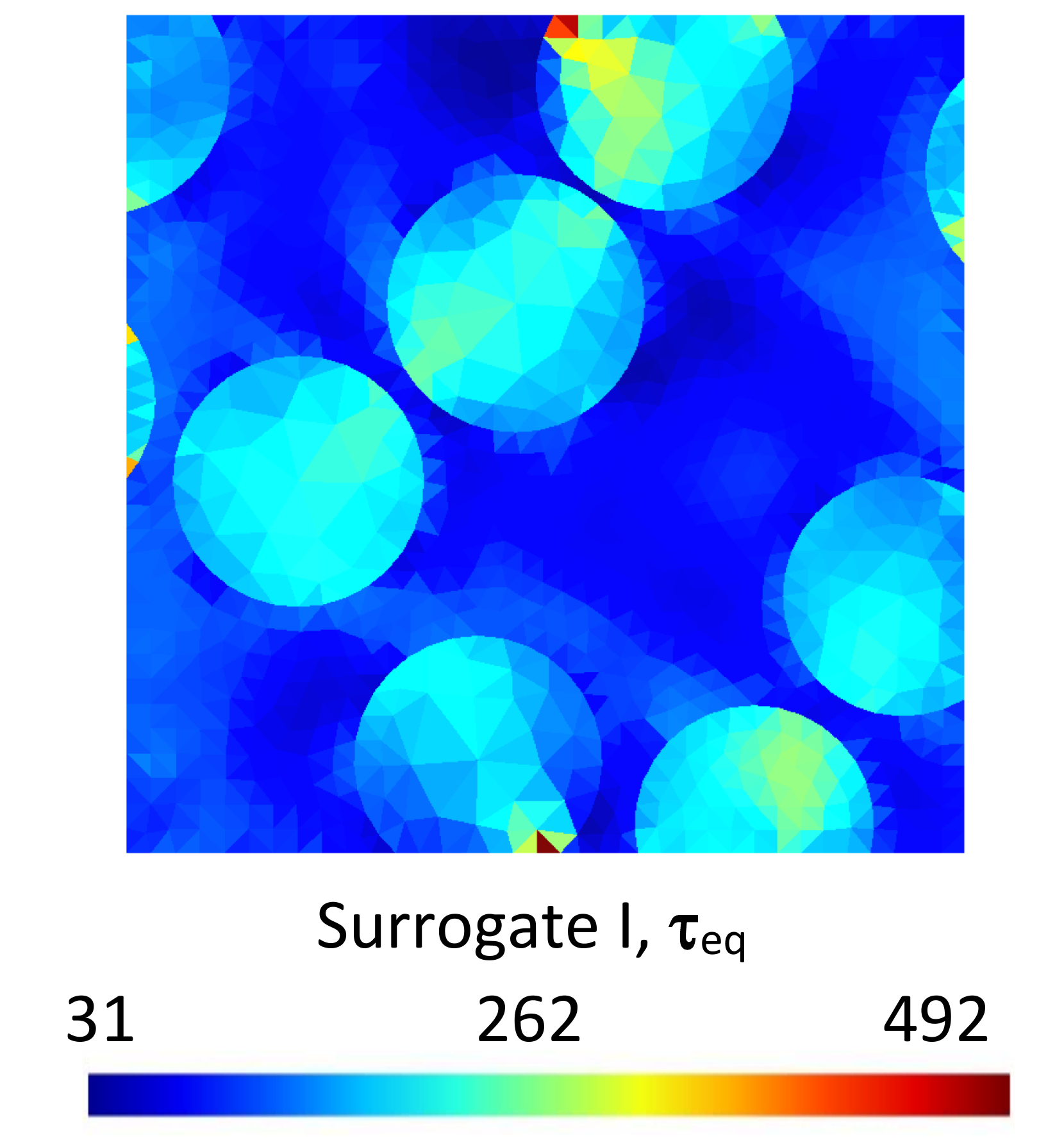}\label{fig:S1_SVM1}}\,
	\subfigure[]{\includegraphics[scale=0.23]{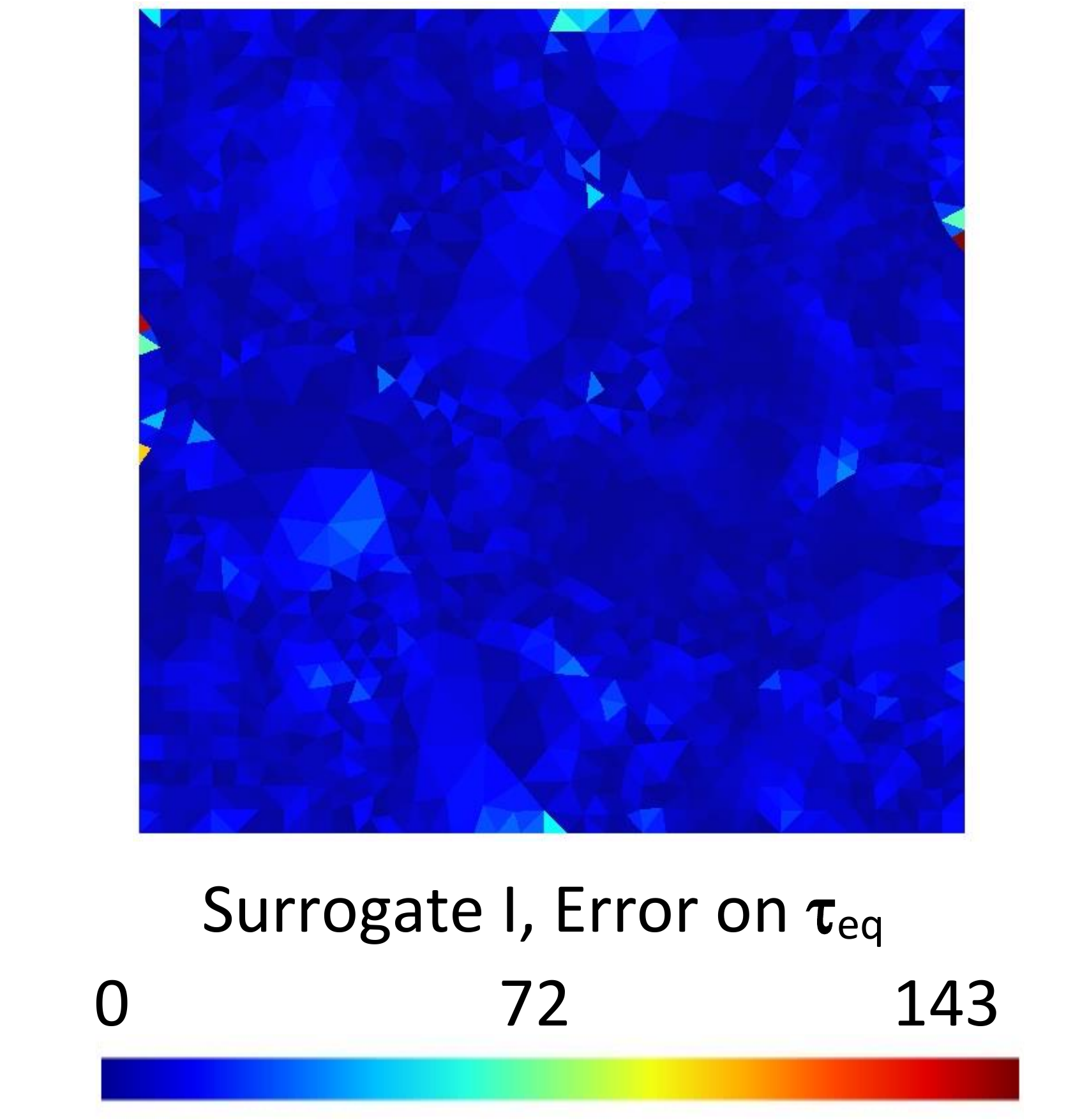}\label{fig:S1_SVM1_error}}\,
	\subfigure[]{\includegraphics[scale=0.23]{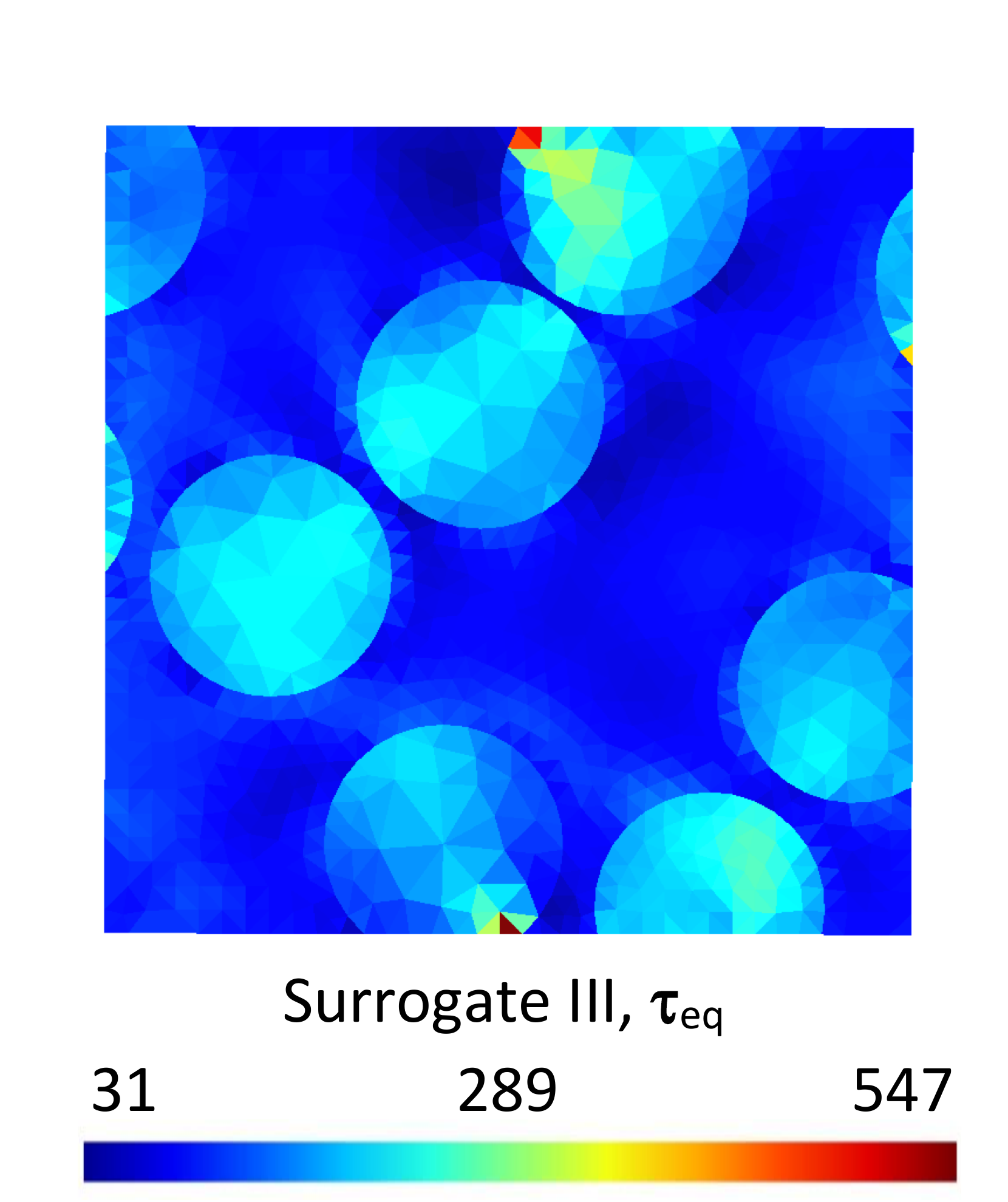}\label{fig:S3_SVM1}}\,
        \subfigure[]{\includegraphics[scale=0.23]{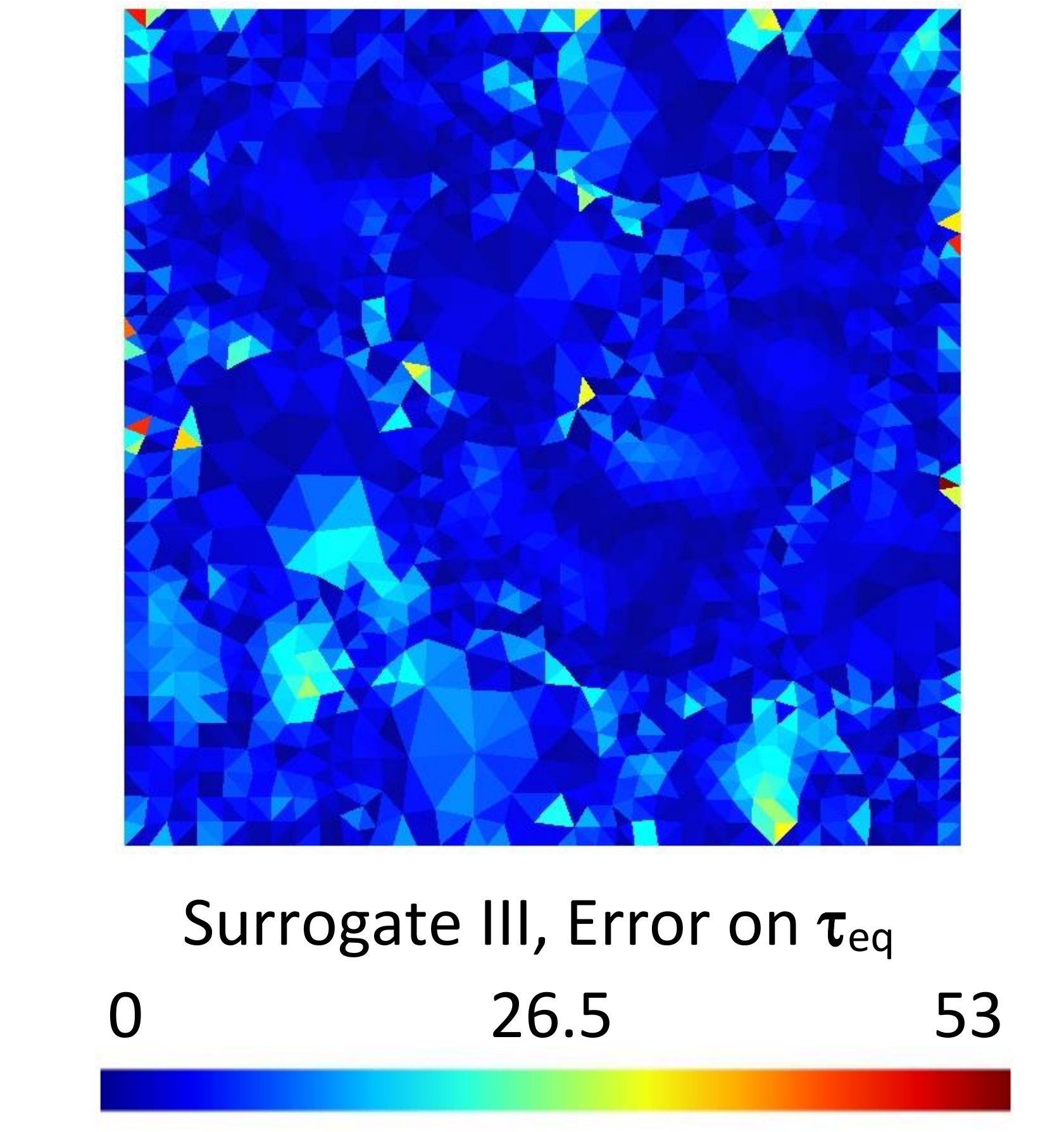}\label{fig:S3_SVM1_error}}\,		
	\caption{Distributions of the equivalent von Mises stress in the RVE at loading point ``1'', see Fig. \ref{fig:LStep}: (a) From finite element simulations; and (b, c) and (d, e) Respectively using Surrogates I and III. Errors are in terms of absolute values.}\label{fig:SVM_FieldP1}
\end{figure}
\begin{figure}[!htb]
	\centering
	\subfigure[]{\includegraphics[scale=0.23]{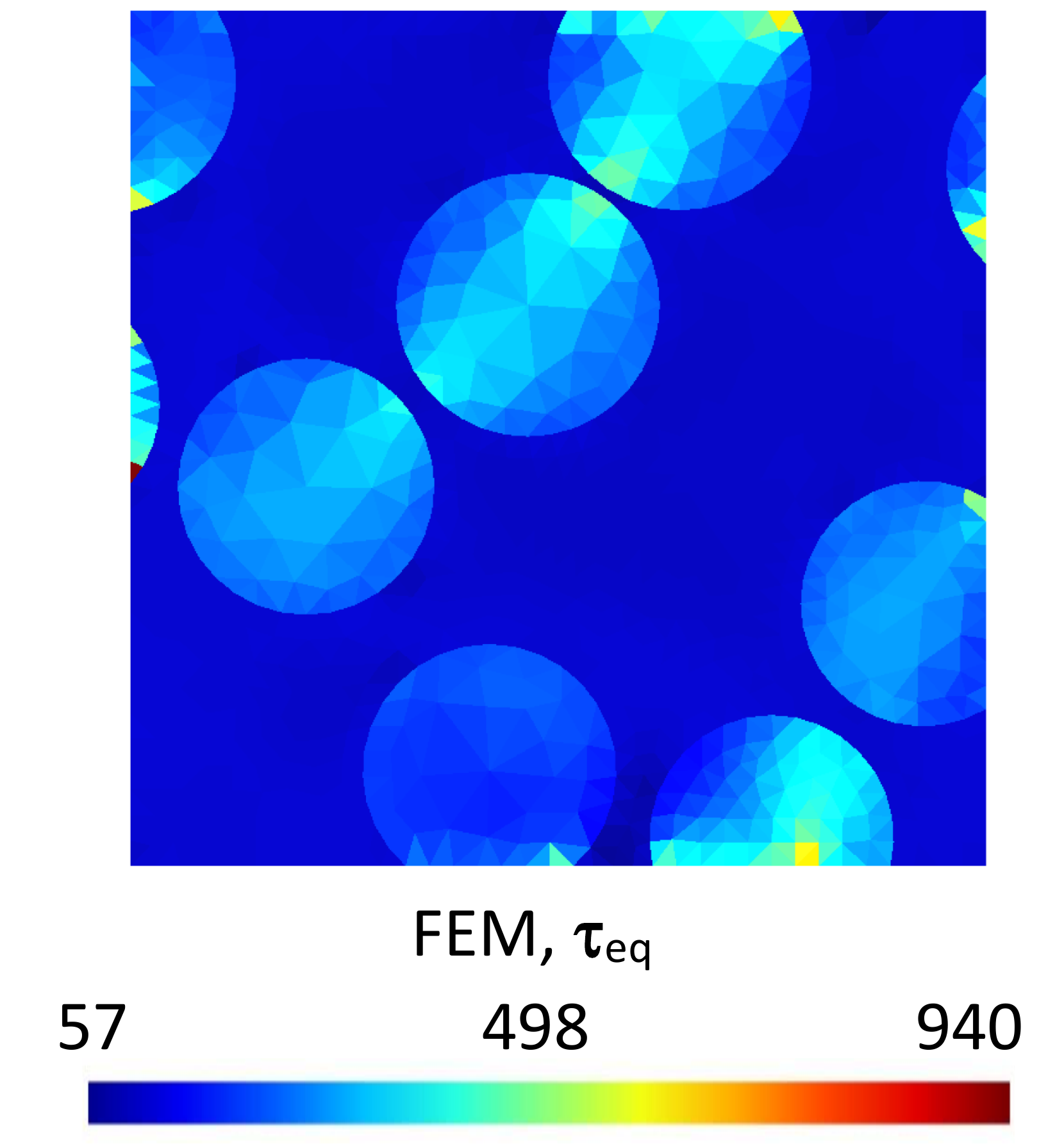}\label{fig:FE_SVM2}}\\
	\subfigure[]{\includegraphics[scale=0.23]{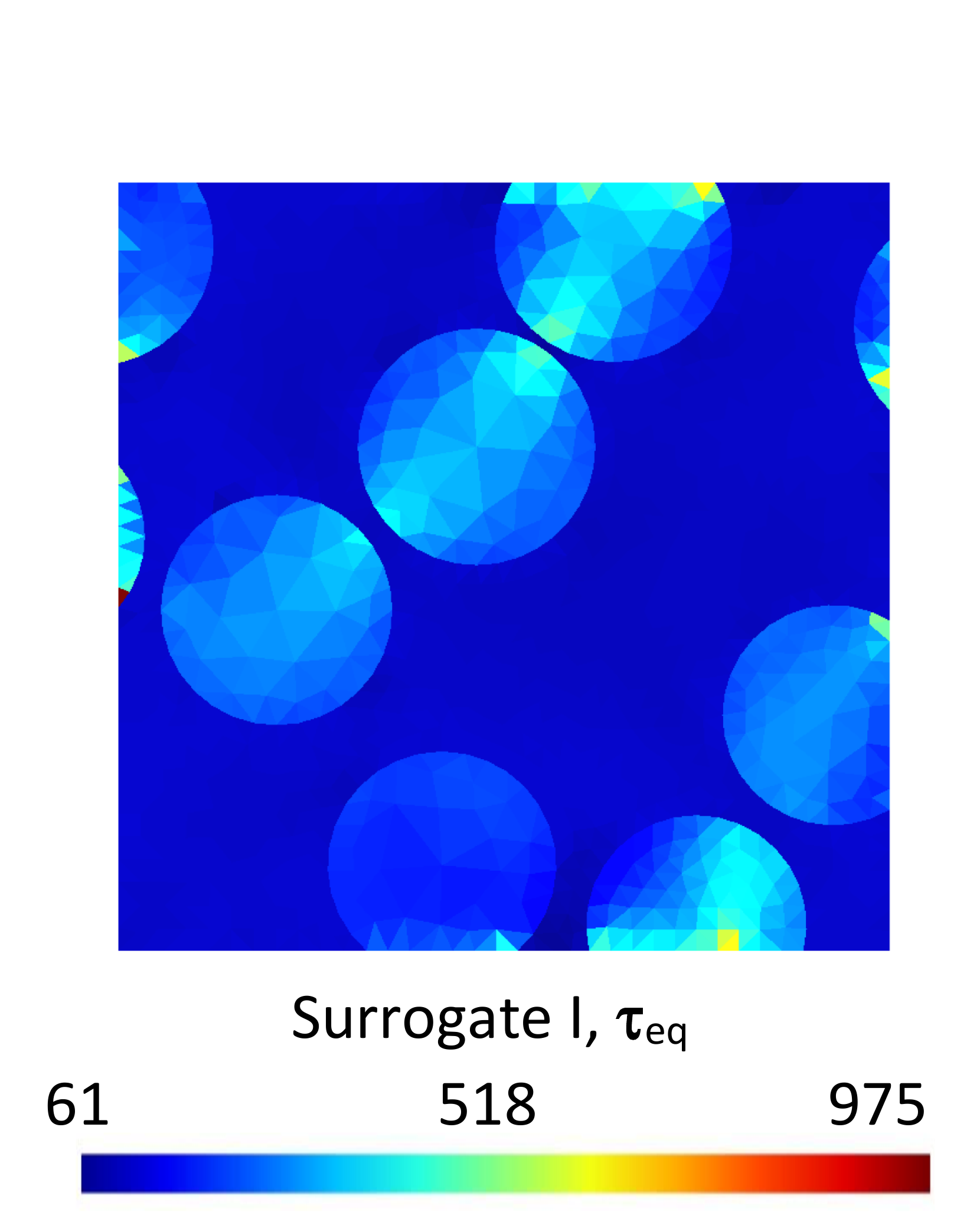}\label{fig:S1_SVM2}}\,
	\subfigure[]{\includegraphics[scale=0.23]{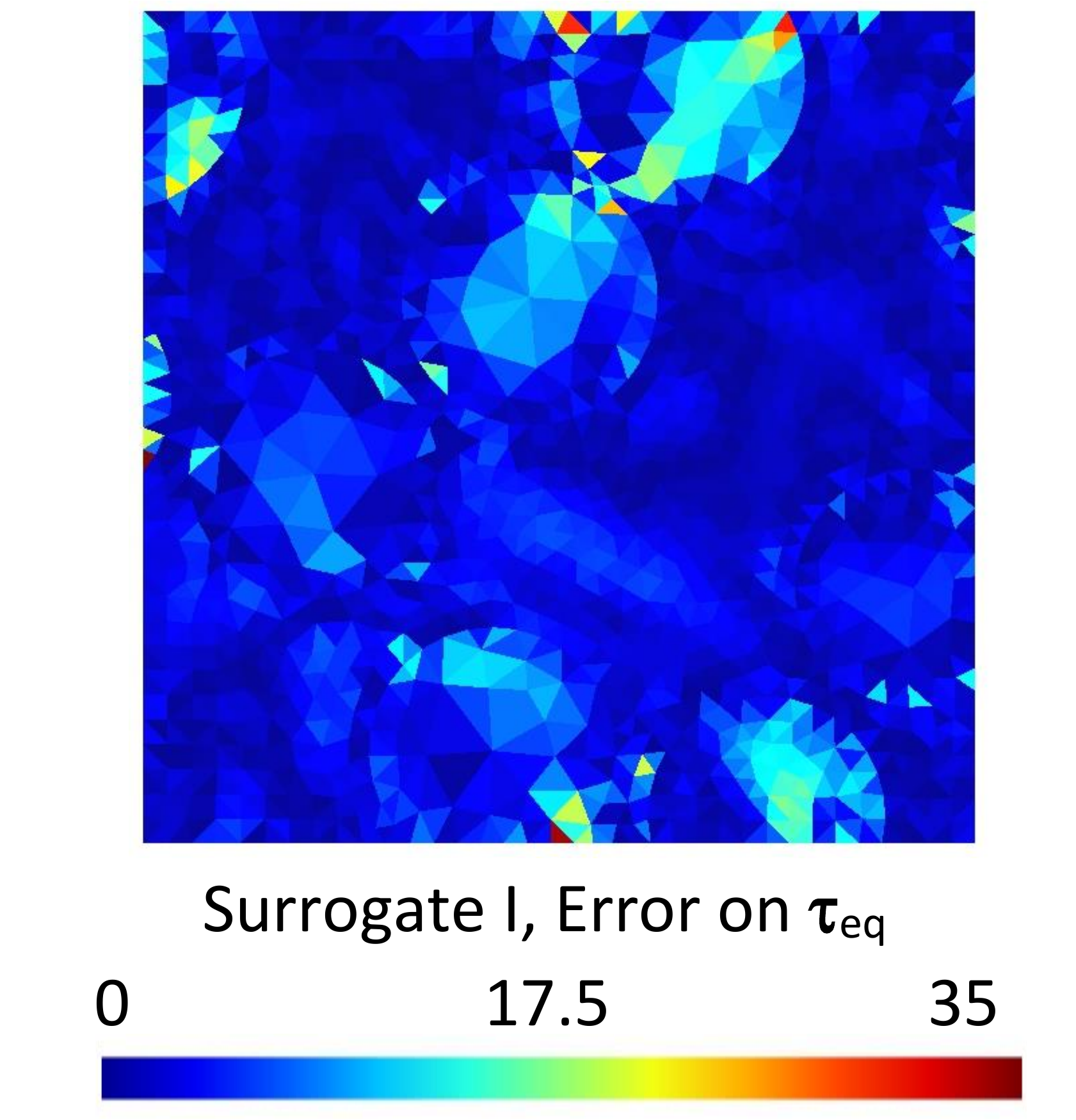}\label{fig:S1_SVM2_error}}\,
	\subfigure[]{\includegraphics[scale=0.23]{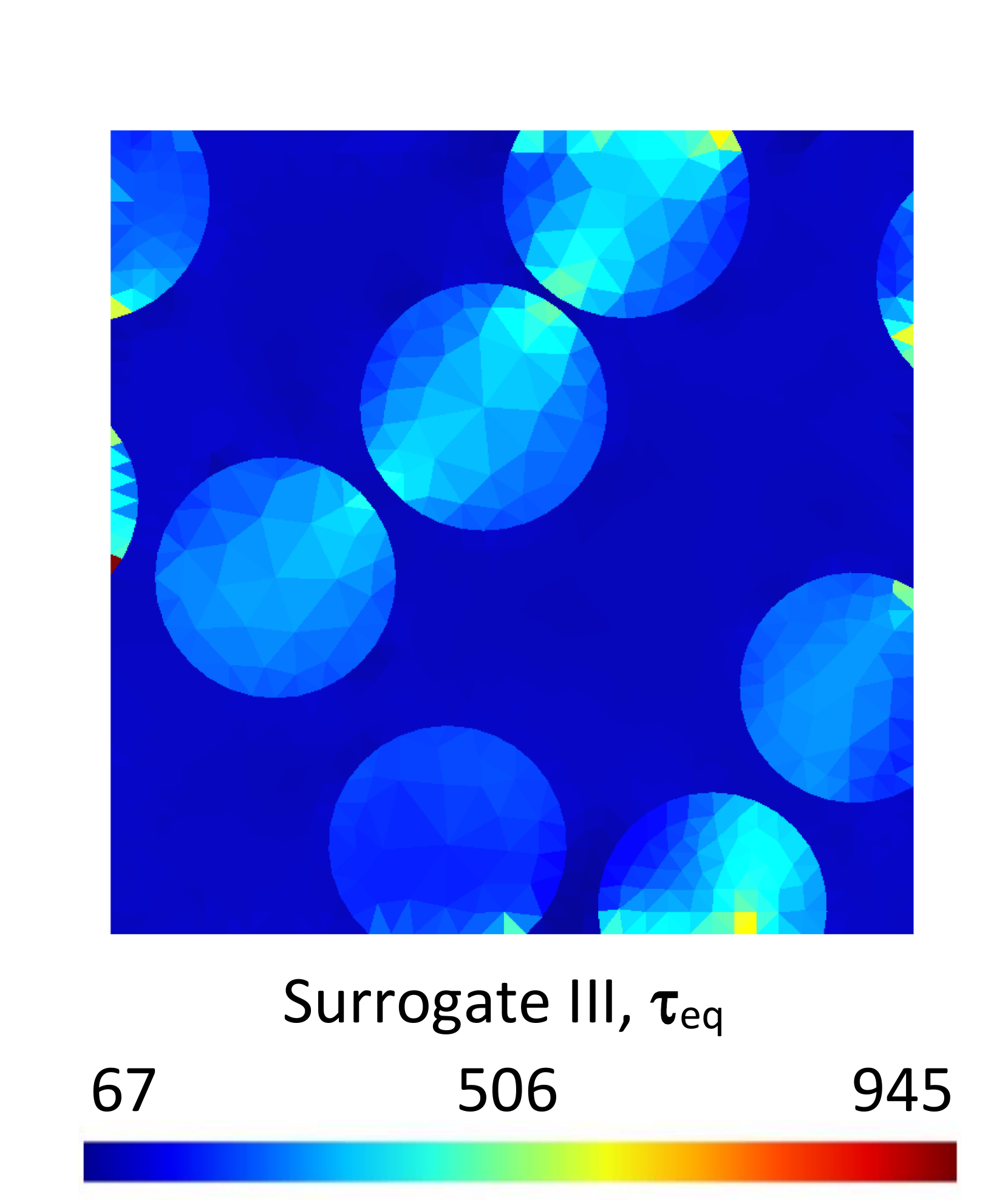}\label{fig:S3_SVM2}}\,
        \subfigure[]{\includegraphics[scale=0.23]{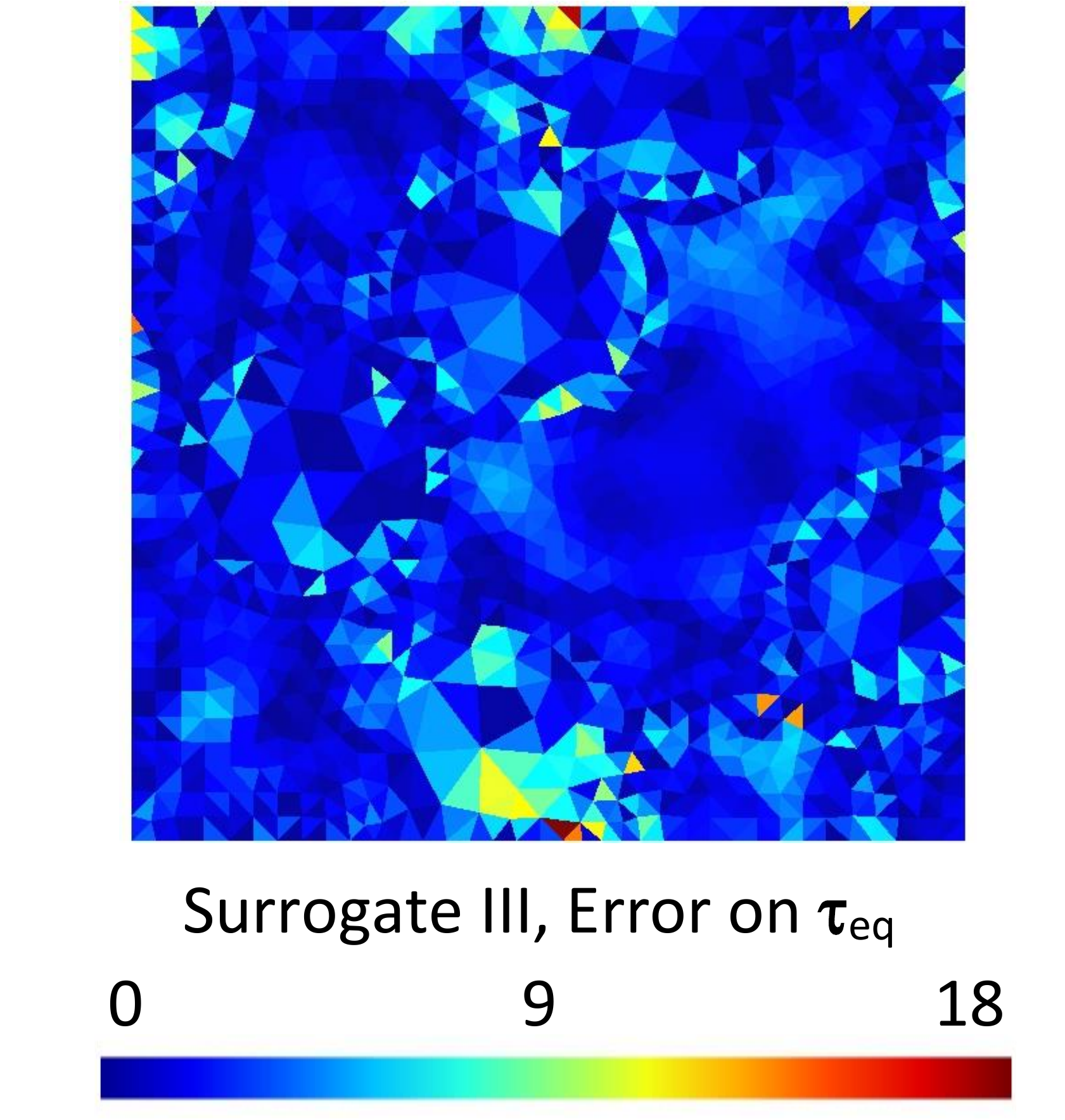}\label{fig:S3_SVM2_error}}\,		
	\caption{Distributions of the equivalent von Mises stress in the RVE at loading point ``2'', see Fig. \ref{fig:LStep}: (a) From finite element simulations; and (b, c) and (d, e) Respectively using Surrogates I and III. Errors are in terms of absolute values.}\label{fig:SVM_FieldP2}
\end{figure}
\begin{figure}[!htb]
	\centering
	\subfigure[]{\includegraphics[scale=0.23]{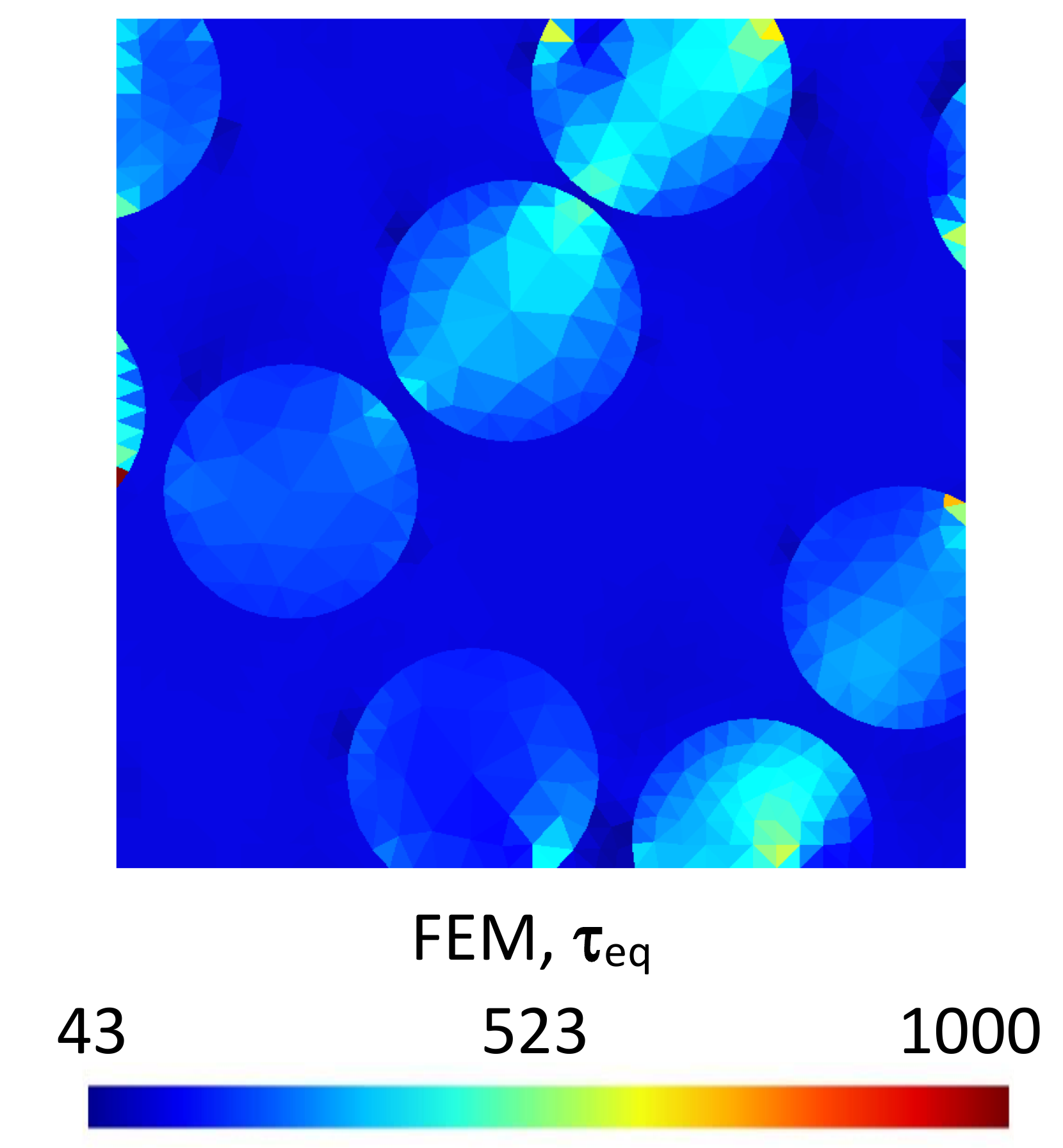}\label{fig:FE_SVM3}}\\
	\subfigure[]{\includegraphics[scale=0.23]{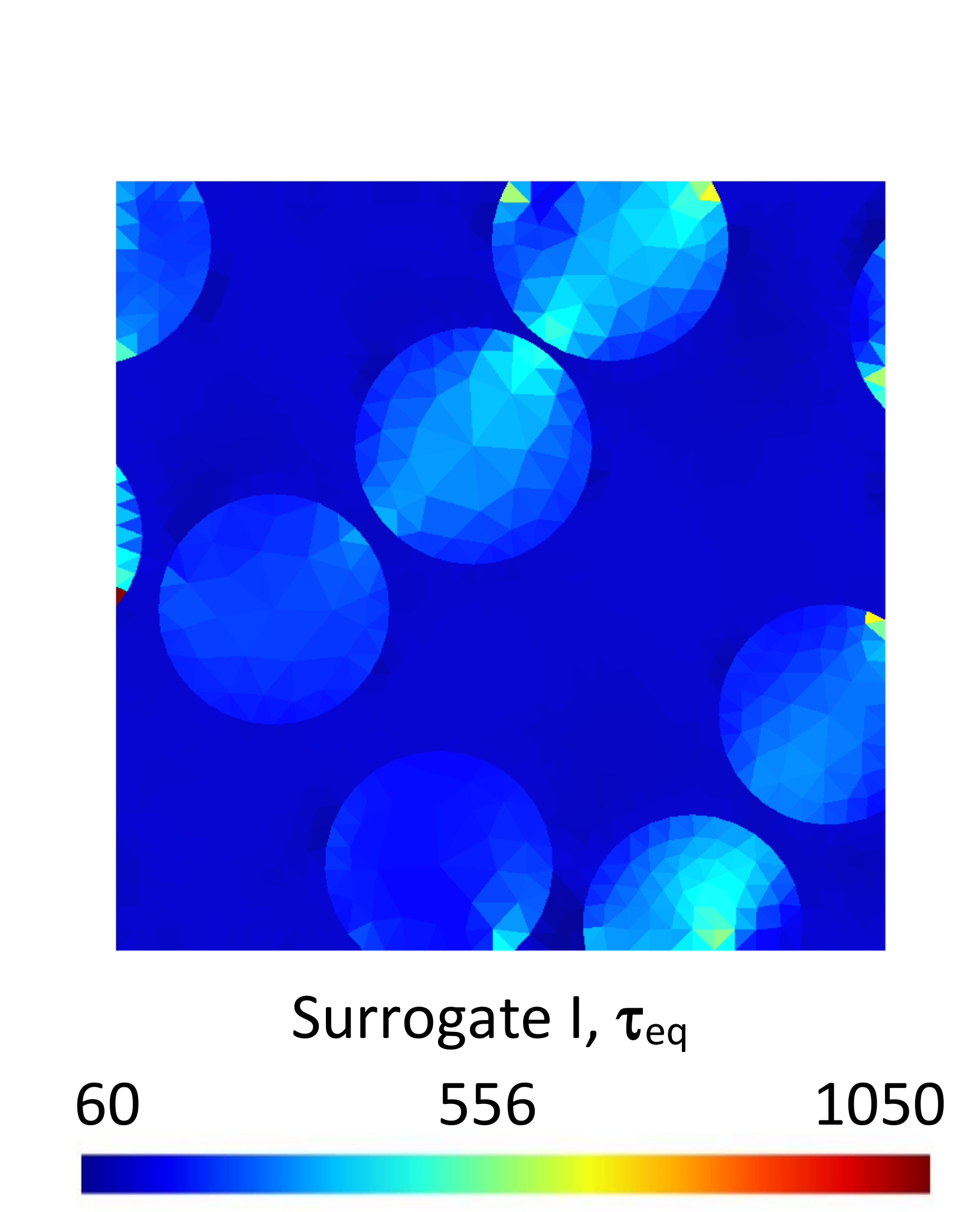}\label{fig:S1_SVM3}}\,
	\subfigure[]{\includegraphics[scale=0.23]{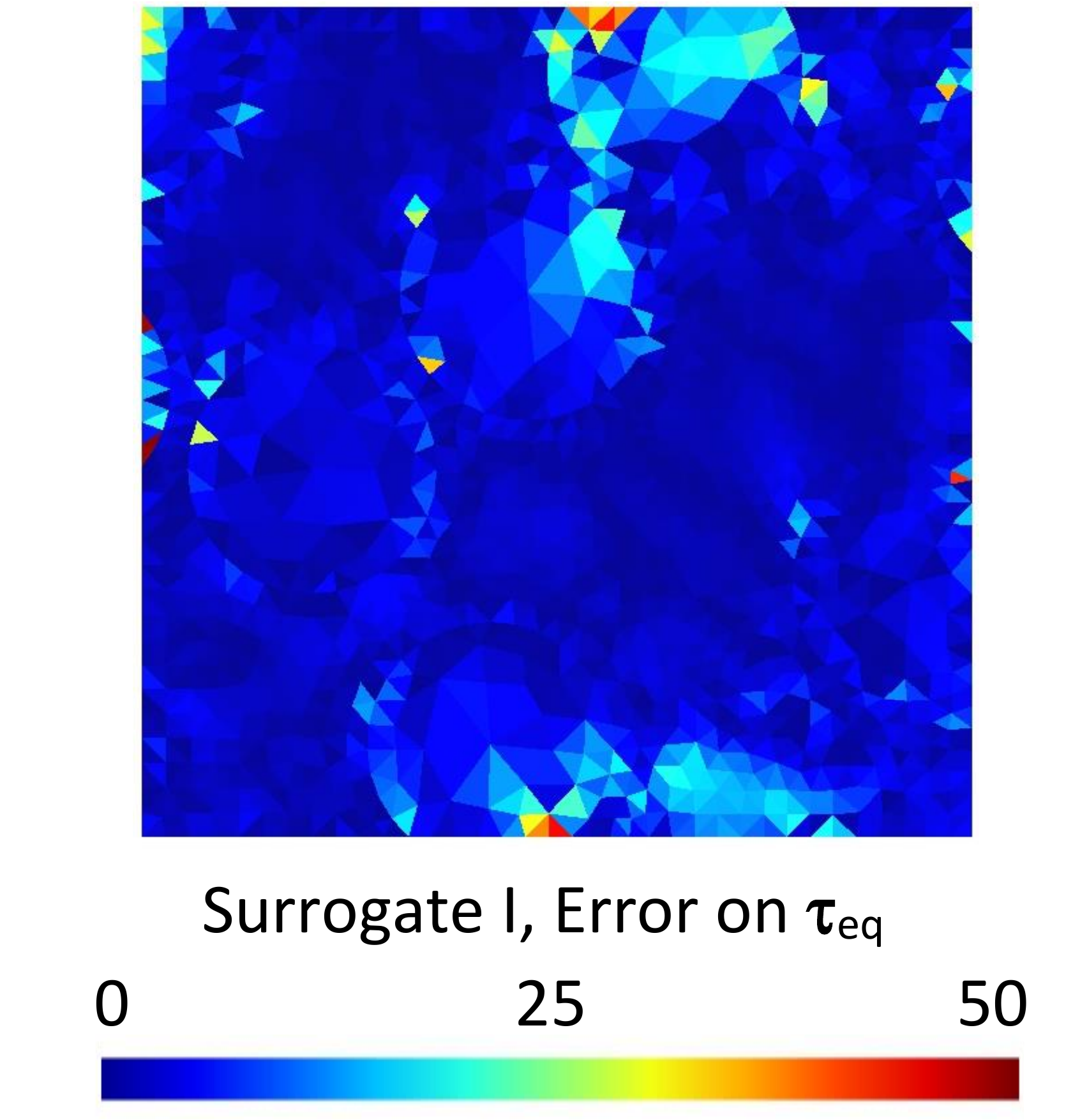}\label{fig:S1_SVM3_error}}\,
	\subfigure[]{\includegraphics[scale=0.23]{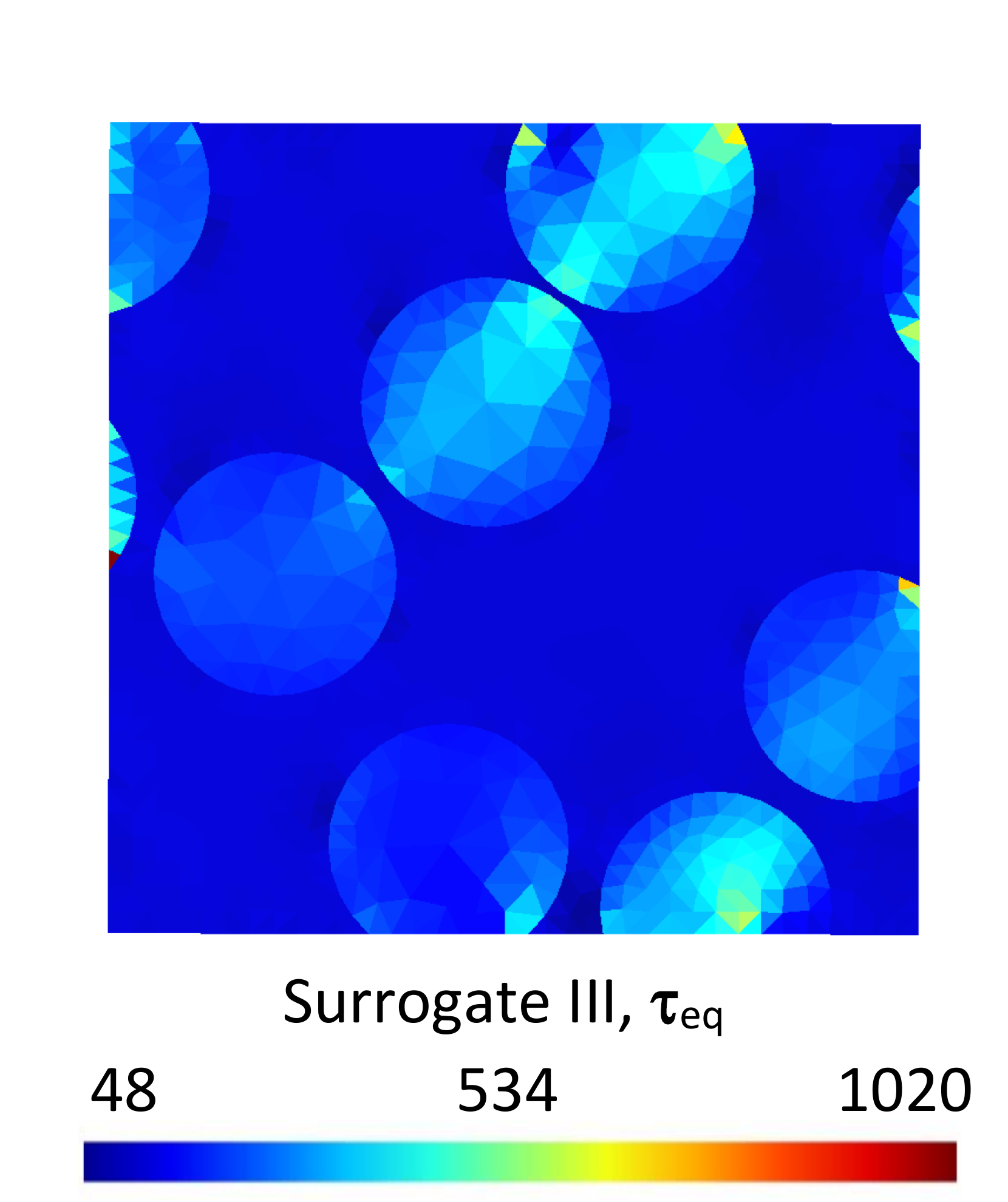}\label{fig:S3_SVM3}}\,
        \subfigure[]{\includegraphics[scale=0.23]{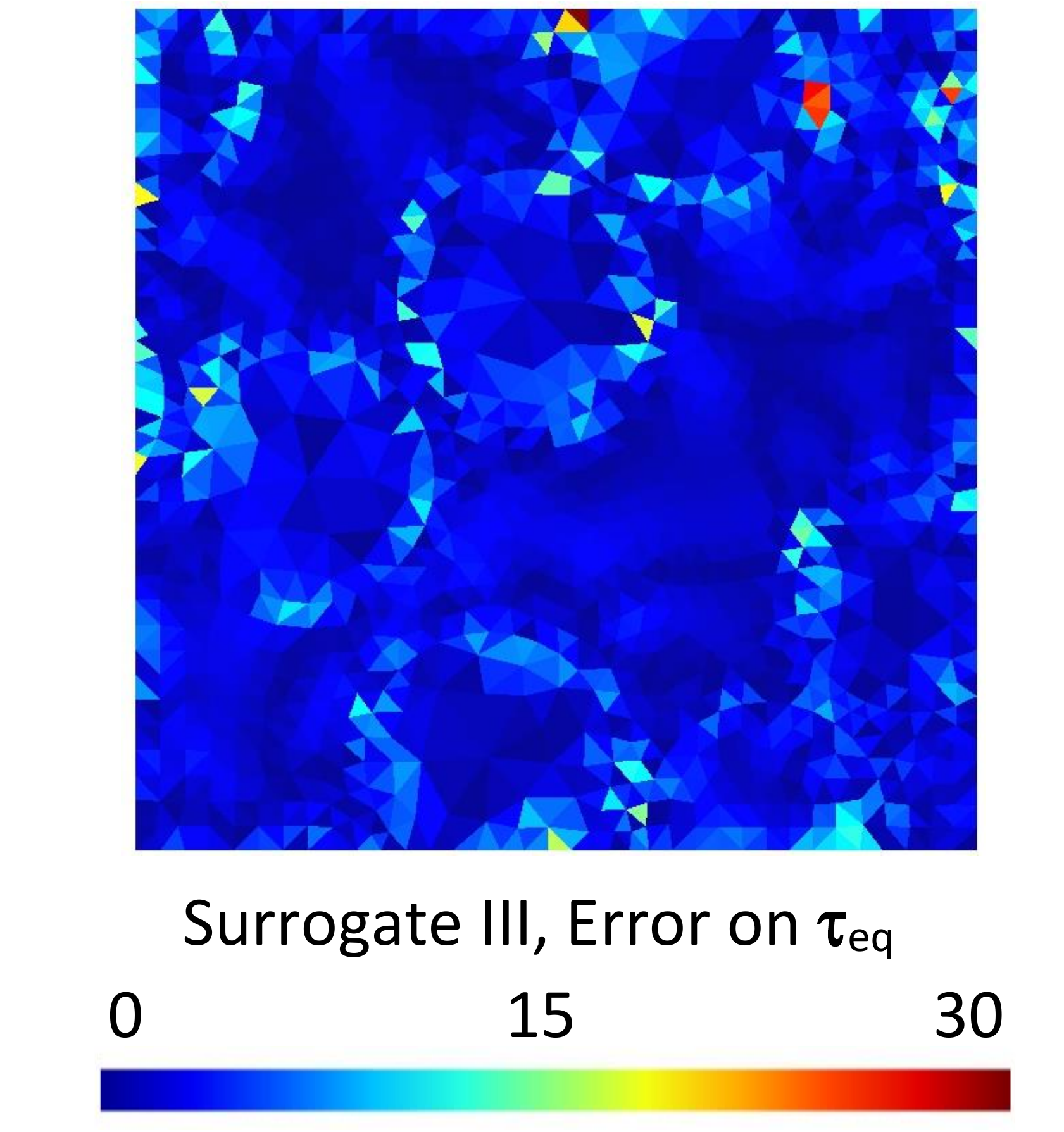}\label{fig:S3_SVM3_error}}\,		
	\caption{Distributions of the equivalent von Mises stress in the RVE at loading point ``3'', see Fig. \ref{fig:LStep}: (a) From finite element simulations; and (b, c) and (d, e) Respectively using Surrogates I and III. Errors are in terms of absolute values.}\label{fig:SVM_FieldP3}
\end{figure}
}

Using the loading paths depicted in Figs. \ref{fig:LPath_R} and \ref{fig:LPath_C}, the evolution of the maximum equivalent von Mises stress in the RVE at each loading step predicted by the surrogate models is compared to the direct finite element simulation results in Fig. \ref{fig:Max_SVM}.
Although Surrogate III required a training of 48 hours while Surrogate I required a training longer than 96 hours, Surrogate III provides better predictions, as already observed for the equivalent plastic strain evolution.
The distributions of the equivalent von Mises stress $\bm{\tau}_\text{eq}$ within the RVE reconstructed by the Surrogates I and III are compared with the direct finite element results in Figs. \ref{fig:SVM_FieldP1}-\ref{fig:SVM_FieldP3}, for the loading steps/configurations marked in Fig. \ref{fig:LStep}.
It can be seen that the different stress distribution patterns can be predicted with Surrogate III, the latter giving more accurate predictions than Surrogate I. \secondreviewer{The larger errors can be found in locations of stress concentration arising within one element, see \emph{e.g.} Fig. \ref{fig:S3_SVM2_error}.}

\section{Conclusions}\label{sec:conclusions}

In order to accelerate the time consuming numerical analysis process of finite element simulations, we use Artificial Neural Networks to substitute to high dimensional and non-linear history-dependent problems in computational mechanics.
Recurrent Neural Networks, in particular Gated Recurrent Units, are adopted to construct surrogate models in this context.
In a previous work, homogenized strain and stress were collected from direct finite element resolutions on RVEs in order to conduct the surrogate training and to construct a surrogate model of a meso-scale BVP in multi-scale simulations.
Since only the macro-stress-macro-strain response of the RVE were considered, the micro-structure information could not be recovered in a so-called localization step.
In this work, we develop RNNs as surrogates of the RVE response while being able to recover the evolution of the local micro-structure state variables. 
Considering the high dimensionality of the micro-structure state variables and the limitation of computing resources, three surrogates models are proposed, i) direct RNN modeling with implicit NNW \secondreviewer{dimensionality reduction}, ii) RNN with PCA \secondreviewer{dimensionality reduction}, and iii) RNN with PCA \secondreviewer{dimensionality reduction} and \secondreviewer{dimensionality break down}.
The three surrogate models are compared through a case study on the equivalent plastic strain and von Mises stress distributions in a micro-structure for different loading scenarios.

On the one hand, it appears that the direct use of RNN with PCA \secondreviewer{dimensionality reduction} is not a wise choice for complex non-linear history-dependent problems, since more hidden variables will be required to reach the same accuracy.
Nevertheless, with the aim of using GPU instead of CPU in order to speed-up the training process, since the use of GPU is limited by its memory, using PCA \secondreviewer{dimensionality reduction} is appealing when followed by a \secondreviewer{dimensionality break down} so that a RNN of large size can be substituted by a few smaller RNNs.
We then develop an \emph{ad hoc} training process for GPU so that the training time of this surrogate model is reduced compared to the surrogate models embedding a larger RNN, while the accuracy remains comparable. 

On the other hand, the architecture of the RNN involves several hyper-parameters to be defined, which can be time consuming for the user if a \secondreviewer{trial and error approach} is conducted on the full surrogate.
Through RNN modeling of the principal components coefficients, the connection between physical state variables and hidden variables of GRU is revealed, and exploited to accelerate the design of the RNN-based surrogate models. 
Indeed, for a targeted error, PCA \secondreviewer{dimensionality reduction} and break down allow performing the \secondreviewer{trial and error approach} of the RNN design on a RNN with a single output: one principal component coefficient.
This accelerated \secondreviewer{trial and error approach} is then applied to design a surrogate model of the evolution of the local micro-structure state variables and is shown to yield accurate predictions.

\secondreviewer{In this work, we have limited ourselves to 2D problems, with in-plane loading. Although the methodology is general and can be used for 3D problems, because of the increase in the input dimensionality, \emph{i.e.} strain tensor size from 3 components to 6 components, it is expected that a larger number of generated loading paths would be necessary to train the surrogates, and that the neural network might require more layers or nodes in order to remain accurate. The increase in time required to generate the loading paths, on the one hand because of the use of 3D finite element meshes and on the other hand because of the requirement to cover a larger strain space, can be handled by using a computer cluster since this step is fully scalable.
  Another difficulty lies in the increase in the dimensionality of the output variables, \emph{i.e.} the history of the distributions of the state variables, for which the dimensionality reduction methods studied in the paper will become even more meaningful.}

\section*{Acknowledgment}

This project has received funding from the European Union’s Horizon 2020 research and innovation programme under grant agreement No 862015 for the project ``Multi-scale Optimisation for Additive Manufacturing of fatigue resistant shock-absorbing MetaMaterials (MOAMMM) of the H2020-EU.1.2.1. - FET Open Programme.

\section*{Data availability}

The raw/processed data required to reproduce these findings are available on\\ https://gitlab.uliege.be/moammm/moammmpublic/tree/master/publicationsData/2022\_CMAME\_RNNPCA under the Creative Commons Attribution 4.0 International (CC BY 4.0) licence and on \cite{wu_ling_dataRNNPCA}.

\appendix

\secondreviewer{
  \section{Notations}\label{app:notations}
  \subsection{Tensors}
\begin{itemize}
\item  We use italic fonts $a$ to denote 0-order tensors or scalars, bold italic fonts $\bm{a}$ to denote first-order tensors, roman bold fonts $\mathbf{A}$ to denote second-order tensors, and blackboard fonts $\mathbb{C}$ to denote fourth-order tensors.
\item The inner product $\cdot$ and double inner product $:$ between two tensors of order $m$ and $n$ yield tensors of order, respectively, $m+n-2$ (if $m+n\geq 2$) and $m+n-4$ (if $m+n \geq 4$). The dyadic product $\otimes$ between two tensors of order $m$ and $n$ yields a tensors or order $m+n$.
\item  The superscript ``$\text{dev}$'' refers to the deviatoric part of a second order tensor: $\bullet^\text{dev}=\bullet-\frac{1}{3}\text{tr}(\bullet)$, where the notation ``$\text{tr}$'' refers to the trace $\text{tr}(\bullet)=\mathbf{I}:\bullet$ with $\mathbf{I}$ the second-order identity tensor.
  \item The subscripts ``$\text{M}$'' and ``$\text{m}$'' refer respectively to the macro- and micro-scales. 
\end{itemize}

  \subsection{Vectors and matrices}
\begin{itemize}
\item  A vector in $\mathbb{R}^n$ is represented by bold italic fonts $\bm{a}$ and constructed following the notation $\bm{a}=[a_1,\,\ldots,\,a_n]$. 
\item  A matrix in $\mathbb{R}^{m\times n}$ is represented by a bold roman fonts $\mathbf{A}$ and constructed following the notation $\mathbf{A}=\left[{\bm{a}}_1\; {\bm{a}}_2 \;\ldots\; {\bm{a}}_n\right]_{m\times n}$ with $A_{ij}$ the $i^\text{th}$ component of $\bm{a}_j$.
\end{itemize}

  \subsection{Operations on tensors, vectors and matrices}
\begin{itemize}
\item   
  A set or sequence of tensors, vectors, or matrices $\bullet$ is represented by $\{ \bullet \,:\; \text{condition}\}$ or $\{ \bullet_1,\,\bullet_2,\,\ldots\}$.
\item Normalized forms of tensors, vectors, or matrices $\bullet$ are represented by $\underline{\bullet}$.
\item Approximations by surrogate or dimensionality reduction of tensors, vectors, or matrices $\bullet$ are represented by $\hat{\bullet}$.
    \end{itemize}
}

\secondreviewer{
\section{Detail on the GRU}\label{app:gru}

\begin{figure}[!h]
        \centering
        \includegraphics[scale=0.5]{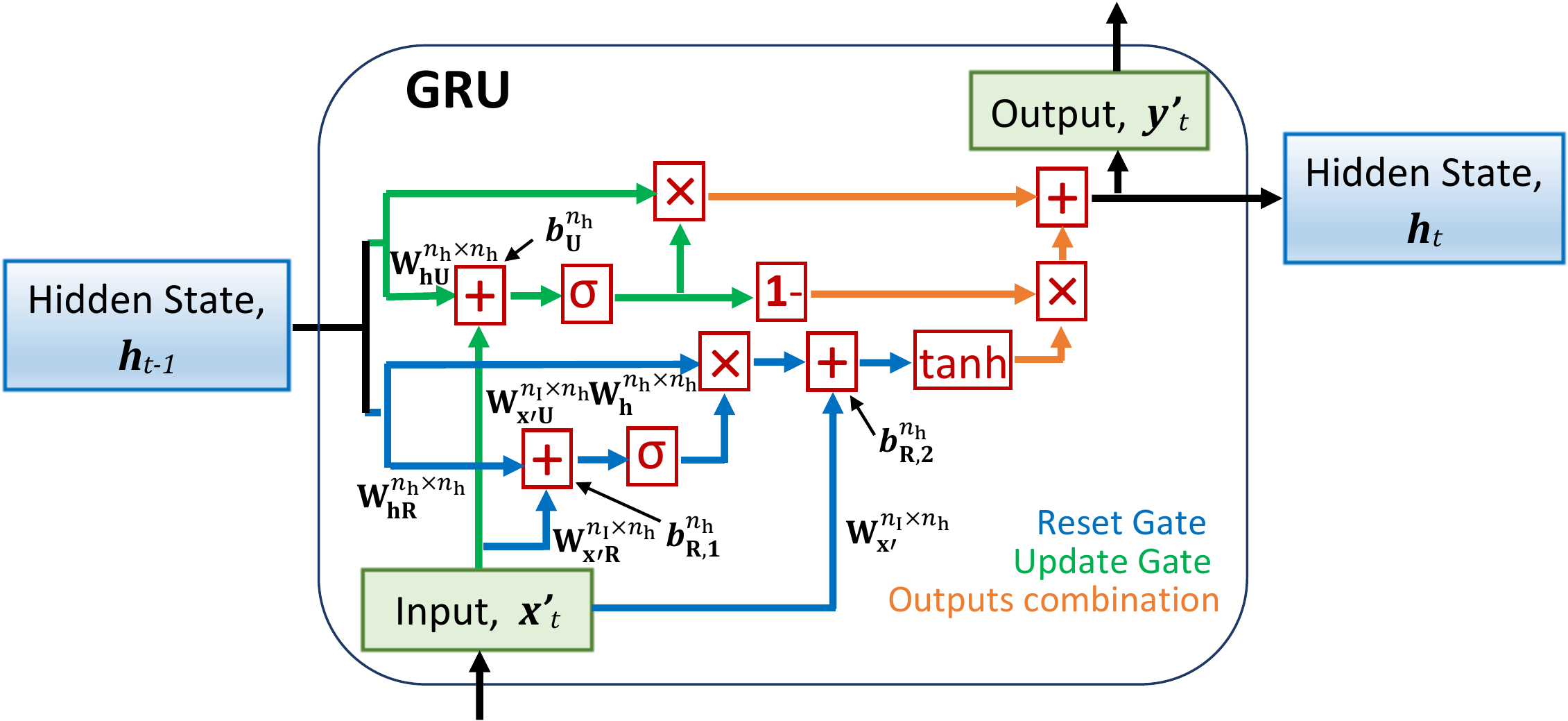}\\
        \caption{Detailed GRU architecture with as trainable model parameters the weights matrices $\mathbf{W}^{n_{\text{I}}\times n_{\text{h}}}_\mathbf{x'U}$, $\mathbf{W}^{n_{\text{I}}\times n_{\text{h}}}_\mathbf{x'R}$, $\mathbf{W}^{n_{\text{I}}\times n_{\text{h}}}_\mathbf{x'}$, $\mathbf{W}^{n_{\text{h}} \times n_{\text{h}}}_\mathbf{hU}$, $\mathbf{W}^{n_{\text{h}} \times n_{\text{h}}}_\mathbf{hR}$, $\mathbf{W}^{n_{\text{h}}\times n_{\text{h}}}_\mathbf{h}$ and bias vectors $\bm{b}^{n_{\text{h}}}_\mathbf{U}$, $\bm{b}^{n_{\text{h}}}_\mathbf{R,1}$ and $\bm{b}^{n_{\text{h}}}_\mathbf{R,2}$. The superscripts $n_{\text{I}}$ and $n_{\text{h}}$ refer respectively to the sizes of the input vector $\bm{x}'$ and of the hidden variables vector $\bm{h}$, the output vector $\bm{y}'$ has the same dimensionality $n_{\text{h}}$. The subscripts $\mathbf{x'}$, $\mathbf{h}$, $\textsf{R}$, $\textsf{U}$ refer respectively to the input variables, hidden variables, and constructed Reset vector and Update vector.} \label{fig:GRU_detail}
\end{figure}

  The structure of the GRU presented in Fig. \ref{fig:GRU} is detailed in Fig. \ref{fig:GRU_detail}, in which the operation symbols correspond to
  \begin{itemize}
        \item $\boxed{+}$: the element-wise sum operator on two vectors, $\bm{x}$ and $\bm{y}$, of same size, \emph{i.e.} $\bm{r}=\bm{x} + \bm{y}$;
        \item  $\boxed{\bm{1}-}$: The element-wise operator on vector $\bm{x}$, \emph{i.e.} $\bm{r}=\bm{1}-\bm{x}$;
        \item $\boxed{\times}$: The element-wise multiplication, or Hadamard product, on two vectors, $\bm{x}$ and $\bm{y}$, of same size, \emph{i.e.} $r_i=x_iy_i$.
        \item $\boxed{\sigma}$: The non-linear activation sigmoid function, \emph{i.e.}
    \begin{equation}
        \sigma(x)=\frac{1}{1+\exp(-x)}\,,
    \end{equation}
    which returns values in the range 0 to 1;
    \item  $\boxed{\text{tanh}}$:  The non-linear activation hyperbolic tangent function.
  \end{itemize}
  Three paths compose the GRU presented in Fig. \ref{fig:GRU_detail}: the reset gate (Blue path), the update gate (Green path) and the combination of the outputs (Orange path), which are respectively summarized by
  \begin{itemize}
  \item The reset gate has as inputs the previous hidden state $\bm{h}_{t-1}$ and the current input data $\bm{x}'_{t}$, which corresponds to the output of the first feed-forward $\text{NNW}_{\text{I}}$, see Fig. \ref{fig:GRU}. A Reset vector is obtained by applying a sigmoid function on the weighted sum vector of these two input vectors so that the values fall in the range 0 and 1. The Reset vector filters the less-important and more-important information for the subsequent steps. After having been multiplied by a trainable weight, the previous hidden state $\bm{h}_{t-1}$ undergoes an element-wise multiplication with the Reset vector in order to decide which information is to be kept from the previous configuration together with the new inputs.
Eventually, the non-linear activation $\boxed{\text{tanh}}$ function is applied on the weighted sum of the current input $\bm{x}'_{t}$ and of the last result, \emph{i.e.} the element-wise multiplication of the Reset vector and of the previous hidden state $\bm{h}_{t-1}$.
\item  The update gate has also as inputs the previous hidden state $\bm{h}_{t-1}$ and the current input data $\bm{x}'_{t}$. The Update vector is obtained by similar operations as the Reset vector, but using different weight matrices. The Update vector undergoes an element-wise multiplication with the unweighted previous hidden state $\bm{h}_{t-1}$ in order to determine how much of the past information stored in the previous hidden state needs to be retained for the subsequent steps.
\item  The outputs combination first applies an element-wise $\boxed{\bm{1}-}$-operation on the Update vector, and the results undergoes an element-wise multiplication with the unweighted output from the reset gate in order for the update gate to determine which portion of the new information should be stored in the hidden state.
  Finally, this last unweighted result is summed with an output of the update gate, \emph{i.e.} the element-wise multiplication of the Update vector with the unweighted previous hidden state $\bm{h}_{t-1}$, in order to provide the updated hidden state vector $\bm{h}_{t}$ and the GRU output vector $\bm{y}'_{t}=\bm{h}_{t}$, which constitutes the input of the second feed-forward $\text{NNW}_{\text{O}}$, see Fig. \ref{fig:GRU}.
\end{itemize}
  All the weight matrices $\mathbf{W_{\bullet}}$ and bias vectors $\bm{b}_\bullet$ used in Fig. \ref{fig:GRU_detail} are updated when the entire network is trained through back-propagation.
}

\bibliographystyle{elsarticle-num}
\bibliography{manuscript}

\end{document}